\documentclass[12pt]{article}
\pdfoutput=1

\usepackage[utf8]{inputenc}

\usepackage{draft}
\usepackage{putex}
 
 \usepackage[table]{xcolor}
 \usepackage{dsfont}
\usepackage[all]{xy}

\usepackage{stackrel,amssymb}

\usepackage{graphicx}
\usepackage{caption}
\usepackage{amsmath,bm}
\usepackage{array}
\usepackage{subcaption}
\usepackage{epstopdf}
\usepackage{enumerate}
\usepackage{enumitem}
\usepackage{cite}
\usepackage{tensor}
\usepackage{slashed}
\usepackage[utf8]{inputenc}
\usepackage{rotating}
\usepackage{bigfoot}
\usepackage[
      colorlinks=true,
      linkcolor=blue,
      urlcolor=blue,
      filecolor=black,
      citecolor=red,
      ]{hyperref}

\usepackage{tabu}

\def\le{\left(}
\def\ri{\right)}

\def\XXint#1#2#3{{\setbox0=\hbox{$#1{#2#3}{\int}$ }
		\vcenter{\hbox{$#2#3$ }}\kern-.6\wd0}}

\usepackage{adjustbox}
\usepackage{multirow}
\usepackage{makecell}

\usepackage{tikz-cd}
\usetikzlibrary{arrows.meta}
\usetikzlibrary{bending}
\usetikzlibrary{decorations.markings}
\usepackage{pifont}
\usepackage{quiver}
\newcommand{\cmark}{\ding{51}}
\newcommand{\xmark}{\ding{55}}

\numberwithin{equation}{section}

\usepackage[normalem]{ulem}

\newcommand{\thetaa}{x} 
\newcommand{\varthetaa}{u} 
\newcommand{\phii}{v}

\def\<{\langle}
\def\>{\rangle}

\def\pa{\partial}

\def\ie{\begin{equation}\begin{aligned}}
\def\fe{\end{aligned}\end{equation}}

\newcommand{\leftrarrows}{\mathrel{\raise.75ex\hbox{\oalign{
				$\scriptstyle\leftarrow$\cr
				\vrule width0pt height.5ex$\hfil\scriptstyle\relbar$\cr}}}}
\newcommand{\lrightarrows}{\mathrel{\raise.75ex\hbox{\oalign{
				$\scriptstyle\relbar$\hfil\cr
				$\scriptstyle\vrule width0pt height.5ex\smash\rightarrow$\cr}}}}
\newcommand{\Rrelbar}{\mathrel{\raise.75ex\hbox{\oalign{
				$\scriptstyle\relbar$\cr
				\vrule width0pt height.5ex$\scriptstyle\relbar$}}}}

\makeatletter
\def\leftrightarrowsfill@{\arrowfill@\leftrarrows\Rrelbar\lrightarrows}
\newcommand{\xleftrightarrows}[2][]{\ext@arrow 3399\leftrightarrowsfill@{#1}{#2}}
\makeatother

\renewcommand{\title}[1]{\vbox{\center\LARGE{#1}}\vspace{5mm}}
\renewcommand{\author}[1]{\vbox{\center#1}\vspace{5mm}}

\newcommand{\email}[1]{\vbox{\center\tt#1}\vspace{5mm}}

\begin{document}

\begin{titlepage}

	  \begin{flushright}
	 \end{flushright} 

\begin{center}
\vspace{-16pt}
\title{\LARGE Chiral Tube Algebras I:
\\
\Large Topological Defect Lines, Twisted Modules, and Finite Gauging
}

~\vskip 0.01 in

\author{
	Nathan Benjamin$^{a}$, Ho Tat Lam$^{a, b, c}$, Conghuan Luo$^{a}$
}
${}^a$\emph{\small Department of Physics and Astronomy, \\ University of Southern California, Los Angeles, CA 90089, USA}
\\
${}^b$\emph{\small Center for Theoretical Physics -- a Leinweber Institute, \\
Massachusetts Institute of Technology,
 Cambridge, MA 02139, USA}
\\
${}^c$\emph{\small Leinweber Institute for Theoretical Physics, 
Stanford University, Stanford, CA 94305, USA}
~\vskip .2 in
\email{
	nathanbe@usc.edu, hotatlam@usc.edu, luocongh@usc.edu
}

\end{center}

\begin{center}
{\bf Abstract}
\end{center}

Chiral algebras and topological defect lines (TDLs) represent two complementary notions of symmetry in 2d conformal field theories. In this paper, we introduce chiral tube algebras to unify and extend these two notions. 
Chiral tube algebras generalize chiral algebras in two ways. First, they extend the action of chiral algebras beyond the local Hilbert space to include defect Hilbert spaces twisted by TDLs. Second, they allow for non-local chiral currents attached by TDLs and thus can map between different defect Hilbert spaces, analogous to the tube algebras of TDLs. Since local chiral currents can become non-local after finite gauging, chiral tube algebras provide a natural framework for describing the image of chiral algebras under such gauging.
We illustrate this framework through a variety of examples that generalize familiar chiral algebras, including Kac-Moody algebras, $\mathcal{W}$ algebras, superconformal algebras, and their orbifolds/bosonizations. We construct their irreducible modules, which are isomorphic to twisted modules of the corresponding chiral algebras, and use them to organize local and defect Hilbert spaces.
In a subsequent paper, we will study chiral tube algebras generated by non-local chiral currents with fractional spins, which have no counterparts in chiral algebras. 

\vfill

\end{titlepage}

\tableofcontents

\section{Introduction and Summary}

Chiral algebras\cite{Belavin:1984vu,zamolodchikov1995infinite} are the cornerstones of 2d conformal field theories (CFTs). They are generated by local chiral currents and encode an infinite number of symmetries that underlie many exact solutions of 2d CFTs. Topological defect lines (TDLs) \cite{Verlinde:1988sn,Petkova:2000ip,Frohlich:2004ef,Frohlich:2006ch,Drukker:2010jp}, on the other hand, furnish a different class of symmetries that generalizes ordinary group symmetries to non-invertible symmetries \cite{Feiguin:2006ydp,Aasen:2016dop,Bhardwaj:2017xup,Chang:2018iay}. Their similarities and differences are summarized in Table~\ref{tab:chiral_algebra_vs_TDL}, which we will further elaborate in the coming subsections.

\begin{table}[!ht]
    \centering
    \begin{tabular}{|c|c|c|c|}
    \hline
        & Chiral Algebra & TDL & Chiral Tube Algebra\\
        \hline
        &&&\\[-0.8em]
       chiral currents $V(z)$& local& \xmark & local $\&$ non-local\\
       &&&\\[-0.7em]
       \makecell{topological \\ symmetry operators} & $V_n$, $n\in\mathbb{Z}$ & $\mathcal{L}$ & $V_{n+\thetaa}$, $n\in\mathbb{Z}$\\
        &&&\\[-0.7em]
       spacetime dependence? & \cmark & \xmark & \cmark\\
       &&&\\[-0.8em]
       commute with $T(z)$? & \xmark & \cmark & \xmark\\
       &&&\\[-0.8em]
       twisted Hilbert space? & \xmark & $\cH_\cL$ & \xmark\\
       &&&\\[-0.9em]
       symmetry action & \ \ \ \tikz[baseline=0]{\draw (0,0) circle (0.75);  
       \node[right] at (0.55,-0.5) {\small $V_n$};
       \node at (0,-1.1) {$\cH\rightarrow\cH$};} 
       &  \tikz[baseline=0]{\draw (0,0)--(0,1.5); \draw (0,0) circle (0.75); \node[right] at (-0.0,0.35) {\small $\cL'$}; \draw[black, -{Stealth[round, length=5pt, width=5pt, bend]}] (0,0)--(0,0.4); \draw[black, -{Stealth[round, length=5pt, width=5pt, bend]}] (0,0)--(0,1.2); \node[right] at (-0.0,1.1) {\small $\cL''$}; \draw[black,-{Stealth[round, length=5pt, width=5pt, bend]}] (0,-0.75) arc (-90:10:0.75); \node[right] at (0.55,-0.5) {\small $\cL$}; 
       \node at (0.05,-1.1) {$\cH_{\cL'}\rightarrow\cH_{\cL''}$} }
       & \quad\ \ \ \ \tikz[baseline=0]{\draw (0,0)--(0,1.5); \draw (0,0) circle (0.75); \node[right] at (-0.0,0.35) {\small $\cL'$}; \draw[black, -{Stealth[round, length=5pt, width=5pt, bend]}] (0,0)--(0,0.4); \draw[black, -{Stealth[round, length=5pt, width=5pt, bend]}] (0,0)--(0,1.2); \node[right] at (-0.0,1.1) {\small $\cL''$}; \draw[black,-{}] (0,-0.75) arc (-90:0:0.75); \node[right] at (0.55,-0.5) {\small $V_{n+x} P_x$}; 
       \node at (0.05,-1.1) {$\cH_{\cL'}\rightarrow\cH_{\cL''}$}} \\
       &&&\\[-0.8em]
       \makecell{preserve\\ conformal weight?}  & \xmark & \cmark & \xmark\\
       &&&\\[-0.9em]
       mathematical structure & \makecell{vertex operator\\ algebra} & \makecell{semisimple\\tensor category}& $\mathbf{???}$\\
       \hline
    \end{tabular}
    \caption{Comparisons between ordinary chiral algebras, TDLs and chiral tube algebras. Here, the symmetry charge $V_n$ are defined as $V_n=\oint \frac{dz}{2\pi i}  V(z) z^{n+h-1}$.
    }
    \label{tab:chiral_algebra_vs_TDL}
\end{table}

The goal of this paper is to investigate the interplay between these two notions of symmetry and combine them into a unifying symmetry structure. In particular, we will address the following three closely related questions:
\begin{itemize}[leftmargin=12pt]
\item How does a chiral algebra act on defect Hilbert spaces twisted by TDLs?
\item What happens to a chiral algebra under finite gauging?
\item Can non-local chiral currents attached by TDLs give rise to a chiral algebra?
\end{itemize}
In answering these questions, a natural unifying symmetry structure emerges, which we refer to as ``chiral tube algebras". 

Chiral tube algebras generalize ordinary chiral algebras by extending their action beyond the local Hilbert space to include defect Hilbert spaces. In this sense, they are analogous to the tube algebras associated with TDLs~\cite{ocneanu1994chirality,Lin:2022dhv}, which describe how TDLs act on both local and defect Hilbert spaces, and hence the name ``chiral tube algebras".
Because of this extended action, chiral tube algebras can incorporate both local chiral currents and non-local chiral currents, thereby expanding the scope of ordinary chiral algebras. In particular, unlike ordinary chiral algebras, chiral tube algebras generated by non-local chiral currents necessarily mix the local and defect Hilbert spaces, similar to the tube algebras of non-invertible TDLs. Since local chiral currents often become non-local under finite gauging, chiral tube algebras also provide a natural framework for describing the image of chiral algebras under finite gauging. In Table~\ref{tab:chiral_algebra_vs_TDL}, we compare chiral tube algebras with ordinary chiral algebras and TDLs. We will explain the details in the following subsections.

Before introducing the formalism of chiral tube algebras, we will briefly review the relevant aspects of TDLs and chiral algebras.

\subsection{Topological Defect Lines and Tube Algebras}
\label{subsec:tdls}

Topological defects are natural generalizations of ordinary global symmetries\cite{Gaiotto:2014kfa}. They are invariant under smooth deformations of their underlying submanifolds and commute with the stress tensor, thereby defining conserved quantities and generalized global symmetries. 

An ordinary global symmetry with symmetry group $G$ is described by a set of co-dimension 1 topological defects $U_g$ labeled by group elements $g\in G$, whose fusions obey the group multiplication law,
\ie
U_g\times U_{g'} = U_{gg'} \,.
\fe
In particular, every topological defect $U_g$ has an inverse given by its orientation reversal $\overline{U_g}=U_{g^{-1}}$, such that $U_g\times U_{g^{-1}}=I$, where $I$ is the identity defect. Sweeping a topological defect $U_g$ past a local point operator $\cO$ induces a symmetry transformation associated with the group element $g$, which maps $\cO$ to another local point operator $g\cdot\cO$.

In 2d, co-dimension 1 topological defects are topological defect lines (TDLs). Their fusions, however, do not always obey a group multiplication law. Instead, they take a more general form,
\ie
\cL_i \times \cL_j = \sum_k N_{ij}^k \cL_k \,,
\label{eq:nijkdef}
\fe
where $N_{ij}^k\in \mathbb{N}$ are known as the fusion coefficients. A particular consequence of these more general fusions is that a TDL $\cL$ may not have an inverse $\cL^{-1}$, such that $\mathcal{L}\times \cL^{-1}=I$. Such TDLs are called non-invertible TDLs and they define non-invertible generalizations of ordinary global symmetries \cite{Feiguin:2006ydp,Aasen:2016dop,Bhardwaj:2017xup,Chang:2018iay}.

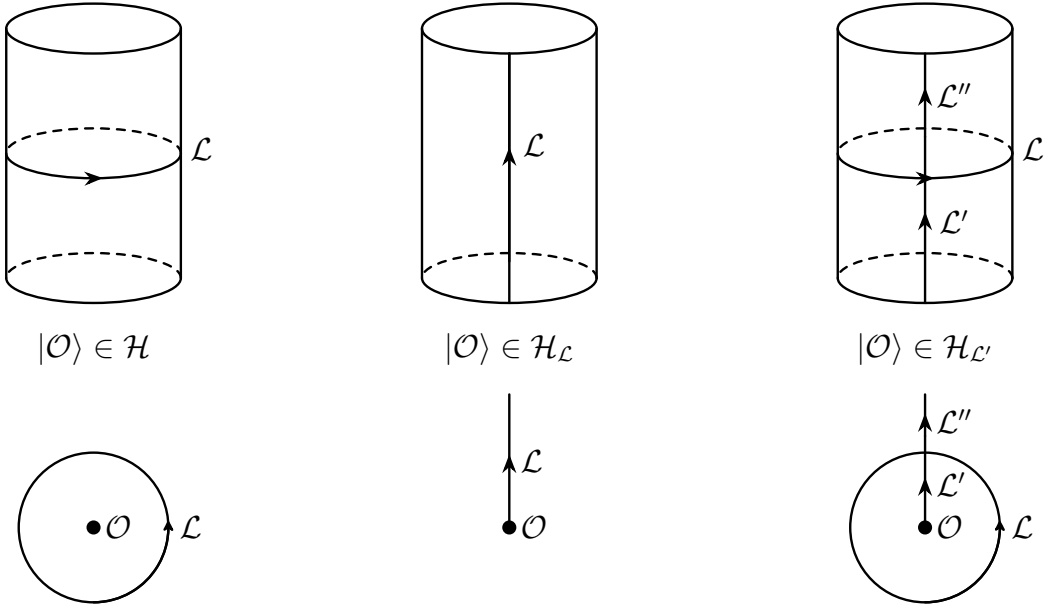
\begin{figure}
    \centering

\usetikzlibrary{decorations.markings,arrows.meta}

\vspace{-10pt}

\begin{tikzpicture}[
    scale=1.1,
    line cap=round,
    line join=round,
    line width=0.9pt,
    >={Stealth[length=5pt,width=5pt]},
    midarrow/.style={
        postaction={decorate},
        decoration={
            markings,
            mark=at position .55 with {\arrow{Stealth}}
        }
    }
]

\def\rx{1.05}
\def\ry{0.3}
\def\H{3}

\begin{scope}[shift={(0,0)}]
    \draw (-\rx,0) -- (-\rx,\H);
    \draw ( \rx,0) -- ( \rx,\H);

    \draw (0,\H) ellipse [x radius=\rx, y radius=\ry];

    \draw[dashed] (\rx,0) arc[start angle=0,end angle=180,
        x radius=\rx,y radius=\ry];
    \draw (-\rx,0) arc[start angle=180,end angle=360,
        x radius=\rx,y radius=\ry];

    \def\ymid{1.5}
    \draw[dashed] (\rx,\ymid) arc[start angle=0,end angle=180,
        x radius=\rx,y radius=\ry];
    \draw[midarrow] (-\rx,\ymid) arc[start angle=180,end angle=360,
        x radius=\rx,y radius=\ry];

    \node at (1.3,\ymid+0.05) {$\cL$};
    \node at (0,-0.85) {$|\cO\rangle \in \mathcal H$};
    
    \draw (0,-3) circle (0.9);
    \draw[black,-{Stealth[round, length=5pt, width=5pt, bend]}] (0,-3.9) arc (-90:5:0.9);
    \node[right] at (0.9,-3) {$\mathcal{L}$};
    \filldraw[black] (0,-3) circle (2pt) node[right] {$\mathcal{O}$};
\end{scope}

\begin{scope}[shift={(5,0)}]
    \draw (-\rx,0) -- (-\rx,\H);
    \draw ( \rx,0) -- ( \rx,\H);

    \draw (0,\H) ellipse [x radius=\rx, y radius=\ry];

    \draw[dashed] (\rx,0) arc[start angle=0,end angle=180,
        x radius=\rx,y radius=\ry];
    \draw (-\rx,0) arc[start angle=180,end angle=360,
        x radius=\rx,y radius=\ry];

    \draw[midarrow] (0,-\ry+0.5) -- (0,\H-\ry);
    \draw (0,-\ry) -- (0,\H-\ry);

    \node at (0.3,1.6) {$\cL$};
    \node at (0,-0.85) {$|\cO\rangle \in \mathcal H_\cL$};

    \draw[midarrow] (0,-3) -- (0,-1.4);
    \node[right] at (0,-2.2) {$\mathcal{L}$};
    \filldraw[black] (0,-3) circle (2pt) node[right] {$\mathcal{O}$};
\end{scope}

\begin{scope}[shift={(10,0)}]
    \draw (-\rx,0) -- (-\rx,\H);
    \draw ( \rx,0) -- ( \rx,\H);

    \draw (0,\H) ellipse [x radius=\rx, y radius=\ry];

    \draw[dashed] (\rx,0) arc[start angle=0,end angle=180,
        x radius=\rx,y radius=\ry];
    \draw (-\rx,0) arc[start angle=180,end angle=360,
        x radius=\rx,y radius=\ry];

    \draw[midarrow] (0,-\ry) -- (0, 1.7);
    \draw[midarrow] (0, 1.8) -- (0,\H-\ry);
    \draw (0, 1.8) -- (0,1.7);

    \def\ymid{1.5}
    \draw[dashed] (\rx,\ymid) arc[start angle=0,end angle=180,
        x radius=\rx,y radius=\ry];
    \draw[midarrow] (-\rx,\ymid) arc[start angle=180,end angle=360,
        x radius=\rx,y radius=\ry];

    \node at (0.35,0.68) {$\cL'$};
     \node at (0.35,2.2) {$\cL''$};
    \node at (1.3,\ymid+0.05) {$\cL$};
    \node at (0,-0.85) {$|\cO\rangle \in \mathcal H_{\cL'}$};

    \draw[midarrow] (0,-3) -- (0,-1.9);
    \draw[midarrow] (0,-1.9) -- (0,-1.4);
    \node[right] at (0.0,-2.48) {$\mathcal{L}'$};
    \node[right] at (0.0,-1.72) {$\mathcal{L}''$};
    \node[right] at (0.9,-3) {$\mathcal{L}$};
    \filldraw[black] (0,-3) circle (2pt) node[right] {$\mathcal{O}$};
    \draw (0,-3) circle (0.9);
    \draw[black,-{Stealth[round, length=5pt, width=5pt, bend]}] (0,-3.9) arc (-90:5:0.9);
\end{scope}

\end{tikzpicture}
\vspace{3pt}\\
    \caption{The left figure is the interpretation of a TDL as an operator acting on the local Hilbert space $\cH$. The middle figure is the interpretation of a TDL as a defect that twists the boundary condition and defines the defect Hilbert space $\cH_\cL$. The right figure illustrates how a TDL maps between different defect Hilbert spaces. The bottom row is related to the top row by the conformal map $z=e^{w}$, where $z$ is the complex coordinate on the plane and $w=\tau+i\phi$, with $\tau$ the Euclidean time and $\phi$ the angular coordinate on the cylinder. }
    \label{fig:TDL_interpretation}
\end{figure}

Given the notion of TDLs, there are two different interpretations (see Figure~\ref{fig:TDL_interpretation} for an illustration). If one inserts a TDL $\cL$ along the spatial direction, it can be interpreted as a  conserved operator $\cL$ that acts on the Hilbert space $\cH$ defined on the spatial slice. In 2d CFTs, a state $|\mathcal{O}\ra$ in the Hilbert space $\cH$ defined on a circle is dual to a local point operator $\cO$ by the state-operator correspondence. We will therefore refer to $\cH$ as the local Hilbert space. Under the state-operator conformal map, the TDL $\cL$ acting on $\cH$ is mapped to a TDL encircling the local point operator $\cO$ as illustrated in Figure~\ref{fig:TDL_interpretation}. Shrinking the TDL towards $\cO$ then produces the image of $\cO$ under the TDL action. If one instead inserts the TDL $\cL$ along the time direction, it can be interpreted as a defect that defines a defect Hilbert space $\cH_\cL$ with a twisted boundary condition. In 2d CFTs, by state-operator correspondence, a state $|\mathcal{O}\ra$ in the defect Hilbert space $\cH_\cL$ is dual to a non-local point operator $O$ attached by the TDL $\cL$. We will refer to such non-local point operators as defect operators. 

The Hilbert space interpretation imposes a stringent constraint on TDLs. In particular,
given a set of TDLs $\cL_i$ that admit well-defined defect Hilbert spaces $\cH_{\cL_i}$, only non-negative integer linear combinations of them,
\ie
\cL = \sum n_i \cL_i \,, \quad n_i \in \mathbb{N} \,,
\label{eqn:tdl_decomposition}
\fe
admit well-defined defect Hilbert spaces. The corresponding defect Hilbert space $\cH_{\cL}$ is naturally given by the direct sum, 
\ie
\cH_\cL = {\bigoplus} \,n_i \cH_{\cL_i} \,, \quad n_i \in \mathbb{N} \,.
\fe

The non-negative integral decomposition \eqref{eqn:tdl_decomposition} naturally leads to the notion of simple TDLs, which are TDLs that cannot be decomposed further. TDLs that can be written as finite direct sums of simple TDLs are called semisimple.
Mathematically, TDLs of this kind are described by semisimple tensor categories (see~\cite{etingof2015tensor} for a comprehensive review). An important piece of data contained in the semisimple tensor category of TDLs is the $F$-symbols, which implement the associativity isomorphisms between different fusion channels:
\vspace{-4pt}
\ie
\begin{tikzpicture}[
    scale=0.7,
    dline/.style={
        line width=1.2pt, 
        color=black, 
        postaction={decorate}, 
        decoration={markings, mark=at position 0.55 with {\arrow{stealth}}}
    },
    dot/.style={
        circle, 
        fill=black, 
        inner sep=0pt, 
        minimum size=4.5pt
    }
]

    \coordinate (L_i) at (-2, -2.4);
    \coordinate (L_j) at (0, -2.4);
    \coordinate (L_k) at (2, -2.4);
    \coordinate (V_jk) at (1, -1.2);
    \coordinate (V_ip) at (0, 0);
    \coordinate (L_ell) at (1, 1.2);

    \draw[dline] (L_i) -- (V_ip);
    \draw[dline] (L_j) -- (V_jk);
    \draw[dline] (L_k) -- (V_jk);
    \draw[dline] (V_jk) -- (V_ip);
    \draw[dline] (V_ip) -- (L_ell);

    \node[dot] at (V_jk) {};
    \node[dot] at (V_ip) {};

    \node[below] at (L_i) {$\mathcal{L}_i$};
    \node[below] at (L_j) {$\mathcal{L}_j$};
    \node[below] at (L_k) {$\mathcal{L}_k$};
    \node[right,xshift=-2pt,yshift=1pt] at (L_ell) {$\mathcal{L}_\ell$};
    \node[left, xshift=-2pt] at (1.6, -0.3) {$\mathcal{L}_p$};
    
    \node at (4.2, -0.7) {
       $\displaystyle = \sum_{q}  \left[ F_{ijk}^\ell \right]_{p,q}$
    };

    \coordinate (R_i) at (6, -2.4);
    \coordinate (R_j) at (8, -2.4);
    \coordinate (R_k) at (10, -2.4);
    \coordinate (V_ij) at (7, -1.2);
    \coordinate (V_qk) at (8, 0);
    \coordinate (R_ell) at (9, 1.2);

    \draw[dline] (R_i) -- (V_ij);
    \draw[dline] (R_j) -- (V_ij);
    \draw[dline] (R_k) -- (V_qk);
    \draw[dline] (V_ij) -- (V_qk);
    \draw[dline] (V_qk) -- (R_ell);

    \node[dot] at (V_ij) {};
    \node[dot] at (V_qk) {};

    \node[below] at (R_i) {$\mathcal{L}_i$};
    \node[below] at (R_j) {$\mathcal{L}_j$};
    \node[below] at (R_k) {$\mathcal{L}_k$};
    \node[right,xshift=-2pt,yshift=1pt] at (R_ell) { $\mathcal{L}_\ell$};
    \node[right, xshift=2pt] at (6.35, -0.3) {$\mathcal{L}_q$};

\end{tikzpicture}
\fe
When the number of simple TDLs is finite, the semisimple tensor category is called a fusion category. The TDLs discussed in most of this paper are described by fusion categories, except in Section \ref{sec:SU(2)}, which discusses continuous symmetries that contain infinitely many TDLs.

A simple TDL $\cL$ has a property that when it fuses with its orientation reversal $\bar \cL$, the fusion always includes the identity line $I$ with multiplicity 1,
\ie
\cL\times \bar \cL = I +\cdots\,.
\fe
If the fusion contains only the identity line, the TDL $\cL$ is invertible and has $\bar \cL$ as its inverse. On the other hand, if the fusion contains more lines, the TDL $\cL$ is non-invertible. A canonical example of non-invertible TDL is given by the duality line $\cN$ in Ising CFT \cite{Oshikawa:1996dj,Frohlich:2004ef,Frohlich:2006ch}, which implements the Kramers-Wannier duality \cite{Kramers:1941kn} and obeys the following non-invertible fusion,
\ie
\cN\times\cN = I+\eta\,.
\fe
Here, $\cN$ is its own orientation reversal and $\eta$ is the TDL associated with the $\mathbb{Z}_2$ symmetry in the Ising CFT, which obeys a $\mathbb{Z}_2$ fusion $\eta\times\eta=I$.

A key characteristic of non-invertible TDLs is that they map local operators into defect operators \cite{Chang:2018iay}. Recall that a TDL $\cL$ implements a symmetry transformation by sweeping past the local operator $\cO$. Sweeping a non-invertible TDL $\cL$ past a local operator $\cO$ can generate non-local operators $\cO_i$ attached by TDLs $\cL_i$ that appear in the fusion $\cL\times\bar{\cL}$\,:
\ie
\begin{tikzpicture}[
    arrowline/.style={
        thick, 
        postaction={decorate}, 
        decoration={markings, mark=at position #1 with {\arrow{stealth}}}
    },
    dashedarrow/.style={
        thick, 
        dashed,
        postaction={decorate}, 
        decoration={markings, mark=at position #1 with {\arrow{stealth}}}
    },
    dot/.style={
        circle, 
        fill=black, 
        inner sep=0pt, 
        minimum size=4pt
    }
]

\begin{scope}[shift={(0,-0.4)}]
\begin{scope}[rotate=90]
    \draw[thick,arrowline=0.55] (-0.4,1.5) -- (-0.4,-1.5);
    \node[below] at (-0.4,-1.5) {$\mathcal{L}$};

    \node[dot] (O1) at (0.4,0) {};
    \node[right,yshift=2pt] at (O1) {$\cO$};
\end{scope}
\end{scope}

\node at (2.4,0) {$=$};

\begin{scope}[shift={(4.8,0.8)}]
\begin{scope}[rotate=90]

    \draw[thick,arrowline=0.55] (0.6,1.5) -- (0.6,0.1);
    \draw[thick,arrowline=0.55] (0.6,-0.1) -- (0.6,-1.5);
    \node[below] at (0.6,-1.5) {$\mathcal{L}$};

     \draw[thick,arrowline=0.55] (0.6,0.1) -- (-0.2,0.1);
     \draw[thick,arrowline=0.55] (-0.2,-0.1) -- (0.6,-0.1);

     \draw[thick] (-1.4,0) arc (180:8:0.6);

     \draw[thick] (-1.4,0) arc (-180:-8:0.6);

     \draw[thick,-{Stealth[round, length=5pt, width=5pt, bend]}] (-1.4,0) arc (-180:-85:0.6);

    \node[dot] (O2) at (-0.8,0) {};
    \node[right,yshift=2pt] at (O2) {$\cO$};
\end{scope}
\end{scope}

\node at (7.8,0) {$=$\quad \Large$\sum\limits_i$};

\begin{scope}[shift={(10.6,0.8)}]
\begin{scope}[rotate=90]

    \node[dot] (Oi) at (-0.8,0) {};
    \node[right,yshift=2pt] at (Oi) {$\cO_i$};

    \draw[thick,dashedarrow=0.5] (-0.6,0) -- (0.6,0);
    \node[right] at (-0.05,0) {$\mathcal{L}_i$};

    \draw[thick,arrowline=0.55] (0.6,1.5) -- (0.6,-1.5);
    \node[below] at (0.6,-1.5) {$\mathcal{L}$};
\end{scope}
\end{scope}
\label{eq:non-invertible_TDL_action}
\end{tikzpicture}
\fe
An example of how a non-invertible TDL acts is given by the duality line in the Ising CFT, which maps the local order operator $\sigma$ to the non-local disorder operator $\mu$\,:
\ie
\begin{tikzpicture}[
    arrowline/.style={
        thick, 
        postaction={decorate}, 
        decoration={markings, mark=at position #1 with {\arrow{stealth}}}
    },
    dashedarrow/.style={
        thick, 
        dashed,
        postaction={decorate}, 
        decoration={markings, mark=at position #1 with {\arrow{stealth}}}
    },
    dot/.style={
        circle, 
        fill=black, 
        inner sep=0pt, 
        minimum size=4pt
    }
]

\begin{scope}[shift={(0,-0.4)}]
\begin{scope}[rotate=90]
    \draw[thick,arrowline=0.55] (-0.4,1.5) -- (-0.4,-1.5);
    \node[below] at (-0.4,-1.5) {$\mathcal{N}$};

    \node[dot] (O1) at (0.4,0) {};
    \node[right,yshift=2pt] at (O1) {$\sigma$};
\end{scope}
\end{scope}

\node at (2.4,0) {$=$};

\begin{scope}[shift={(4.8,0.8)}]
\begin{scope}[rotate=90]

    \node[dot] (Oi) at (-0.8,0) {};
    \node[right,yshift=2pt] at (Oi) {$\mu$};

    \draw[thick,dashedarrow=0.5] (-0.6,0) -- (0.6,0);
    \node[right] at (-0.05,0) {$\eta$};

    \draw[thick,arrowline=0.55] (0.6,1.5) -- (0.6,0);
    \draw[thick,arrowline=0.55] (0.6,0) -- (0.6,-1.5);
    \node[below] at (0.6,-1.5) {$\mathcal{N}$};
\end{scope}
\end{scope}

\end{tikzpicture}
\fe

More generally, a TDL, either invertible or non-invertible, can map between different defect Hilbert spaces (see the illustration in Figure \ref{fig:TDL_interpretation}).
By the state-operator conformal map, the symmetry action of $\mathcal{L}$ that maps $\cH_{\cL'}$ to $\cH_{\cL''}$ is implemented by the following lasso operator,
\ie
  \tikz{
  \draw[thick] (0,) -- (0,1.6);
  \draw[thick,-{Stealth[round, length=5pt, width=5pt, bend]}] (0,0) -- (0,0.6);
    \draw[thick,-{Stealth[round, length=5pt, width=5pt, bend]}] (0,0.6) -- (0,1.3);
    \node[right] at (0.0,3-2.48) {$\mathcal{L}'$};
    \node[right] at (0.0,3-1.72) {$\mathcal{L}''$};
    \node[right] at (0.9,3-3) {$\mathcal{L}$};
    \draw[thick] (0,3-3) circle (0.9);
    \draw[thick,-{Stealth[round, length=5pt, width=5pt, bend]}] (0,3-3.9) arc (-90:5:0.9);}
    \label{eq:lasso_operator_tube}
\fe
where the four-way junction is labeled by the different fusion channels from $\cL'\times \cL$  to $\cL\times \cL''$. 
Mathematically, these lasso operators form an algebra known as the tube algebra ${\rm Tube}(\cC)$ associated with the fusion category $\cC$ that describes the TDLs\cite{ocneanu1994chirality,Lin:2022dhv,Bartsch:2023wvv}.
The representations of this algebra are called tube representations, which form a modular tensor category ${\rm Rep}({\rm Tube}(\cC))$ that is braided equivalent to the Drinfeld center of the fusion category $Z(\cC)$\cite{evans1995ocneanu,Izumi:2000qa,Mueger:2001crc}. In physics terms, the Drinfeld center is the set of anyons in the 3d SymTFT \cite{Gaiotto:2014kfa,kong2015boundarybulkrelationtopologicalorders,Kong_2017,Pulmann:2019vrw,Thorngren:2019iar,Ji:2019jhk,Lichtman:2020nuw,Kong_2020,Gaiotto:2020iye,Aasen:2020jwb,Apruzzi:2021nmk,Burbano:2021loy,Chatterjee:2022kxb,Freed:2022qnc,Kaidi:2022cpf} of the TDLs, which is defined as the Turaev-Viro topological quantum field theory~\cite{TURAEV1992865,Barrett:1993ab,kirillov2010turaevviroinvariantsextendedtqft} associated with fusion category $\mathcal{C}$. The tube representations are in one-to-one correspondence with these anyons~\cite{Lin:2022dhv,Bartsch:2023wvv,Bhardwaj:2023ayw}.

\subsection{Chiral Algebras as Holomorphic Symmetries}
We now turn to review chiral algebras.\footnote{In the mathematics literature, chiral algebras are usually formulated as vertex operator algebras (VOAs). Throughout this paper, we follow the common physics terminology and use the term ``chiral algebra'' synonymously with VOA.
} In particular, we will emphasize the similarities and distinctions between chiral algebras and TDLs as global symmetries.

A chiral algebra is generated by local chiral currents $V^a(z)$ that depend on only the holomorphic coordinate $z$. In unitary compact CFTs, these chiral currents have conformal weight $(h,\bar h)=(h^a,0)$ with $h^a\in\mathbb{Z}$ in bosonic theories and $h^a\in\frac{1}{2}\mathbb{Z}$ in fermionic theories.\footnote{In a non-unitary CFT, a holomorphic operator $V(z)$ does not necessarily have $\bar{h} = 0$. Conversely, an operator $O(z,\bar{z})$ with conformal weight $\bar{h}=0$ is not necessarily holomorphic, i.e.~it may not satisfy $\bar\pa O(z,\bar z) = 0$. Because in this paper we focus on unitary compact CFTs, we will not worry about this distinction. 
} They are closed under the operator product expansion (OPE),
\ie
V^a(z) V^b(w)=\sum_c \frac{C^{ab}{}_c\, V^c(w)}{(z-w)^{h_a+h_b-h_c}}\,.
\label{eqn:chiral_algebra}
\fe
The OPE of two chiral currents cannot contain non-chiral operators, because otherwise, there would be a factor of $(\bar z-\bar w)^{\bar h_c}$ on the right hand side, contradicting with $\bar\partial V^a(z)=0$. This closed subalgebra generated by chiral currents is called the chiral algebra.

Each chiral current results in an infinite set of continuous symmetries, with the associated symmetry charges given by the Laurent modes of the chiral current,
\ie
V^a_n=\oint_\gamma \frac{dz}{2\pi i}\, z^{n+h_a-1}V^a(z)\,,\quad V^a(z)=\sum_n\frac{V^a_n}{z^{n+h_a}}\,.
\fe
Similar to TDLs, these symmetry charges are topological in the sense that they are invariant under smooth deformation of the integration contour $\gamma$. However, an important difference is that they carry explicit spacetime dependence through the factor of $z^{n+h_a-1}$, so the symmetries they generate are holomorphic symmetries, which can be parametrized by a holomorphic function $f(z)$.\footnote{This is reminiscent of modulated symmetries \cite{Gromov:2018nbv,sala2021dynamicssystemsmodulatedsymmetries,Gorantla:2022eem,Pace:2024tgk,Pace:2025hpb} in lattice systems and non-relativistic field theories. The difference is that modulated symmetries generally depend only on space and are independent of time, while the symmetries generated by chiral algebras depend on the holomorphic coordinate $z$.} A consequence of this explicit spacetime dependence is that, although the symmetry charges $V^a_n$ are topological, they may not commute with the stress tensor. This is to be contrasted with TDLs, which by definition commute with the stress tensor. As we will discuss below, this distinction leads to an important difference between the representations of chiral algebras and TDLs. Another distinction is that unlike TDLs, the symmetry charges of chiral algebras do not admit a defect Hilbert space interpretation. These similarities and distinctions are summarized in Table~\ref{tab:chiral_algebra_vs_TDL}.

From the OPEs of chiral currents \eqref{eqn:chiral_algebra}, we can derive a mode algebra formed by the symmetry charges $V^a_n$. This mode algebra acts naturally on the local Hilbert space $\cH$, which can therefore be organized into its representations/modules.
In the physics literature, the term ``chiral algebra'' is often used interchangeably for both the OPE algebra of chiral currents and the corresponding mode algebra. We will adopt the same terminology. 

Let us illustrate the above discussion using the stress tensor in a 2d CFT, which has a holomorphic component $T(z)$ and an anti-holomorphic component $\bar T(\bar z)$. The holomorphic component $T(z)$ has conformal weight $(h,\bar h)=(2,0)$, and its OPE with itself is given by
\ie
T(z) T(w) \sim \frac{c}{2(z-w)^4} + \frac{2T(w)}{(z-w)^2} + \frac{\pa T(w)}{z-w} \,.
\fe
The symmetry charges associated with $T(z)$ are given by its Laurent modes,
\ie
L_n = \oint\frac{dz}{2\pi i} z^{n+1} T(z)\,,\quad T(z)=\sum_n\frac{L_n}{z^{n+2}}\,.
\fe
They generate the conformal transformation $z\mapsto f(z)$ and form the Virasoro algebra with central charge $c$,
\ie
{[L_n,L_m]} = (n-m) L_{n+m} + \frac{c}{12}n(n^2-1) \delta_{n+m,0} \,.
\fe
Similarly, the anti-holomorphic component $\bar T(\bar z)$ of the stress tensor gives rise to an anti-chiral copy of the Virasoro algebra, generated by the charges $\bar L_n$. 

Local operators can be organized into modules of the chiral and anti-chiral Virasoro algebras. Physically, we are interested in highest weight modules, whose highest weight states $|\cO\ra$ have the lowest conformal weights within their modules and are therefore annihilated by the modes that would lower its conformal weights. Since 
\ie
{[L_0,L_{n}]}=-nL_{n}\,,
\fe
the modes $L_{n}$ shift the conformal weights $h$ by $-n$. A highest weight state must then satisfy
\ie
L_0 |\cO\ra= h |\cO\ra\,,\quad
L_n |\cO\ra= 0 \,, \quad \forall\ n>0 \,.
\fe
By the state-operator correspondence, such a state corresponds to a Virasoro primary operator $\cO$. The full Virasoro module is generated by acting on $|\cO\ra$ with the lowering modes $L_{-n}$ for $n>0$. The resulting states correspond to descendant operators, whose conformal weights are higher than that of the primary operator $\cO$. Some of these descendants appear in the OPE between the stress tensor $T(z)$ and the primary operator $\cO$,
\ie
T(z)\cO(w)=\frac{h \cO(w)}{(z-w)^2}+\frac{\partial\cO(w)}{z-w}+\sum_{n\geq2}(z-w)^{n-2}{\left(L_{-n}\cO\right)\!(w)}\,.
\fe

Importantly, this example illustrates that the symmetry charges $V_n^a$ of a chiral algebra generally shift the conformal weights, unlike TDLs, which preserve them. In this sense, chiral algebras are spectrum generating symmetries. Their modules are generally infinite-dimensional and include states of different conformal weights. This is the consequence of the explicit spacetime dependence of the symmetry charges.\footnote{See \cite{Apruzzi:2025hvs} for a similar discussion for the global conformal group.}

Let us explain this in more details. Consider the CFT on a cylinder with Euclidean time $\tau$ and angular coordinate $\phi$. The cylinder is related to the complex plane by the conformal transformation $z=e^{\tau+i\phi}$.
For a chiral current $V^a(z)$ that is a Virasoro primary, the corresponding symmetry charges on the cylinder are given by
\ie
V^a_n=\oint\frac{d\phi}{2\pi } \,V^a(\tau,\phi)\,e^{n(\tau+i\phi)}\,.
\fe
Here, we used the state-operator conformal map $z=e^{\tau+i\phi}$, under which the chiral current transforms as $V^a(\tau,\phi)=z^{h_a} V^a(z)$.
The stress tensor zero modes $L_0$ and $\bar L_0$ are related to the Hamiltonian $H$ and angular momentum $J$ on the cylinder by 
\ie
L_0=\frac{1}{2}(H+J)+\frac{c}{24}\,,\qquad \bar L_0=\frac{1}{2}(H-J)+\frac{c}{24}\,.
\fe
Since the symmetry charges $V_n^a$ are invariant under smooth deformations of the integration contour, we have the following Heisenberg equation of motion
\ie
\left(-\frac{d}{d\tau} +i\frac{d}{d\phi}\right)V_n^a= 2[V_n^a,L_0]+\left(-\frac{\partial}{\partial\tau} +i\frac{\partial}{\partial\phi}\right) V_n^a=0\,,
\fe
where the second term comes from the explicit spacetime dependence of the charges. This explicit dependence leads to a non-trivial commutator between $L_0$ and $V_n^a$,
\ie
{[L_0,V_n^a]}=\frac{1}{2}\left(-\frac{\partial}{\partial\tau} +i\frac{\partial}{\partial\phi}\right) V_n^a=-n V_n^a\,,
\fe
which exactly reproduces the expectation that $V_n^a$ shifts the conformal weight $h$ by $-n$.

When there are more chiral currents than the stress tensor, the chiral algebra extends the Virasoro algebra and is sometimes referred to as an extended chiral algebra. In bosonic theories, the chiral currents must have integer spins. If the additional chiral currents are spin 1 currents, the extended chiral algebra is a Kac-Moody algebra discussed in Section \ref{sec:SU(2)},
\ie
{[J_n^a, J_m^b]} = \sum_c if^{abc} J_{n+m}^c + k n \delta^{ab} \delta_{n+m,0} \,.
\label{eqn:kac_moody_algebra}
\fe
The Kac-Moody algebra is realized in the Wess-Zumino-Witten model, which can be described as a nonlinear sigma model with field $g(z,\bar z)$ valued in a Lie group $G$. In this realization, the Kac-Moody algebra generates a
holomorphic symmetry that acts as
\ie
g(z,\bar z) \mapsto h(z) g(z,\bar z) \, .
\fe
If the additional chiral currents have spin $2$, the extended chiral algebra can be decomposed into a direct product of multiple copies of Virasoro algebras. 
More generally, extensions by higher spin currents lead to $\cW$ algebras. The simplest example is the $\cW_3$ algebra discussed in Section \ref{sec:w3}, which contains an additional chiral spin 3 current. In fermionic theories, the chiral currents can also have half-integer spins. Spin $\frac{1}{2}$ chiral currents correspond to free chiral fermions, while spin $\frac{3}{2}$ chiral currents extend the Virasoro algebra to a superconformal algebra discussed in Section \ref{sec:susy}.

\subsection{Chiral Tube Algebras}
\label{subsec:chiral_tube_algebra}

As discussed in Section \ref{subsec:tdls}, TDLs give rise to a rich spectrum of
defect operators, which naturally motivates the questions raised at the beginning
of the paper. We will explain that chiral tube algebras provide a natural
framework for addressing all of these questions.

\subsubsection*{How does a chiral algebra act on defect Hilbert spaces?}
TDLs, by definition, commute with the stress tensor, so their associated defect Hilbert
spaces always preserve the Virasoro algebra and therefore can be organized into Virasoro modules. However, these defect Hilbert spaces may not preserve an extended chiral algebra generated by additional chiral currents $V^a(z)$. This happens when the TDLs act nontrivially on the chiral currents. Since TDLs implement global symmetries, they must preserve the OPE structure of
the chiral algebra and hence act as automorphisms of the chiral algebra (assuming the TDLs do not map the chiral currents to non-local chiral currents)\footnote{More generally, TDLs act on chiral algebra in the form of a hypergroup\cite{Bischoff:2016jmy,Bischoff_2022,riesen2022fusion,Dong:2025ttr,Gannon:2026ttf}. }.
As we explain below,
although the original extended chiral algebra is no longer preserved on the defect Hilbert spaces, a version of it twisted by these automorphisms still survives. The defect Hilbert spaces can then be organized into modules of this twisted chiral algebra.

Suppose a TDL $\cL$ acts on the chiral currents $V^a(z)$ by a phase as
\vspace{-6pt}
\ie
\cL \cdot V^a(z)= \tikz[baseline=-3.5pt]{
    \filldraw[black] (0,0) circle (2pt) node[above,xshift=1pt] {$V^a(z)$};
    \draw[thick,-{Stealth[round, length=5pt, width=5pt, bend]}] (0.9,0) arc (0:365:0.9);
    \node[right] at (0.9,0) {$\cL$};
    }=e^{2\pi ix_a} D_\cL\, V^a(z)\,,
\fe
where $D_{\cL}$ is the quantum dimension of the TDL $\cL$, which is defined as the expectation value of $\cL$.
If the TDL mixes different chiral currents, we can choose a basis of currents that diagonalizes this action. 
Because of the TDL action, the chiral current $V^a(z)$ acquires a nontrivial monodromy when winding around a $\cL$ defect operator $\cO_\cL(0)$
\vspace{-4pt}
\ie
  V^a(z e^{2\pi i}) \cO_\cL(0) =\tikz[baseline=0pt]{
  \draw[thick] (0,0) -- (0,1.6);
  \draw[thick,-{Stealth[round, length=5pt, width=5pt, bend]}] (0,0) -- (0,0.6);
    \node[right] at (0.0,3-2.48) {$\mathcal{L}$};
    \filldraw[black] (0,0) circle (2pt) node[right,yshift=0pt,xshift=-0.1pt] {$\cO_{\!\cL}$};
    \filldraw[black] (0,-0.9) circle (2pt) node[below,xshift=1pt] {$V^a(z)$};
    \draw[thick, dashed,-{Stealth[round, length=5pt, width=5pt, bend]}] (0.23293714059,-0.86933324366) arc (15-90:345-90:0.9);
    }=  e^{-2\pi i x_a}\tikz[baseline=0pt]{
  \draw[thick] (0,0) -- (0,1.6);
  \draw[thick,-{Stealth[round, length=5pt, width=5pt, bend]}] (0,0) -- (0,0.6);
    \node[right] at (0.0,3-2.48) {$\mathcal{L}$};
    \filldraw[black] (0,0) circle (2pt) node[right,yshift=0pt,xshift=-0.1pt] {$\cO_{\!\cL}$};
    \filldraw[black] (0,-0.9) circle (2pt) node[below,xshift=1pt] {$V^a(z)$};
    }= e^{-2\pi i x_a} V^a(z) \cO_\cL(0) \,.
    \label{eq:lasso_operator_chiral_tube_First}
\fe
This shifts the mode number in the mode expansion of $V^a(z)$ from $\mathbb{Z}$ to $\mathbb{Z}+x_a$,
\ie
V^a(z)=\sum_{n}\frac{V^a_{n+x_a}}{z^{n+h_a+x_a}}\,.
\label{eq:shiftedVax}
\fe
Thus, in the defect Hilbert space $\cH_\cL$, the symmetry charges are twisted into
\ie
V^a_{n+x_a}=\oint_\gamma \frac{dz}{2\pi i}\, z^{n+h_a+x_a-1}V^a(z)\,.\label{eq:twisted_symmetry_charge}
\fe
We can conveniently summarize the action of these twisted symmetry charges on defect Hilbert
spaces using the following lasso operators, analogous to the lasso operators \eqref{eq:lasso_operator_tube} that generate the
tube algebra of TDLs:
\ie
  \tikz[baseline=0pt]{
  \draw[thick] (0,) -- (0,1.6);
  \draw[thick,-{Stealth[round, length=5pt, width=5pt, bend]}] (0,0) -- (0,0.6);
    \draw[thick,-{Stealth[round, length=5pt, width=5pt, bend]}] (0,0.6) -- (0,1.3);
    \node[right] at (0.0,3-2.48) {$\mathcal{L}$};
    \node[right] at (0.0,3-1.72) {$\mathcal{L}$};
    \node[right] at (0.9,3-3) {$V^a_{n+x_a}$};
    \draw[thick] (0,3-3) circle (0.9);
    }\,.
\label{eq:lasso_operator_chiral_tube}
\fe
Here, the vertical lines are TDLs and the horizontal circle is the symmetry charge.
We refer to the algebra generated by these lasso operators as the chiral tube
algebra. It extends the chiral algebra action beyond the local Hilbert space to include defect Hilbert spaces.

Here, the chiral tube algebra takes a particularly simple form. Each lasso operator preserves the defect Hilbert space on which it acts, and only lasso operators acting on the same defect Hilbert space can be composed. This implies that the chiral tube algebra decomposes into a direct sum of subalgebras, each acting independently on a given defect Hilbert space. Below, we will encounter more
sophisticated chiral tube algebras, whose lasso operators can map between
different defect Hilbert spaces. 

For a given defect Hilbert space, the corresponding subalgebra is a twisted version of the original chiral algebra, with the mode numbers shifted as in \eqref{eq:shiftedVax}. The defect Hilbert space can then be organized into modules of this twisted chiral algebra, which is referred to as twisted modules \cite{dong1995twistedrepresentationsvertexoperator}. An example of twisted chiral algebra is given by the twisted $\mathfrak{su}(2)_1$ Kac-Moody algebra, discussed in Section \ref{sec:SU(2)},
\ie
&[J^3_{n}, J^3_{m}] = \frac{n}{2} \delta_{n+m,0} \,, \qquad\ \ \ \ \, [J^+_{n+x}, J^+_{m+x}] = [J^-_{n-x}, J^-_{m-x}] = 0 \,, \\
&[J^3_{n}, J^{\pm}_{m\pm x}] = \pm J^{\pm}_{n+m\pm x}\,,\qquad [J^+_{n+x}, J^-_{m-x}] = (n + x) \delta_{n+m,0} + 2J^3_{n+m}  \,.
\label{eqn:First_twisted_su2_chiral_algebra}
\fe
Other examples of twisted chiral algebras and twisted modules are discussed in Section \ref{sec:w3} for the $\cW_3$ algebra and in Section \ref{sec:susy} for superconformal algebras. 

\subsubsection*{What happens to a chiral algebra under finite gauging?}
Gauging a finite symmetry, also known as orbifolding \cite{Dixon:1985jw,Dixon:1986qv,Dijkgraaf:1989hb}, amounts to summing over all possible topologically inequivalent insertions of the corresponding TDLs.\footnote{In this paper, we focused on gauging finite abelian groups.} At the operator level, this reshuffles the spectrum of local and defect operators. For example, gauging a $\mathbb{Z}_2$ symmetry maps the operators as follows:
\ie
\renewcommand{\arraystretch}{1} 
\begin{tabular}{|c|c|c|}
\hline
     & $\mZ_2$-even & $\mZ_2$-odd \\
     \hline
    {\rm untwisted} & \colorbox{blue!20}{A} & \colorbox{green!20}{B}\\
    \hline
    $\mZ_2$-{\rm twisted} & \colorbox{red!20}{C}& \colorbox{orange!20}{D}
    \\
    \hline
\end{tabular}
\quad\Longleftrightarrow\quad
\renewcommand{\arraystretch}{1} 
\begin{tabular}{|c|c|c|}
\hline
     & $\hat{\mZ}_2$-even & $\hat{\mZ}_2$-odd \\
     \hline
    {\rm untwisted} & \colorbox{blue!20}{A}& \colorbox{red!20}{C}\\
    \hline
    $\hat{\mZ}_2$-{\rm twisted} & \colorbox{green!20}{B}& \colorbox{orange!20}{D}
    \\
    \hline
\end{tabular}
\label{eq:Z2_gauging_shuffle}
\fe
Here, the $\hat{\mathbb{Z}}_2$ symmetry in the gauged theory is generated by the topological Wilson line of the $\mathbb{Z}_2$ gauge field, and is known as the quantum symmetry \cite{Vafa:1989ih}, or equivalently the dual symmetry. Gauging the $\mathbb{Z}_2$ symmetry makes the original $\mathbb{Z}_2$ TDL transparent, while in the meantime attaching $\mathbb{Z}_2$ Wilson lines, i.e.~the $\hat{\mathbb{Z}}_2$ TDLs, to the $\mathbb{Z}_2$-odd operators. Therefore, after gauging, the original $\mathbb{Z}_2$-even operators becomes the untwisted sector operators, whereas the original $\mathbb{Z}_2$-odd operators become $\hat{\mathbb{Z}}_2$-twisted sector operators. Moreover, because a $\mathbb{Z}_2$ Wilson line picks up a $-1$ phase when it crosses the original $\mathbb{Z}_2$ TDL, operators coming from the original untwisted sector are $\hat{\mathbb{Z}}_2$-even in the gauged theory, while those coming from the original $\mathbb{Z}_2$-twisted sector are $\hat{\mathbb{Z}}_2$-odd. The operator spectrum is thus reshuffled as in \eqref{eq:Z2_gauging_shuffle}.
Conversely, the original theory can be recovered by gauging the dual $\hat \mZ_2$ symmetry in the gauged theory.

This example illustrates that the theories before and after gauging encode the same physical information, although their local and defect operators are reorganized. This suggests that the chiral algebra of the original theory should also survive after gauging, but possibly in a different form. This is reminiscent to what happens to ordinary global symmetries under finite gauging \cite{Bhardwaj:2017xup,Tachikawa_2020,Benini_2019,Hsin_2021,Kaidi:2021xfk,Choi:2022zal}. If all chiral currents are neutral under the gauged symmetry, they remain local operators after gauging and the chiral algebra is unchanged\footnote{In fact, gauging can often lead to the extension of the original chiral algebra because some non-local chiral currents can become local after gauging. Note that given a chiral algebra $\cV$, the extensions are classified by the 2-Morita classes of condensable algebras $\cA$ in ${\rm Rep}(\cV)$, which also classify the inequivalent gauging preserving $\cV$ in the diagonal model. }. On the other hand, if some chiral currents are charged under the gauged symmetry, they become non-local operators attached by TDLs. Naively, this suggests that the gauged theory has fewer local chiral currents and hence a smaller chiral algebra, unless we can also incorporate these non-local chiral currents. This naturally leads to the question of whether non-local chiral currents can be used to construct a ``chiral algebra".

As we explain below, the answer is yes and chiral tube algebras provide a natural formulation. To illustrate the idea, let us consider gauging a $\mathbb{Z}_2$ symmetry that acts nontrivially on a local chiral current $V(z)$. In the gauged theory, $V(z)$ becomes a non-local chiral current attached by the dual $\hat{\mathbb{Z}}_2$ TDL, which we denote by $\eta$. 

Now consider winding this non-local current around a local operator $\mathcal O$ by $2\pi$:
\ie
\tikz[baseline=-0.75]{\draw[thick] (0,0.75)--(0,1.5); \draw[thick] (0,0.75) arc (90:45:0.75); \draw[thick] (0.53,0.53)--(0.73,0.73); \node[right] at (-0.05,1.35) {\small $\eta$};\node[right] at (0.63,0.83) {\small $V(z)$};\filldraw[black] (0,0) circle (2pt) node[right,yshift=0pt,xshift=-0.1pt] {$\cO$};} 
\tikz[baseline=-0.75]{\draw[black,-{Stealth[round, length=5pt, width=5pt, bend]}] (-2.8,0)--(-1.25,0);
\node[above] at (-2,0.1) {\small $z\rightarrow z e^{2\pi i}$};\draw[thick] (0,0.75)--(0,1.5); \draw[thick] (0,0.75) arc (90:405:0.75); \draw[thick] (0.53,0.53)--(0.73,0.73); \node[right] at (0.63,0.83) {\small $V(z)$}; \node[right] at (-0.05,1.35) {\small $\eta$}; \filldraw[black] (0,0) circle (2pt) node[right,yshift=0pt,xshift=-0.1pt] {$\cO$};}
=\quad\ \,
\tikz[baseline=-0.75]{\draw[thick] (0,0.75)--(0,1.5); \draw[thick] (0,0.75) arc (90:45:0.75); \draw[thick] (0.53,0.53)--(0.73,0.73); \node[right] at (-0.05,1.35) {\small $\eta$};\node[right] at (0.63,0.83) {\small $V(z)$};\filldraw[black] (0,0) circle (2pt) node[right,yshift=0pt,xshift=-0.1pt] 
{$\cO$};\draw[thick] (0,0.6) arc (90:450:0.6);
} 
\,.
\fe
After the $2\pi$ winding, the current returns to itself, but in the process it drags the attached $\hat{\mathbb{Z}}_2$ TDL $\eta$ around the operator $\mathcal O$. The resulting holonomy therefore depends on the $\hat{\mathbb{Z}}_2$ charge of $\mathcal O$. For a $\hat{\mathbb{Z}}_2$-even operator, the holonomy is trivial, while for a $\hat{\mathbb{Z}}_2$-odd operator, the holonomy is $-1$. 

To give the non-local current a definite monodromy, we introduce the projectors onto the $\eta$ eigenspaces,
\ie
P_\pm=\frac{1}{2}(1\pm\eta)\,,
\fe
and push the non-local chiral currents onto these projectors:
\ie
V(z) P_{\pm}=
\tikz[baseline=-0.75]{\draw[thick] (0,0.75)--(0,1.5); \draw[thick] (0,0.75) arc (90:45:0.75); \draw[thick] (0.53,0.53)--(0.73,0.73); \node[right] at (-0.05,1.35) {\small $\eta$};\node[right] at (0.63,0.83) {\small $V(z)$};\draw[thick] (0,0.6) arc (90:450:0.6);
\node[right] at (0.50,-0.4) {\small $P_{\pm}$};
}  = \frac{1}{2} \le \ \quad \ \
\tikz[baseline=-0.75]{\draw[thick] (0,0.75)--(0,1.5); \draw[thick] (0,0.75) arc (90:45:0.75); \draw[thick] (0.53,0.53)--(0.73,0.73); \node[right] at (-0.05,1.35) {\small $\eta$};\node[right] at (0.63,0.83) {\small $V(z)$};}\  \pm \quad 
\tikz[baseline=-0.75]{
\draw[thick] (0,0.75)--(0,1.5); \draw[thick] (0,0.75) arc (90:405:0.75); \draw[thick] (0.53,0.53)--(0.73,0.73); \node[right] at (0.63,0.83) {\small $V(z)$}; \node[right] at (-0.05,1.35) {\small $\eta$}; } \ri \,.
\fe
After projection, the current has trivial monodromy on $P_+$ and $-1$ monodromy on $P_-$. This means that the currents should be expanded into integer modes on $P_+$ and half-integer modes on $P_-$. We thus arrive at the following generators of the chiral tube algebra:
\ie
{[V_n]}^{\eta}_I =\ \tikz[baseline=-0.75]{\draw[thick] (0,0.75)--(0,1.5); \draw[thick] (0,0) circle (0.75); \node[right] at (0,1.1) {\small $\eta$}; \node[right] at (0.6,-0.5) {\small $V_n P_{+}$}}\,,
\qquad 
[V_{n+\frac{1}{2}}]^{\eta}_I =\ \tikz[baseline=-0.75]{\draw[thick] (0,0.75)--(0,1.5); \draw[thick] (0,0) circle (0.75); \node[right] at (0,1.1) {\small $\eta$}; \node[right] at (0.6,-0.5) {\small $V_{n+\frac{1}{2}} P_{-}$}}\, .
\label{eq:lasso_operator_Vn_up}
\fe
where $V_n$ and $V_{n+\frac{1}{2}}$ are the contour integrals defined in \eqref{eq:twisted_symmetry_charge}.\footnote{Our construction of lasso operators is reminiscent of the construction of continuous non-invertible symmetry \cite{Chang:2012.01429} from non-local spin 1 currents in \cite{Thorngren:2021yso, Antinucci:2025uvj,Delmastro:2025ksn} and the construction of translation invariant defects from non-local chiral currents in \cite{Ambrosino:2025myh,Ambrosino:2025pjj}, except that we do not exponentiate our generators and our integral contains an explicit $z$ dependence. Non-local currents have also been used to construct conformal manifolds of conformal boundaries and interfaces \cite{Antinucci:2025uvj,Choi:2025ebk} and to compute transmission coefficients of conformal interfaces \cite{Furuta:2511.00356}.} 

These generators are similar to the lasso operators in \eqref{eq:lasso_operator_chiral_tube}, with the important difference that the vertical TDL changes across the horizontal chiral symmetry charges. As a result, these lasso operators map the local Hilbert space $\mathcal H$ to the $\eta$ defect Hilbert space $\mathcal H_\eta$. This is exactly what we expect:~since the non-local chiral current $V(z)$ should be a descendant of the vacuum module of the chiral tube algebra, the corresponding chiral tube algebra generators must mix the local Hilbert space $\mathcal H$ with the defect Hilbert space $\mathcal H_\eta$. This mixing of different defect Hilbert spaces is reminiscent of the action of non-invertible TDLs illustrated in \eqref{eq:non-invertible_TDL_action}.

Similarly, we can push the non-local current $V(z)$ onto the projectors from the inside and integrate it to obtain the following lasso operators that map from the defect Hilbert space $\cH_{\eta}$ back the local Hilbert space $\cH$ 
\ie
{[V_n]}^I_{\eta} = \tikz[baseline=-0.75]{\draw[thick] (0,0)--(0,0.75); \draw[thick] (0,0) circle (0.75); \node[right] at (0,0.375) {\small $\eta$}; \node[right] at (0.6,-0.5) {\small $V_{n} P_{+}$}} \,,
\qquad
[V_{n+\frac{1}{2}}]^I_{\eta} = \tikz[baseline=-0.75]{\draw[thick] (0,0)--(0,0.75); \draw[thick] (0,0) circle (0.75); \node[right] at (0,0.375) {\small $\eta$}; \node[right] at (0.6,-0.5) {\small $V_{n+\frac{1}{2}} P_{-}$}}\,.
\label{eq:lasso_operator_Vn_down}
\fe
The full chiral tube algebra is generated by the lasso operators in \eqref{eq:lasso_operator_Vn_up} and \eqref{eq:lasso_operator_Vn_down}. Because of the projectors, only lasso operators acting on the same $\hat{\mZ}_2$ eigenspace can be composed. In particular, ${[V_n]}_I^{\eta}$ and ${[V_n]}_{\eta}^I$ can be composed, and they generate an algebra isomorphic to the untwisted chiral algebra. Similarly, ${[V_{n+\frac12}]}_I^{\eta}$ and ${[V_{n+\frac12}]}^I_{\eta}$ can be composed, and they generate an algebra isomorphic to the twisted chiral algebra.

This is consistent with how the spectrum is reshuffled under the $\mathbb{Z}_2$ gauging as illustrated in \eqref{eq:Z2_gauging_shuffle}. Below, we illustrate how the operators are organized into modules under the chiral tube algebra before and after gauging,
\ie
\renewcommand{\arraystretch}{1} 
\begin{tabular}{|c|c|c|}
\hline
     & $\mZ_2$-even & $\mZ_2$-odd \\
     \hline
    {\rm untwisted} & \colorbox{blue!20}{$\cO,\,(V_{-1})^2\cO,\,\cdots$} & \colorbox{green!20}{$V_{-1}\cdot \cO,\, V_{-2}\cO,\,\cdots$}\\
    \hline
    $\mZ_2$-{\rm twisted} & \colorbox{red!20}{$\tilde\cO,\,(V_{-\frac{1}{2}})^2\tilde\cO,\,\cdots$}& \colorbox{orange!20}{$V_{-\frac{1}{2}}\tilde\cO,\, V_{-\frac{3}{2}}\tilde\cO,\,\cdots$}
    \\
    \hline
\end{tabular}
\\
\Big\Updownarrow\qquad\qquad\qquad\qquad\qquad\quad\,
\\
\renewcommand{\arraystretch}{1} 
\begin{tabular}{|c|c|c|}
\hline
     & $\hat{\mZ}_2$-even & $\hat{\mZ}_2$-odd \\
     \hline
    {\rm untwisted} & \colorbox{blue!20}{$\cO,\,(V_{-1})^2\cO,\,\cdots$}& \colorbox{red!20}{$\tilde\cO,\,(V_{-\frac{1}{2}})^2 \tilde\cO,\,\cdots$}\\
    \hline
    $\hat{\mZ}_2$-{\rm twisted} & \colorbox{green!20}{$V_{-1} \cO,\,V_{-2}\cO,\,\cdots$}& \colorbox{orange!20}{$V_{-\frac{1}{2}}\tilde\cO,\,V_{-\frac{3}{2}}\tilde\cO,\,\cdots$}
    \\
    \hline
\end{tabular}\,
\fe
In particular, because the original $\mathbb{Z}_2$ symmetry acts nontrivially on the chiral current $V(z)$, the $\mathbb{Z}_2$-twisted Hilbert space must form a twisted module of the original chiral algebra.

More generally, one can also consider the action of the orbifold chiral tube algebra on defect Hilbert spaces twisted by TDLs other than those attached to the non-local chiral current. Such chiral tube algebras are generated by lasso operators of the form
\ie
  {[V_{n+x}]}^{\cL''}_{\cL'}=\ \tikz[baseline=0pt]{
  \draw[thick] (0,) -- (0,1.6);
  \draw[thick,-{Stealth[round, length=5pt, width=5pt, bend]}] (0,0) -- (0,0.6);
    \draw[thick,-{Stealth[round, length=5pt, width=5pt, bend]}] (0,0.6) -- (0,1.3);
    \node[right] at (0.0,3-2.48) {$\mathcal{L}'$};
    \node[right] at (0.0,3-1.72) {$\mathcal{L}''$};
    \node[right] at (0.9,3-3) {$V_{n+x} P_{x}$};
    \draw[thick] (0,3-3) circle (0.9);
    }\,,
\label{eq:lasso_operator_chiral_tube_general}
\fe
where $\mathcal L''$ must appear in the fusion product $\mathcal L'\times \eta$.

In summary, chiral tube algebras provide a natural framework for describing the image of chiral algebras after finite gauging. In Sections~\ref{sec:w3} and~\ref{sec:SU(2)}, we discuss the $\mathbb{Z}_2$ orbifold of the $\mathcal W_3$ algebra and the $\mathbb{Z}_N$ orbifold of the $\mathfrak{su}(2)_1$ Kac-Moody algebra, respectively. Similar considerations apply to bosonizations of fermionic chiral algebras, since bosonization can be viewed heuristically as gauging the $(-1)^F$ symmetry. In Section~\ref{sec:susy}, we discuss the bosonization of the $\mathcal N=1$ superconformal algebra.

\subsubsection*{Can non-local chiral currents attached by TDLs give rise to a chiral algebra?}

This question was partly answered above, when we explained how to construct the orbifold chiral tube algebra from non-local chiral currents.
However, the non-local chiral currents that arise from gauging chiral algebras always have integer or half-integer spin.
It is then natural to wonder if the construction can be generalized to intrinsically non-local currents that do not arise from gauging. Examples of such intrinsically non-local currents are given by non-local chiral currents with more general fractional spins. The resulting chiral tube algebras necessarily have no counterparts in conventional chiral algebras.

In fact, an example of ``chiral algebras" constructed from non-local fractional currents has already appeared in the literature, which is the $\mZ_N$ parafermionic algebra\cite{Fateev:1985mm} generated by non-local chiral currents living at the ends of $\mZ_N$ lines. Such chiral currents can have fractional spin, for example $h=\frac{k}{N}$ with integer $k$. For certain choices of the spins, the parafermionic algebra is sufficiently constraining to uniquely fix the corresponding rational CFT. As we have learned from the example of orbifold chiral tube algebras, these parafermionic algebras necessarily mix local and defect Hilbert spaces. It is therefore more appropriate to formulate them as a chiral tube algebra.

In a companion paper \cite{PartTwo} to appear, we will reformulate parafermionic algebras in the framework of chiral tube algebras. We then take this framework further and construct the full chiral tube algebra of the simplest CFT, i.e.~the Ising CFT. The Ising CFT has no extended chiral algebra. Its only local chiral currents are polynomials of $T(z)$ and their derivatives, which are the Virasoro descendants of the identity operator. However, at the end of the $\mathbb{Z}_2$ TDL $\eta$, there is a chiral fermion of spin $h=\frac{1}{2}$ and at the end of the non-invertible duality line $\cN$, there is a chiral current of spin $h=\frac{1}{16}$. We will explain how to construct a chiral tube algebra from these non-local currents.

Generally, to construct a chiral tube algebra, one needs to specify the following data:~a set of TDLs described by a tensor category; a set of local or non-local chiral currents attached by a subcategory of these TDLs, together with their OPEs; and the action of the TDLs on these chiral currents, or equivalently, their tube representations. Importantly, the TDLs need not preserve the chiral algebra generated by local chiral currents. As a result, the associated non-local chiral currents may transform as twisted modules of the chiral algebra. 
The chiral tube algebra simplifies in the special case when the relevant TDLs are Verlinde lines that preserve the chiral algebra. These Verlinde lines themselves form a modular tensor category. In this case, the non-local chiral currents are untwisted modules under the chiral algebra, and the chiral tube algebra obtained by including all the non-local chiral currents at the end of Verlinde lines becomes the vertex tensor category formulated in \cite{Huang:1993xy,Huang:1993df,Huang:1995yc,Huang:1994xa,Huang:1995yw} (see also the survey in \cite{Huang:2013jza}) and their OPEs are encoded in the intertwining operators of the vertex tensor category. Physically, the vertex tensor category encodes representations of the chiral algebra.

\subsection{Outline}

We have introduced the general notion of chiral tube algebras in the introduction. In the rest of the paper, we will illustrate this framework in various concrete examples. 

Section \ref{sec:w3} focuses on the three-state Potts CFT and its $\mZ_2$ orbifold, i.e.~the tetracritical Ising CFT. The three-state Potts CFT is the simplest CFT that hosts a $\cW_3$ chiral algebra. 
We construct the corresponding $\cW_3$ chiral tube algebra that acts on all defect Hilbert spaces twisted by TDLs. We then construct the modules of this chiral tube algebra and show that each of them is isomorphic to either an untwisted or a twisted $\cW_3$ module. Interestingly, the Virasoro modules not contained in the untwisted $\cW_3$ modules naturally combine into twisted $\cW_3$ modules. Thus, the untwisted and twisted $\cW_3$ modules together cover all Virasoro modules at this central charge. In the tetracritical Ising CFT, the $\cW_3$ current becomes non-local. We explain how to use this non-local current to construct the orbifold of the $\cW_3$ chiral tube algebra and describe how the local and defect Hilbert spaces are organized under it.

Section \ref{sec:SU(2)} 
is devoted to chiral tube algebras that generalize the $\mf{su}(2)_1$ Kac-Moody algebra. Since the spin 1 currents of the $\mf{su}(2)_1$ algebra also generate an ordinary $SU(2)$ symmetry, there is a natural $\mf{su}(2)_1$ chiral tube algebra describing how the $\mf{su}(2)_1$ Kac-Moody algebra acts on the defect Hilbert spaces twisted by these $SU(2)$ TDLs. Its modules are isomorphic to untwisted or twisted $\mf{su}(2)_1$ modules, which provide an alternative derivation of spectral flow \cite{Schwimmer:1986mf}. We discuss an explicit realization of this $\mathfrak{su}(2)_1$ chiral tube algebra in the $SU(2)_1$ WZW model, and then study its $\mathbb{Z}_N$ orbifold, realized in the corresponding $\mathbb{Z}_N$ orbifold of the $SU(2)_1$ WZW model, which is a compact boson CFT.

Section \ref{sec:susy} studies the $\cN=1$ superconformal algebra and its bosonization in the example of the tricritical Ising CFT and its fermionization, which is an $\cN=1$ minimal model. In the tricritical Ising CFT, the supercurrents are non-local and attached by a $\mathbb{Z}_2$ TDL. Unlike the non-local currents encountered in the previous examples, these supercurrents are themselves charged under the attached $\mathbb{Z}_2$ TDL. This leads to new features in the corresponding chiral tube algebras.

Finally, in Section \ref{sec:discussion} we conclude with some open questions and possible future directions. 

\section{\texorpdfstring{$\mathcal{W}_3$}{W3} Algebra and Orbifold}
\label{sec:w3}

In this section, we consider the extensions of the Virasoro algebra to $\mathcal{W}$ algebras by including chiral higher spin currents \cite{zamolodchikov1995infinite} (see \cite{Bouwknegt:1992wg} for a review). We focus on the simplest example, $\mathcal{W}_3$ algebra, which is generated by the stress tensor $T(z)$ together with a chiral spin $3$ current $V(z)$ that is itself a Virasoro primary.\footnote{In this section, we will discuss how the three-state Potts CFT and the tetracritical Ising model organize themselves under the $\mathcal{W}_3$ algebra and its orbifold. We expect there to be a similar story for general minimal models, with other $\mathcal{W}$ algebras (that we will specify). We outline how this should work in Appendix \ref{app:generalminimalmodel}. We also expect our analysis generalizes to the $\cW_N$ algebra, which has chiral currents of spin $2, 3,\ldots,N$.}

In the $\cW_3$ algebra, the OPEs of $V(z)$ and the stress tensor $T(z)$ take the form\cite{zamolodchikov1995infinite}:
\ie\label{eq:OPE_V}
T(z) V(z')=\,&\frac{3\, V(z')}{(z-z')^2}+\frac{\partial V(z')}{z-z'}+\cdots\,,
\\
V(z)V(z')=\,&\frac{c}{3(z-z')^6}+\frac{2\,T(z')}{(z-z')^4}+\frac{\partial T(z')}{(z-z')^3}+\frac{1}{(z-z')^2}\left[\frac{3\, \partial^2T(z')}{10}+\frac{32\, \Lambda(z')}{22+5c}\right]
\\
&+\frac{1}{z-z'}\left[\frac{\partial^3T(z')}{15}+\frac{16\, \partial\Lambda(z')}{22+5c}\right]+\cdots\,,
\fe
where $\Lambda(z)$ is defined by the regular term in the stress tensor OPE,
\ie\label{eq:OPE_Lambda}
T(z)T(z')=\frac{c}{2(z-z')^4}+\frac{2\,T(z')}{(z-z')^2}+\frac{\partial T(z')}{z-z'}+\frac{3}{10}\partial^2T(z')+\Lambda(z')+\cdots\,.
\fe
Expanding the current $V(z)$ into modes, we obtain the following charges
\ie
V(z)=\sum_{n \in \mathbb{Z} }\frac{W_n}{z^{n+3}}\,,\quad W_n=\oint \frac{dz}{2\pi i}\, z^{n+2}\,{V(z)}\,.
\fe
These charges obey the following commutation relations with the Virasoro generators $L_n$\cite{zamolodchikov1995infinite}
\ie\label{eq:W3_algebra}
{\left[L_m, L_n\right]=}\, & (m-n) L_{m+n}+\frac{c}{12} m\left(m^2-1\right) \delta_{m+n, 0} \,, \\
{\left[L_m, W_n\right]=}\, & (2 m-n) W_{m+n} \,, \\
{\left[W_m, W_n\right]=}\, & (m-n)\left[\frac{1}{15}(m+n+3)(m+n+2)-\frac{1}{6}(m+2)(n+2)\right] L_{m+n} \\
& +\frac{16}{22+5c}(m-n) \Lambda_{m+n}+\frac{c}{360} m\left(m^2-1\right)\left(m^2-4\right) \delta_{m+n, 0} \,,
\fe
where $\Lambda_n$ is a quadratic operator defined as
\ie
\Lambda_n\,&=\sum_{p \leq-2} L_p L_{n-p}+\sum_{p \geq-1} L_{n-p} L_p-\frac{3}{10}(n+2)(n+3) L_n\,.
\fe
Because $\Lambda_n$ is quadratic in $L_n$, the $\mathcal{W}_3$ algebra closes non-linearly and is therefore not a Lie algebra unlike the Virasoro algebra or Kac-Moody algebras.

Note that the OPEs in \eqref{eq:OPE_V} and \eqref{eq:OPE_Lambda} have a $\mathbb{Z}_2$ automorphism:
\ie
T(z) \mapsto T(z)\,, \quad 
V(z) \mapsto -V(z)\,, \quad 
\Lambda(z) \mapsto \Lambda(z)\,.
\fe
If a TDL does not mix $V(z)$ with other operators, then it must act on the $\cW_3$ algebra either trivially or via this $\mathbb{Z}_2$ automorphism.
Using this automorphism, we can define a twisted version of the $\mathcal{W}_3$ algebra, where $V(z)$ acquires a nontrivial monodromy around the origin
\ie
V(e^{2\pi i} z) = -\,V(z)\,.
\fe
This modifies the mode expansion of $V(z)$ from integer to half-integer modes,
\ie
V(z)=\sum_{r \in \mathbb{Z} + \frac{1}{2}}\frac{W_r}{z^{r+3}} \,,\quad W_r=\frac{1}{2\pi i}\oint dz\, z^{r+2}\,{V(z)}\,. 
\fe
We use $n$ to denote integer modes and $r$ to denote half-integer modes. Since the $\mathbb{Z}_2$ automorphism acts trivially on the stress tensor, the twist does not affect the mode expansion of $T(z)$. The twisted algebra generated by $L_n$ and $W_r$ takes the same form as the untwisted algebra \eqref{eq:W3_algebra}, with the integer $n$ replaced by half-integer $r$. We will refer to this twisted algebra as the $\mathcal{W}_3'$ algebra. The twisted $\cW_3'$ algebra has been studied in for example \cite{Ho-Kim:1988ast,Bouwknegt:1992wg}. It will play an important role in the examples discussed below.

\subsection{Three-State Potts CFT}
\label{subsec:3potts}

There is a family of $\mathcal{W}_3$ minimal models at central charge~\cite{Fateev:1987vh}
\ie
c = 2\le 1 - \frac{12(p - q)^2}{pq}\ri \,,
\fe
where $p,q \geq 3$ are coprime integers. These minimal models contain only finitely many $\cW_3$ modules. The simplest unitary $\mathcal{W}_3$ minimal model is the three-state Potts CFT, corresponding to $(p,q)=(5,4)$ and central charge $c=\frac{4}{5}$. It is also the non-diagonal $D$-series Virasoro minimal model with $(p,q)=(6,5)$. We will focus on this theory in the rest of this subsection.

\subsubsection{Local Operator Spectrum}
At central charge $c=\frac{4}{5}$, the Virasoro algebra has 10 distinct irreducible modules. They are labeled by the conformal weights $h$ of their primary fields, which take values in
\ie
h \in \left\{0,\,3,\,\frac{2}{5},\,\frac{7}{5},\,\frac{2}{3},\,\frac{1}{15},\,\frac{1}{8},\,\frac{13}{8},\,\frac{1}{40},\,\frac{21}{40}\right\}\,.
\label{eqn:conformal_weight_tetra}
\fe
We denote their Virasoro characters by $\chi_h$.  
The modular $S$-matrix acting on them is displayed in Table~\ref{tab:Smatrix}. 

\begin{table}
\renewcommand{\arraystretch}{1.2}
\scalebox{0.95}{
\begin{tabular}{|c|c|c|c|c|c|c|c|c|c|c|} 
\hline
$\Delta$&$\chi_0$ & $\chi_3$ & $\chi_{\frac{2}{5}}$ & 
$\chi_{\frac{7}{5}}$& 
$\chi_{\frac{2}{3}}$
&$\chi_{\frac{1}{15}}$&$\chi_{\frac{1}{8}}$&$\chi_{\frac{13}{8}}$&$\chi_{\frac{1}{40}}$&$\chi_{\frac{21}{40}}$
\\
&&&&&&&&&&\vspace{-15pt}
\\
\hline
&&&&&&&&&&\vspace{-15pt}
\\
$\chi_0$&$1$&$1$&$\lambda^2$&$\lambda^2$&$2$&$2\lambda^2
$&
$\sqrt{3}$&$\sqrt{3}$&$\sqrt{3}\lambda^2$&$\sqrt{3}\lambda^2$\\
$\chi_3$&$1$&$1$&$\lambda^2$&$\lambda^2$&$2$&$2\lambda^2
$&
$-\sqrt{3}$&$-\sqrt{3}$&$-\sqrt{3}\lambda^2$&$-
\sqrt{3}\lambda^2$\\
$\chi_{\frac{2}{5}}$&$\lambda^2$&$\lambda^2$&$-1$&$-
1$&$2\lambda^2$&$-2$&
$-\sqrt{3}\lambda^2$&$-
\sqrt{3}\lambda^2$&$\sqrt{3}$&$\sqrt{3}$\\
$\chi_{\frac{7}{5}}$&$\lambda^2$&$\lambda^2$&$-1$&$-
1$&$2\lambda^2$&$-2$&
$\sqrt{3}\lambda^2$&$\sqrt{3}\lambda^2$&$-\sqrt{3}$&$-
\sqrt{3}$\\
$\chi_{\frac{2}{3}}$&$2$&$2$&$2\lambda^2$&$2\lambda^2$&$-2$&$-
2\lambda^2$&
$0$&$0$&$0$&$0$\\
$\chi_{\frac{1}{15}}$&$2\lambda^2$&$2\lambda^2$&$-2$&$-2$&$-
2\lambda^2$&$2$&
$0$&$0$&$0$&$0$\\
$\chi_{\frac{1}{8}}$&$\sqrt{3}$&$-\sqrt{3}$&$-\sqrt{3}\lambda^2$ 
&$\sqrt{3}\lambda^2$&$0$&$0$&$-\sqrt{3}$&$\sqrt{3}$
&$\sqrt{3}\lambda^2$&$-\sqrt{3}\lambda^2$\\
$\chi_{\frac{13}{8}}$&$\sqrt{3}$&$-\sqrt{3}$&$-\sqrt{3}\lambda^2$ 
&$\sqrt{3}\lambda^2$&$0$&$0$&$\sqrt{3}$&$-\sqrt{3}$
&$-\sqrt{3}\lambda^2$&$\sqrt{3}\lambda^2$\\
$\chi_{\frac{1}{40}}$&$\sqrt{3}\lambda^2$&$-\sqrt{3}\lambda^2$&$\sqrt{3}$ 
&$-
\sqrt{3}$&$0$&$0$&$\sqrt{3}\lambda^2$&$-\sqrt{3}\lambda^2$
&$\sqrt{3}$&$-\sqrt{3}$\\
$\chi_{\frac{21}{40}}$&$\sqrt{3}\lambda^2$&$-
\sqrt{3}\lambda^2$&$\sqrt{3}$ &$-
\sqrt{3}$&$0$&$0$&$-\sqrt{3}\lambda^2$&$\sqrt{3}\lambda^2$
&$-\sqrt{3}$&$\sqrt{3}$\vspace{-15pt}\\
&&&&&&&&&&
\\
\hline
\end{tabular} }
\caption{The modular S-matrix for Virasoro characters 
(multiplied by $\sqrt{30+6\sqrt{5}}$) at central charge $c=\frac{4}{5}$.  Here, we use the shorthand $\lambda^2=\frac{1}{2}(1+\sqrt{5})$. (See e.g. \cite{DiFrancesco:1997nk}.)} \label{tab:Smatrix}
\end{table}

On the other hand, at central charge $c=\frac{4}{5}$, the $\mathcal{W}_3$ algebra has 6 irreducible modules. They are labeled by the conformal weights $h$ and the $W_0$ eigenvalues $w$ of their primary fields, which take values in 
\ie\label{eq:W3_module}
\big(h,w\big)=\left\{\big(0,0\big),\, \left(\frac{1}{15},\pm\frac{1}{9}\sqrt{\frac{2}{195}}\right),\,  \left(\frac{2}{3},\pm\frac{2}{9}\sqrt{\frac{26}{15}}\right),\, \left(\frac{2}{5},0\right)\right\}
\fe
We denote the corresponding $\mathcal{W}_3$ characters by $\chi^\mathcal{W}_h$ when there is only one irreducible module with conformal weight $h$, and by $\chi^\mathcal{W}_{h,\pm}$ if there are two irreducible modules of the same conformal weight $h$ but opposite $W_0$ eigenvalue $w$. These $\mathcal{W}_3$ characters can be decomposed into the Virasoro characters as
\begin{alignat}{2}
\chi^\mathcal{W}_0&=\chi_0+\chi_3\,,\quad\ \chi^\mathcal{W}_{\frac{2}{5}}&=\chi_{\frac{2}{5}}+\chi_{\frac{7}{5}}\,, \quad\ \chi^\mathcal{W}_{\frac{2}{3},\pm}=\chi_{\frac{2}{3}}\,,\quad\ \chi^\mathcal{W}_{\frac{1}{15},\pm}=\chi_{\frac{1}{15}}\,.
\label{eqn:w3_module}
\end{alignat}
They obey the following fusion rules:
\ie\label{eq:W3_fusion_rule}
&\chi^\cW_{\frac{2}{3},+} \otimes \chi^\cW_{\frac{2}{3},-}
= \chi^\cW_0 \,,
\quad \ \ \, \chi^\cW_{\frac{2}{5}}\otimes \chi^\cW_{\frac{2}{3},\pm}
=
\chi^\cW_{\frac{2}{3},\mp}\otimes \chi^\cW_{\frac{1}{15},\mp}
=
\chi^\cW_{\frac{1}{15},\pm}\,,
\\
&
\chi^\cW_{\frac{2}{3},\pm} \otimes \chi^\cW_{\frac{2}{3},\pm}
= \chi^\cW_{\frac{2}{3},\mp}\,,\quad \ \chi^\cW_{\frac{2}{5}}\otimes \chi^\cW_{\frac{1}{15},\pm}
=
\chi^\cW_{\frac{1}{15},\mp}\otimes \chi^\cW_{\frac{1}{15},\mp}
=
\chi^\cW_{\frac{2}{3},\pm}\oplus \chi^\cW_{\frac{1}{15},\pm}\,,
\\
&\chi^\cW_{\frac{2}{5}}\otimes \chi^\cW_{\frac{2}{5}}
=
\chi^\cW_{\frac{1}{15},+}\otimes \chi^\cW_{\frac{1}{15},-}
=
\chi^\cW_{\frac{1}{15},\pm}\otimes \chi^\cW_{\frac{2}{3},\mp}
=
\chi^\cW_0\oplus \chi^\cW_{\frac{2}{5}}\,.
\fe

The torus partition function of the three-state Potts CFT can be decomposed into $\cW_3$ characters and Virasoro characters as
\ie
Z=
\,&|\chi^\mathcal{W}_{0}|^2+|\chi^\mathcal{W}_{\frac{1}{15},+}|^2+|\chi^\mathcal{W}_{\frac{1}{15},-}|^2
+|\chi^\mathcal{W}_{\frac{2}{3},+}|^2+|\chi^\mathcal{W}_{\frac{2}{3},-}|^2+|\chi^\mathcal{W}_{\frac{2}{5}}|^2
\\
=\,&|\chi_0+\chi_3|^2+|\chi_{\frac{2}{5}}+\chi_{\frac{7}{5}}|^2+2|\chi_{\frac{2}{3}}|^2+2|\chi_{\frac{1}{15}}|^2\,.
\fe
From the torus partition function, we can read off the $\mathcal{W}_3$ primaries, together with the Virasoro primaries contained in the corresponding $\mathcal{W}_3$ modules: 
\begin{alignat}{3}
&1=\left\{1_{0,0},V_{3,0},\bar V_{0,3}, Y_{3,3}\right\}\,,\qquad &&Z=\left\{Z_{\frac{2}{3},\frac{2}{3}}\right\}
\,,\qquad &&\sigma=\left\{\sigma_{\frac{1}{15},\frac{1}{15}}\right\}
\nonumber
\\
&\varepsilon=\left\{\varepsilon_{\frac{2}{5},\frac{2}{5}},\Phi_{\frac{7}{5},\frac{2}{5}},\bar\Phi_{\frac{2}{5},\frac{7}{5}},X_{\frac{7}{5},\frac{7}{5}}\right\}
\,,
\quad\ \,
&&Z^\dagger=\left\{Z^\dagger_{\frac{2}{3},\frac{2}{3}}\right\}
\,,
\quad
&&\sigma^\dagger=\left\{\sigma^\dagger_{\frac{1}{15},\frac{1}{15}}\right\}
\,,
\end{alignat}
where the subscripts are the conformal weights $(h,\bar h)$. The spectrum in particular contains the spin 3 currents $V_{3,0}(z)$ and $\bar{V}_{0,3}(\bar z)$, which generate the chiral and anti-chiral $\cW_3$ algebra. 

\subsubsection{Topological Defect Lines}
The three-state Potts CFT has in total 16 TDLs. Among them, 6 are associated with the $S_3$ global symmetry,
\ie
\{I,\eta,\bar\eta,C,C\eta,C\bar\eta\}\,,
\fe
where $\{I,\eta,\bar\eta=\eta^2\}$ forms the $\mathbb{Z}_3$ subgroup and $\{I,C\}$ forms the $\mathbb{Z}_2$ subgroup. They obey the fusion rules
\ie
\eta^3=\bar\eta^3=I\,,\quad C^2=I\,,\quad C\eta  C= \bar\eta \,.
\fe
There is also a non-invertible duality line $N$, which together with the $\mathbb{Z}_3$ lines $\{I,\eta,\bar\eta\}$ forms a $\mathbb{Z}_3$ Tambara-Yamagami category. It obeys the fusion rule
\ie
N^2=I+\eta+\bar\eta\,.
\fe
Since $N$ absorbs the $\mathbb{Z}_3$ lines, fusing $N$ with the $S_3$ lines produces only two distinct TDLs,
\ie
\{N,N'=CN\}\,.
\fe
In addition, there is a Fibonacci line $W$ which obeys the fusion rule
\ie
W^2=I+W\,.
\fe
The full fusion category of TDLs is the tensor product of the Fibonacci category generated by $W$ with the fusion category generated by the $S_3$ symmetry lines and the duality lines $\cN$, $\cN'$. It consists of the following 16 lines:
\ie
\{I,\eta,\bar\eta,C,C\eta,C\bar\eta,\cN,\cN',W,\eta W,\bar\eta W, CW,C\eta W, C\bar\eta W, \cN W, \cN'W\}\,.
\label{eqn:tdls_3states}
\fe
These TDLs act on the Virasoro primaries as
\begin{equation}
\begin{tabu}{cccccccccccccc}
  &  1 & V &\bar V & Y & \varepsilon & \Phi & \bar\Phi & X & \sigma & \sigma^\dagger & Z & Z^\dagger
\\
\eta :   & 1 & 1 & 1 & 1 & 1 & 1 & 1 & 1 & \omega & \omega^2 & \omega & \omega^2
\\
W :   & \zeta & \zeta & \zeta & \zeta & -\zeta^{-1} & -\zeta^{-1} & -\zeta^{-1} & -\zeta^{-1} & -\zeta^{-1} & -\zeta^{-1} & \zeta & \zeta
\\
 \cN :   & \sqrt3 & -\sqrt3 & \sqrt3 & -\sqrt3 & -\sqrt3 & \sqrt3 & -\sqrt3 & \sqrt3 & 0 & 0 & 0 & 0
 \\
 C: & 1 & -1 & -1 & 1 & 1 & -1 & -1 & 1 & \sigma^\dagger & \sigma & Z^\dagger & Z
\end{tabu}
\end{equation}
where $\omega \equiv e^{\frac{2\pi i}3}$ and $\zeta \equiv \frac{1+\sqrt{5}}{2}$. 
In particular, $C$ exchanges $\sigma$ with $\sigma^\dagger$ and $Z$ with $Z^\dagger$. 

\subsubsection{Chiral Tube Algebra}
From the symmetry actions, we find that 
\ie
\{I, \eta, \bar\eta, W, \eta W, \bar\eta W\}
\fe
commute with the spin 3 currents $V(z), \bar V(\bar z)$, and thus preserve the $\mathcal{W}_3 \times \overline{\mathcal{W}_3}$ algebra. Since the three-state Potts CFT is the diagonal $\mathcal{W}_3$ CFT, these TDLs are the Verlinde lines of the $\mathcal{W}_3$ algebra, which are in one-to-one correspondence with the $\cW_3$ characters:
\ie
\begin{tabular}{cccccc}
$I$ & $W$ & $\eta$  & $\bar\eta$ & $W\eta$ & $W\bar\eta$
\\
$\chi_0^{\cW}$ &$\chi_{\frac{2}{5}}^{\cW}$ &$\chi_{\frac{2}{3},+}^{\cW}$ & $\chi_{\frac{2}{3},-}^{\cW}$ & $\chi_{\frac{1}{15},+}^{\cW}$ & $\chi_{\frac{1}{15},-}^{\cW}$
\end{tabular}\,.
\fe
The other TDLs all have nontrivial interactions with the spin 3 currents $V(z), \bar V(\bar z)$. For example,
\ie
\{C, C\eta, C\bar\eta, CW, C\eta W, C\bar\eta W\}
\fe
act on the spin 3 currents $V(z),\bar V(\bar z)$ both by a $-1$ phase. Therefore, in the corresponding defect Hilbert space, the spin 3 currents acquire a nontrivial monodromy when passed through the TDL $C$ as illustrated below:
\ie
\tikz[baseline=-0.75]{\draw (0,0)--(0,1.5);  \node[right] at (0,0.35) {\small }; \node[right] at (0,1.1) {\small $C$}; \draw[black,dashed,-{Stealth[round, length=5pt, width=5pt, bend]}] (0.649,0.375) arc (30:330:0.75); \node at (0.75,0) [circle,fill,inner sep=1.5pt]{}; \node[right] at (0.75,0.2) {\small $V$}}
\quad = \quad\ \ \tikz[baseline=-0.75]{\draw (0,0)--(0,1.5);  \node[right] at (0,0.35) {\small }; \node[right] at (0,1.1) {\small $C$};  \node at (0.75,0) [circle,fill,inner sep=1.5pt]{}; \node[right] at (0.75,0.2) {\small $-V$}}\,.
\fe
This modifies the mode expansions to the half-integer ones and twists the chiral algebra to the twisted $\mathcal{W}_3' \times \overline{\mathcal{W}_3'}$ algebra. Similarly, 
\ie
\{N,NW\}
\fe
act on $V(z)$ by a $-1$ phase but commute with $\bar V(\bar z)$, so they twist the chiral algebra to the $\mathcal{W}_3'\times \overline{\mathcal{W}_3}$ algebra. Likewise,
\ie
\{N',N'W\}
\fe
act on $\bar V(\bar z)$ by a $-1$ phase but commute with $V(z)$, thereby twisting the chiral algebra to the $\mathcal{W}_3\times \overline{\mathcal{W}_3'}$ algebra.

In summary, the chiral tube algebra of the three-state Potts CFT is generated by the lasso operators, where the vertical TDLs are identical across the horizontal mode operators while the horizontal mode operators are twisted appropriately by the vertical TDLs as described above, such as
\ie
\tikz[baseline=-0.75]{\draw (0,0)--(0,1.5); \draw (0,0) circle (0.75); \node[right] at (0,0.35) {\small $\eta$}; \node[right] at (0,1.1) {\small $\eta$}; \draw[black,-{Stealth[round, length=5pt, width=5pt, bend]}] (-0.75,0) arc (180:360:0.75); \node[right] at (0.55,-0.5) {\small $W_n$}}\,, \ \qquad \tikz[baseline=-0.75]{\draw (0,0)--(0,1.5); \draw (0,0) circle (0.75); \node[right] at (0,0.35) {\small $C$}; \node[right] at (0,1.1) {\small $C$}; \draw[black,-{Stealth[round, length=5pt, width=5pt, bend]}] (-0.75,0) arc (180:360:0.75); \node[right] at (0.55,-0.5) {\small $W_r$}}
\fe
where $n\in\mathbb{Z}$ and $r\in\mathbb{Z}+\frac{1}{2}$. Only lasso operators with the same vertical defects can be composed and they form algebras isomorphic to either the $W_3$ algebra or the twisted $W_3'$ algebra depending on whether the twist is trivial or nontrivial.

\subsubsection{Highest Weight Modules of the Twisted \texorpdfstring{$\cW_3'$}{W3'} Algebra}
We have discussed the 6 irreducible modules of the $\mathcal{W}_3$ algebra at $c=\frac{4}{5}$. What about the irreducible modules of the twisted $\mathcal{W}_3'$ algebra? These twisted modules were studied in~\cite{Ho-Kim:1988ast}. We will provide an alternative derivation for $c=\frac{4}{5}$. A $\mathcal{W}_3'$ module can be generated by successive actions of $L_{-n}$, $W_{-r}$ with $n>0$, $r>0$ on a primary state $|h\rangle$ that satisfies
\ie
L_0|h\ra &= h|h\ra\,,
\\
W_r |h \ra &= L_n |h \ra = 0, \quad n>0,\ r>0\,.
\fe
Note that unlike the $\mathcal{W}_3$ algebra, the $\mathcal{W}_3'$ algebra does not include $W_0$, so its primaries are only labeled by the eigenvalue $h$ of $L_0$.
Since the $\mathcal{W}_3'$ algebra includes the Virasoro algebra as a sub-algebra, the $\cW_3'$ primary is also a Virasoro primary listed in \eqref{eqn:conformal_weight_tetra}. Furthermore, the $\mathcal{W}_3'$ modules can be decomposed into a direct sum of Virasoro modules. In the following, we will argue that there are only two possible $\cW_3'$ modules and their primaries have $h=\frac{1}{8}$ and $h=\frac{1}{40}$, respectively. 

To identify which Virasoro primaries are also $\mathcal{W}_3'$ primaries, let us consider the norm of the $\cW_3'$ descendant $W_{-\frac{1}{2}}|h\ra$. A useful relation we will use repeatedly is that, for a Virasoro primary $|h\ra$, we have
\ie\label{eq:useful_relation}
[W_{\frac{1}{2}}, W_{-\frac{1}{2}}] |h \ra=\left(-\tfrac{9}{40}L_{0}+
\tfrac{8}{13} \Lambda_{0}+\tfrac{1}{320}\right)  |h\ra = \tfrac{8}{13}\left(h-\tfrac{1}{8}\right)\left(h-\tfrac{13}{320}\right) |h\ra \equiv \lambda_h |h\ra \,.
\fe
Using this relation, the norm can be computed easily:
\ie
|W_{-\frac{1}{2}}|h\ra|^2 &= \la h | [W_{\frac{1}{2}}, W_{-\frac{1}{2}}] |h \ra 
=\lambda_h\,.
\fe
It vanishes for $h=\frac{1}{8}$ and $h=\frac{13}{320}$. In these cases, $W_{-\frac{1}{2}}|h\rangle$ is a null state and is therefore absent from the $\cW_3'$ module. However, for the other values of $h$, $W_{-\frac{1}{2}}|h\rangle$ should be included. Importantly, $W_{-\frac{1}{2}}|h\rangle$ is itself a Virasoro primary of conformal weight $h+\frac{1}{2}$\,:
\ie
&L_0(W_{-\frac{1}{2}}|h\rangle)=W_{-\frac{1}{2}}(L_0+\tfrac{1}{2})|h\rangle=(h+\tfrac{1}{2})W_{-\frac{1}{2}}|h\rangle\,,
\\
&L_n (W_{-\frac{1}{2}}|h\rangle)=W_{-\frac{1}{2}}L_n|h\rangle+(2n+\tfrac{1}{2})W_{n-\frac{1}{2}}|h\rangle=0,\quad n>0\,.
\fe
Among the Virasoro primaries listed in \eqref{eqn:conformal_weight_tetra}, the only pair whose conformal weights are $\frac{1}{2}$-apart are $h=\frac{1}{40}$ and $h=\frac{21}{40}$. This means that, aside from $|\frac{1}{8}\rangle$, which satisfies the null-state condition, $|\frac{1}{40}\rangle$ is the only other valid candidate for $\mathcal{W}_3'$ primary.

Since consecutive actions of $W_r$ with $r<0$ on the $\cW_3'$ primary increases the conformal weight by half-integers and the only Virasoro primary whose conformal weight is half-integer apart from $h = \frac{1}{40}$ is $h = \frac{21}{40}$, if $|\frac{1}{40}\rangle$ is a $\mathcal{W}_3'$ primary, then the $\mathcal{W}_3'$ module must be a direct sum of Virasoro modules with $h = \frac{1}{40}$ and $h = \frac{21}{40}$. Similarly, if $|\frac{1}{8}\rangle$ is a $\cW_3'$ primary, the  $\mathcal{W}_3'$ module can at most be a direct sum of Virasoro modules with $h = \frac{1}{8}$ and $h = \frac{13}{8}$. It is however possible that the Virasoro module with $h=\frac{1}{8}$ forms the $\cW'_3$ module itself with all the $h=\frac{13}{8}$ states either null states or Virasoro descendants. We will exclude this possibility below by analyzing the descendants  at level $\frac{3}{2}$.

At level $\frac{3}{2}$, there are 3 descendants of the $\cW_3'$ primary $|\frac{1}{8}\ra$
\ie
W_{-\frac{1}{2}}^3|\tfrac{1}{8}\ra\,, \quad L_{-1}W_{-\frac{1}{2}}|\tfrac{1}{8}\ra\,,\quad W_{-\frac{3}{2}}|\tfrac{1}{8}\ra\,.
\fe
Because $W_{-\frac{1}{2}}|\frac{1}{8}\ra=0$, both $W_{-\frac{1}{2}}^3 |\frac{1}{8}\ra$ and $L_{-1} W_{-\frac{1}{2}} |\frac{1}{8}\ra$ are null and the only nontrivial state to consider is $W_{-\frac{3}{2}}|\frac{1}{8}\rangle$. One can easily verify that  it has non-zero norm:
\ie
|W_{-\frac{3}{2}}|\tfrac{1}{8}\ra|^2 &= \la \tfrac{1}{8} | [W_{\frac{3}{2}}, W_{-\frac{3}{2}}] |\tfrac{1}{8} \ra 
= \la \tfrac{1}{8}| \left(\tfrac{13}{40} L_{0}+
\tfrac{24}{13} \Lambda_{0}-\tfrac{7}{960}\right)  |\tfrac{1}{8}\ra
=\tfrac{13}{120}\,.
\fe
Furthermore, it is a Virasoro primary with conformal weight $h=\frac{13}{8}$: 
\ie
&L_0(W_{-\frac{3}{2}}|h\rangle)=W_{-\frac{3}{2}}(L_0+\tfrac{3}{2})|h\rangle=(h+\tfrac{3}{2})W_{-\frac{3}{2}}|h\rangle\,,
\\
&L_n (W_{-\frac{3}{2}}|h\rangle)=W_{-\frac{3}{2}}L_n|h\rangle+(2n+\tfrac{3}{2})W_{n-\frac{3}{2}}|h\rangle=0,\quad n>0\,.
\fe
Therefore, the $W_3'$ module of $h=\frac{1}{8}$ must consist of
Virasoro modules of $h=\frac{1}{8}$ and $h=\frac{13}{8}$. 

In conclusion, the $\mathcal{W}_3'$ algebra at $c=\frac{4}{5}$ has two irreducible modules:~one with primary $|\frac{1}{40}\ra$ and the other with primary $|\frac{1}{8}\ra$. Their $\mathcal{W}_3'$ characters are
\ie\label{eq:W3'_character}
\chi^{\mathcal{W}'}_{\frac{1}{40}}=\chi_{\frac{1}{40}}+\chi_{\frac{21}{40}}\,,
\\
\chi^{\mathcal{W}'}_{\frac{1}{8}}=\chi_{\frac{1}{8}}+\chi_{\frac{13}{8}}\,.
\fe
Strictly speaking, we have only shown that these combinations of Virasoro modules are consistent with the expected structure for $\mathcal{W}_3'$ modules. We have not yet proven that they actually assemble into $\mathcal{W}_3'$ modules. In next subsubsection, we will verify this expectation by showing that these $\mathcal{W}'$ characters indeed appear in the defect partition function of the $C,N,NW$ TDLs.

\subsubsection{Defect Partition Functions}
Using the modular $S$ transformation, we can compute the defect partition functions for all the 16 TDLs. 
For the $\cW_3 \times \overline{\cW_3}$ preserving Verlinde lines, we have
\ie
Z_{\cL_i} = \sum_{j,k} N_{ik}{}^j \chi^\mathcal{W}_j \bar\chi^\mathcal{W}_k \,,
\fe
where $N_{ik}{}^j$ is the multiplicity appearing in the fusion algebra, $\phi_i\times\phi_j=\sum N_{ij}{}^k\phi_k$, of the $\mathcal{W}_3$ primaries in \eqref{eq:W3_fusion_rule}.
The defect partition functions of these Verlinde lines can be decomposed into $\cW_3\times\overline{\cW_3}$ characters as
\ie
Z_\eta =\,& \chi_{\frac{2}{3},+}^{\cW}\bar\chi_{0}^\mathcal{W}+\chi_{\frac{1}{15},+}^\cW \bar\chi_{\frac{2}{5}}^\mathcal{W}+\chi_{0}^\mathcal{W}\bar \chi_{\frac{2}{3},-}^\cW +\chi_{\frac{2}{5}}^\mathcal{W}\bar \chi_{\frac{1}{15},-}^\cW +\chi_{\frac{2}{3},-}^\cW \bar\chi_{\frac{2}{3},+}^{\cW} +\chi_{\frac{1}{15},-}^{\cW} \bar \chi_{\frac{1}{15},+}^\cW~,
\\
Z_{\bar\eta} = \,&\chi_{\frac{2}{3},-}^{\cW}\bar\chi_{0}^\mathcal{W}+\chi_{\frac{1}{15},-}^\cW \bar\chi_{\frac{2}{5}}^\mathcal{W}+\chi_{0}^\mathcal{W}\bar \chi_{\frac{2}{3},+}^\cW +\chi_{\frac{2}{5}}^\mathcal{W}\bar \chi_{\frac{1}{15},+}^\cW +\chi_{\frac{2}{3},+}^\cW \bar\chi_{\frac{2}{3},-}^{\cW} +\chi_{\frac{1}{15},+}^{\cW} \bar \chi_{\frac{1}{15},-}^\cW~,
\\
Z_W=\,& \chi_0^\mathcal{W} \bar\chi_{\frac{2}{5}}^\mathcal{W}  + \chi_{\frac{2}{5}}^\mathcal{W} \bar\chi_0^\mathcal{W} + |\chi_{\frac{2}{5}}^\mathcal{W}|^2 + \chi_{\frac{2}{3},+}^\cW \bar\chi_{\frac{1}{15},+}^\cW + \chi_{\frac{2}{3},-}^\cW \bar\chi_{\frac{1}{15},-}^\cW + \chi_{\frac{1}{15},+}^\cW \bar\chi_{\frac{2}{3},+}^\cW + \chi_{\frac{1}{15},-}^\cW \bar\chi_{\frac{2}{3},-}^\cW\\
& + |\chi_{\frac{1}{15},+}^\cW|^2 + |\chi_{\frac{1}{15},-}^\cW|^2~,
\\
Z_{\eta W}= \,& (\chi_0^\mathcal{W} + \chi_{\frac{2}{5}}^\mathcal{W}) \bar\chi_{\frac{1}{15},-}^{\cW} + \chi_{\frac{1}{15},+}^\cW (\bar\chi_0^\mathcal{W} + \bar\chi_{\frac{2}{5}}^\mathcal{W})  + \chi_{\frac{2}{5}}^\mathcal{W} \bar\chi_{\frac{2}{3},-}^\cW + \chi_{\frac{2}{3},+}^\cW \bar\chi_{\frac{2}{5}}^\mathcal{W}+ \chi_{\frac{2}{3},-}^\cW \bar\chi_{\frac{1}{15},+}^\cW   \\
& +\chi_{\frac{1}{15},-}^\cW \bar\chi_{\frac{2}{3},+}^\cW + \chi_{\frac{1}{15},-}^\cW \bar\chi_{\frac{1}{15},+}^\cW ~, \\
Z_{\bar\eta W} =\,& (\chi_0^\mathcal{W} + \chi_{\frac{2}{5}}^\mathcal{W}) \bar\chi_{\frac{1}{15},+}^{\cW} + \chi_{\frac{1}{15},-}^\cW (\bar\chi_0^\mathcal{W} + \bar\chi_{\frac{2}{5}}^\mathcal{W})  + \chi_{\frac{2}{5}}^\mathcal{W} \bar\chi_{\frac{2}{3},+}^\cW + \chi_{\frac{2}{3},-}^\cW \bar\chi_{\frac{2}{5}}^\mathcal{W}+ \chi_{\frac{2}{3},+}^\cW \bar\chi_{\frac{1}{15},-}^\cW \\
& +\chi_{\frac{1}{15},+}^\cW \bar\chi_{\frac{2}{3},-}^\cW + \chi_{\frac{1}{15},+}^\cW \bar\chi_{\frac{1}{15},-}^\cW ~.
\label{eq:DPF_threePotts_W}
\fe
For defects that preserve the $\mathcal{W}_3' \times \overline{\mathcal{W}}_3$ algebra, the defect partition function can be decomposed into $\mathcal{W}_3' \times \overline{\mathcal{W}_3}$ characters as 
\ie
&Z_N = \chi_{\frac{1}{8}}^{\cW'} (\bar\chi_0^\cW  + \bar\chi_{\frac{2}{3},+}^\cW + \bar\chi_{\frac{2}{3},-}^\cW) + \chi_{\frac{1}{40}}^{\cW'} (\bar\chi_{\frac{2}{5}}^\cW + \bar\chi_{\frac{1}{15},+}^\cW + \bar\chi_{\frac{1}{15},-}^\cW) \,, \\
&Z_{NW} = \chi_{\frac{1}{8}}^{\cW'} (\bar\chi_0^\cW  + \bar\chi_{\frac{2}{3},+}^\cW + \bar\chi_{\frac{2}{3},-}^\cW) + \chi_{\frac{1}{40}}^{\cW'} (\bar\chi_0^\cW  + \bar\chi_{\frac{2}{5}}^\cW + \bar\chi_{\frac{2}{3},+}^\cW + \bar\chi_{\frac{2}{3},-}^\cW + \bar\chi_{\frac{1}{15},+}^\cW + \bar\chi_{\frac{1}{15},-}^\cW) \,.
\label{eqn:ZN}
\fe
The defect partition functions for defects preserving the $\mathcal{W}_3 \times \overline{\mathcal{W}'_3}$ algebra are given by the transpose of the above expressions.
For defects that preserve the $\mathcal{W}_3' \times \overline{\mathcal{W}'_3}$ algebra, the defect partition function can be decomposed into $\mathcal{W}_3' \times \overline{\mathcal{W}'_3}$ characters as
\ie
&Z_C = Z_{C\eta} = Z_{C\bar\eta} = |\chi^{\cW'}_{\frac{1}{8}}|^2 + |\chi^{\cW'}_{\frac{1}{40}} |^2 \,, \\
&Z_{CW} = Z_{\eta CW} = Z_{\bar\eta CW} = \chi_{\frac{1}{8}}^{\mathcal{W}'} \bar\chi_{\frac{1}{40}}^{\cW'} + \chi_{\frac{1}{40}}^{\cW'} \bar\chi_{\frac{1}{8}}^{\cW'} + |\chi_{\frac{1}{40}}^{\cW'}|^2 \,.
\fe
The appearance of the $\mathcal{W}_3'$ characters in these defect partition functions verifies our claim that the $\cW_3'$ algebra has 2 distinct modules with $\cW_3'$ characters given by \eqref{eq:W3'_character}.

\subsection{Tetracritical Ising CFT}
\label{sec:tetracritical_Ising}

\subsubsection{Local Operator Spectrum}
The tetracritical Ising CFT corresponds to the $A$-series Virasoro minimal model with $(p,q)=(6,5)$. It can be obtained from the three-state Potts CFT by gauging the 
$\mathbb{Z}^C_2$ symmetry generated by
$C$. The local and defect operators related to the $\mathbb{Z}^C_2$ symmetry in the three-state Potts CFT are summarized in Table~\ref{tab:3potts_Z2}. In particular, the $C$ defect partition function graded by the $\mathbb{Z}^C_2$ symmetry is given by
\ie
Z_C^C = \cT \cdot Z_C = \bigl|\chi_{\frac{1}{8}} - \chi_{\frac{13}{8}}\bigr|^2 + \bigl|\chi_{\frac{1}{40}} - \chi_{\frac{21}{40}}\bigr|^2 \,,
\fe
where $\cT$ represents the modular $T$ transformation and $Z_C$ denotes the $C$ defect partition function.
From $Z_C^C$, we can read off the $\mathbb{Z}^C_2$ charge of the operators in $C$ defect Hilbert space.

\begin{table}[h]
    \centering
    \renewcommand{\arraystretch}{1.4} 
    \begin{tabular}{|c|c|c|}
    \hline
         & $\mZ_2$-even & $\mZ_2$-odd \\
         \hline
        {\rm untwisted} & $1_{0,0}, Y_{3,3}, \varepsilon_{\frac{2}{5},\frac{2}{5}}, X_{\frac{7}{5},\frac{7}{5}}, \sigma_{\frac{1}{15},\frac{1}{15}}, Z_{\frac{2}{3},\frac{2}{3}}$ & $V_{3,0}, \bar{V}_{0,3}, \Phi_{\frac{7}{5},\frac{2}{5}}, \bar\Phi_{\frac{2}{5},\frac{7}{5}}, \sigma'_{\frac{1}{15},\frac{1}{15}}, Z'_{\frac{2}{3},\frac{2}{3}}$ \vspace{-18pt}\\
    &&\\
        \hline
        $C$-{\rm twisted} & $\zeta_{\frac{1}{8},\frac{1}{8}}, \tilde\zeta_{\frac{13}{8},\frac{13}{8}}, \varphi_{\frac{1}{40},\frac{1}{40}}, \tilde\varphi_{\frac{21}{40},\frac{21}{40}}$ & $\xi_{\frac{1}{8},\frac{13}{8}}, \bar\xi_{\frac{13}{8},\frac{1}{8}}, 
        \phi_{\frac{1}{40},\frac{21}{40}}, \bar\phi_{\frac{21}{40}, \frac{1}{40}}$
        \vspace{-18pt}\\
    &&\\
    \hline
    \end{tabular}
    \caption{Local and defect operator spectrum of the three-state Potts CFT in sectors related to the $\mathbb{Z}^C_2$ symmetry. We use a rotated basis of local operators defined by $\sigma = \tfrac{1}{\sqrt{2}}(\sigma + \sigma^\dagger)$ and $\sigma' = \tfrac{1}{\sqrt{2}}(\sigma - \sigma^\dagger)$, with a similar definition for $Z$.}
    \label{tab:3potts_Z2}
\end{table}

After the $\mZ_2^C$ gauging, we obtain a dual $\mathbb{Z}_2^{\tilde C}$ symmetry generated by $\tilde C$. The $\mZ_2^C$ gauging reshuffles the local and defect operator spectrum as in Table~\ref{tab:tetra_Z2}. 

\begin{table}[h]
    \centering
    \renewcommand{\arraystretch}{1.4} 
    \begin{tabular}{|c|c|c|}
    \hline
         & $\mZ_2$-even & $\mZ_2$-odd \\
         \hline
        {\rm untwisted} & $1_{0,0}, Y_{3,3}, \varepsilon_{\frac{2}{5},\frac{2}{5}}, X_{\frac{7}{5},\frac{7}{5}}, \sigma_{\frac{1}{15},\frac{1}{15}}, Z_{\frac{2}{3},\frac{2}{3}}$ & $\zeta_{\frac{1}{8},\frac{1}{8}}, \tilde\zeta_{\frac{13}{8},\frac{13}{8}}, \varphi_{\frac{1}{40},\frac{1}{40}}, \tilde\varphi_{\frac{21}{40},\frac{21}{40}}$ 
        \vspace{-18pt}\\
    &&\\
        \hline
        $\tilde{C}$-{\rm twisted} & $V_{3,0}, \bar{V}_{0,3}, \Phi_{\frac{7}{5},\frac{2}{5}}, \bar\Phi_{\frac{2}{5},\frac{7}{5}}, \sigma'_{\frac{1}{15},\frac{1}{15}}, Z'_{\frac{2}{3},\frac{2}{3}}$ & $\xi_{\frac{1}{8},\frac{13}{8}}, \bar\xi_{\frac{13}{8},\frac{1}{8}}, \phi_{\frac{1}{40},\frac{21}{40}}, \bar\phi_{\frac{21}{40}, \frac{1}{40}}$
        \vspace{-18pt}\\
    &&\\
    \hline
    \end{tabular}
    \caption{Local and defect operator spectrum of the tetracritical Ising CFT in sectors related to the dual $\mathbb{Z}_2^{\tilde C}$ symmetry generated by $\tilde C$. }
    \label{tab:tetra_Z2}
\end{table}

\noindent Importantly, the spin 3 currents $V(z)$, $\bar V(\bar z)$ are no longer local operators. They are now defect operators attached by the dual $\mathbb{Z}_2^{\tilde C}$ line $\tilde C$. This affects the chiral tube algebra in the tetracritical Ising CFT.

\subsubsection{Topological Defect Lines}
There are in total 10 TDLs in the tetracritical Ising CFT. They are all Verlinde lines with respect to the Virasoro algebra:
\ie
\{1, \tilde{C}, \tilde{C}W, W, M, MW, N, \tilde{C}N, \tilde{C}WN, WN\}\,,
\label{eqn:tdl_tetra}
\fe
which are in one-to-one correspondence with the Virasoro modules. The TDLs in~\eqref{eqn:tdl_tetra} are presented in the same order as the Virasoro modules in~\eqref{eqn:conformal_weight_tetra}. 
The $M$ line is the image of the original TDLs $\eta$, $\bar\eta$, $C\eta$, $C\bar\eta$ after gauging.
The lines $\{1,\tilde{C}, M\}$ form the Rep$(S_3)$ category:
\ie
M^2 = 1+ \tilde{C} + M,\quad M\tilde{C} = \tilde{C}M = M \,,
\label{eq:S3}
\fe
and $N$ is the duality defect associated to the non-invertible gauging described by the algebra object $A = 1+M$:
\ie
N^2 = 1+ M,\quad MN = N + \tilde{C}N \,.
\fe
The Fibonacci category generated by $W$ again does not interact with other TDLs and simply forms a tensor product with the above lines.

\subsubsection{Orbifold of \texorpdfstring{$\mathcal{W}_3$}{W3} Algebra}
\label{sec:orbifold_W3}

Although the spin 3 current $V(z)$ becomes a defect operator attached by the TDL $\tilde C$ in the tetracritical Ising CFT, we can still use it to construct a chiral tube algebra. To this end, we define the projectors
\ie
P_{\tilde{C}\pm} = \frac{1}{2} (1 \pm \tilde{C}) \,,
\fe
which absorb the $\tilde C$ line.  We can then push the non-local current $V(z)$ onto these projectors
\ie
V P_{\tilde C\pm}=\tikz[baseline=-0.75]{\draw[black] (0,0.75)--(0,1.5); \draw (0,0.75) arc (90:45:0.75); \draw[black] (0.53,0.53)--(0.73,0.73); \node[right] at (-0.05,1.35) {\small $\tilde C$}; \node[right] at (0.73,0.73) {\small $V$}; \draw[black] (0,0.5) arc (90:-90:0.5); \node[right] at (0.35,-0.4) {\small $P_{\tilde C\pm}$}; \draw[black] (0,-0.5) arc (-90:-270:0.5)} = \frac{1}{2} \le \ \tikz[baseline=-0.75]{\draw[black] (0,0.75)--(0,1.5); \draw (0,0.75) arc (90:45:0.75); \draw[black] (0.53,0.53)--(0.73,0.73); \node[right] at (-0.05,1.35) {\small $\tilde C$};\node[right] at (0.73,0.73) {\small $V$}} \pm \ 
\tikz[baseline=-0.75]{\draw[black] (0,0.75)--(0,1.5); \draw (0,0.75) arc (90:45:0.75); \draw[black] (0.53,0.53)--(0.73,0.73); \node[right] at (-0.05,1.35) {\small $\tilde C$};\node[right] at (0.73,0.73) {\small $V$}; \draw[black] (0,0.5) arc (90:-90:0.5); \draw[black] (0,-0.5) arc (-90:-270:0.5); \node[right] at (0.35,-0.4) {\small ${\tilde C}$}; } \ri = \frac{1}{2} \le \ 
\tikz[baseline=-0.75]{\draw[black] (0,0.75)--(0,1.5); \draw (0,0.75) arc (90:45:0.75); \draw[black] (0.53,0.53)--(0.73,0.73); \node[right] at (-0.05,1.35) {\small $\tilde C$};\node[right] at (0.73,0.73) {\small $V$}} \pm \ 
\tikz[baseline=-0.75]{\draw[black] (0,0.75)--(0,1.5); \draw (0,0.75) arc (90:405:0.75); \draw[black] (0.53,0.53)--(0.73,0.73); \node[right] at (0.73,0.73) {\small $V$}; \node[right] at (-0.05,1.35) {\small $\tilde C$}; } \ri \,.
\fe
The current $V(z)$ now lives freely on these projectors and has a well-defined monodromy. On $P_{\tilde C+}$, it obeys periodic boundary condition and has an integer mode expansion, whereas on $P_{\tilde C-}$, it satisfies anti-periodic boundary condition giving rise to a half-integer mode expansion. Expanding the current into appropriate modes, we arrive at the following lasso operators that map the local Hilbert space $\cH$ to the $\tilde C$ defect Hilbert space $\cH_{\tilde C}$
\ie
{[W_n]}_I^{\tilde{C}} = \tikz[baseline=-0.75]{\draw[black] (0,0.75)--(0,1.5); \draw (0,0) circle (0.75); \node[right] at (0,1.1) {\small $\tilde{C}$}; \node[right] at (0.6,-0.5) {\small $W_n P_{\tilde{C}+}$}}\,,\quad\
[W_r]_I^{\tilde{C}} = \tikz[baseline=-0.75]{\draw[black] (0,0.75)--(0,1.5); \draw (0,0) circle (0.75); \node[right] at (0,1.1) {\small $\tilde{C}$}; \node[right] at (0.6,-0.5) {\small $W_{r} P_{\tilde{C}-}$}}\, .\label{eqn:tube_w3_1}
\fe
We can similarly push the non-local current $V(z)$ onto the projectors from the other direction 
\ie
V P_{\tilde C\pm}=\tikz[baseline=-0.75]{\draw[black] (0,0.75)--(0,0); \draw (0,0.75) arc (90:45:0.75); \draw[black] (0.53,0.53)--(0.73,0.73); \node[right] at (-0.4,1.05) {\small $\tilde C$}; \node[right] at (0.73,0.73) {\small $V$}; \draw[black] (0,0.5) arc (90:-90:0.5); \node[right] at (0.35,-0.4) {\small $P_{\tilde C\pm}$}; \draw[black] (0,-0.5) arc (-90:-270:0.5)} = \frac{1}{2} \le \ \tikz[baseline=-0.75]{\draw[black] (0,0.75)--(0,0); \draw (0,0.75) arc (90:45:0.75); \draw[black] (0.53,0.53)--(0.73,0.73); \node[right] at (-0.4,1.05) {\small $\tilde C$};\node[right] at (0.73,0.73) {\small $V$}} \pm \ 
\tikz[baseline=-0.75]{\draw[black] (0,0.75)--(0,0); \draw (0,0.75) arc (90:45:0.75); \draw[black] (0.53,0.53)--(0.73,0.73); \node[right] at (-0.4,1.05) {\small $\tilde C$};\node[right] at (0.73,0.73) {\small $V$}; \draw[black] (0,0.5) arc (90:-90:0.5); \draw[black] (0,-0.5) arc (-90:-270:0.5); \node[right] at (0.35,-0.4) {\small ${\tilde C}$}; } \ri = \frac{1}{2} \le \ 
\tikz[baseline=-0.75]{\draw[black] (0,0.75)--(0,0); \draw (0,0.75) arc (90:45:0.75); \draw[black] (0.53,0.53)--(0.73,0.73); \node[right] at (-0.4,1.05) {\small $\tilde C$};\node[right] at (0.73,0.73) {\small $V$}} \pm \ 
\tikz[baseline=-0.75]{\draw[black] (0,0.75)--(0,0); \draw (0,0.75) arc (90:405:0.75); \draw[black] (0.53,0.53)--(0.73,0.73); \node[right] at (0.73,0.73) {\small $V$}; \node[right] at (-0.4,1.05) {\small $\tilde C$}; } \ri \,.
\fe
Expanding it into modes gives the following lasso operators that map from the $\tilde C$ defect Hilbert space $\cH_{\tilde C}$ to the local Hilbert space $\cH$ 
\ie
{[W_n]}_{\tilde{C}}^I = \tikz[baseline=-0.75]{\draw[black] (0,0)--(0,0.75); \draw (0,0) circle (0.75); \node[right] at (0,0.375) {\small $\tilde{C}$}; \node[right] at (0.6,-0.5) {\small $W_{n} P_{\tilde{C}+}$}} \,,\quad\ [W_r]^I_{\tilde{C}} = \tikz[baseline=-0.75]{\draw[black] (0,0)--(0,0.75); \draw (0,0) circle (0.75); \node[right] at (0,0.375) {\small $\tilde{C}$}; \node[right] at (0.6,-0.5) {\small $W_{r} P_{\tilde{C}-}$}}\,.
\label{eqn:tube_w3}
\fe
The chiral tube algebra generated by these lasso operator now mixes the local and defect operators. The lasso operators $[W_n]_I^{\tilde{C}}$ and $[W_n]^I_{\tilde{C}}$ form an algebra isomorphic to $\cW_3$ algebra, while $[W_r]_I^{\tilde{C}}$ and $[W_r]^I_{\tilde{C}}$ form an algebra isomorphic to the twisted $\cW_3'$ algebra. Importantly, these two sets of lasso operators are independent as they compose to zero due to the projectors. Therefore, the chiral tube algebra is isomorphic to $\cW_3\oplus \cW_3'$. This chiral tube algebra can be interpreted as the orbifold of the $\cW_3$ algebra.

Similarly, the anti-chiral tube algebra consists of the same lasso operators built from $\bar V(\bar z)$ and it is isomorphic to $\overline{\cW_3}\oplus\overline{\cW_3'}$. As a result, the local and $\tilde C$ defect Hilbert space, $\mathcal{H}\oplus\mathcal{H}_{\tilde C}$, together can be decomposed into modules under these chiral and anti-chiral tube algebras.
It is straightforward to check that these algebras act consistently on the Hilbert spaces listed in Table \ref{tab:tetra_Z2}.

\subsubsection{Chiral Tube Algebra in Other Defect Hilbert Spaces}
Let us consider the chiral tube algebra acting on the other defect Hilbert spaces. Consider first the $W$ and $ W\tilde C$ defect Hilbert spaces. Since the  $W$ line does not interact with the other TDLs and acts trivially on the spin 3 currents $V(z),\bar V(\bar z)$, the lasso operators acting on these Hilbert spaces is similar to the ones in~\eqref{eqn:tube_w3_1} and~\eqref{eqn:tube_w3}, with the only difference being an additional $W$ line insertion in the vertical direction. For this reason, we shall not elaborate on it further. 

Next, let us move on to consider the chiral tube algebra acting on the $M$ defect Hilbert space. The defect partition function is given by 
\ie
\tilde{Z}_M =\ &\frac{1}{2}(Z_\eta+Z_{\bar\eta}+Z_{C\eta}+Z_{C\bar\eta})
\\
=\ &\chi_0^\cW\bar\chi_{\frac{2}{3}}^\cW + \chi_{\frac{2}{3}}^\cW \bar\chi_0^\cW + \chi_{\frac{2}{5}}^\cW \bar\chi_{\frac{1}{15}}^\cW + \chi_{\frac{1}{15}}^\cW \bar\chi_{\frac{2}{5}}^\cW +|\chi_{\frac{2}{3}}^\cW|^2 + |\chi_{\frac{1}{15}}^\cW|^2+|\chi_{\frac{1}{8}}^{\cW'}|^2 + |\chi_{\frac{1}{40}}^{\cW'} |^2 \,,
\fe
where we use $\tilde{Z}$ to denote the torus partition function of the tetracritical Ising CFT, and $Z$ to denote that of the three-state Potts CFT. Here, we do not distinguish $W_3$ characters by their $W_0$ eigenvalues. As we will explain later, the sign of the eigenvalues is ambiguous.
The $M$ defect partition function graded by the dual $\mathbb{Z}^{\tilde C}_2$ symmetry is given by
\ie
\tilde{Z}_M^{\tilde{C}} =\ &\frac{1}{2}(Z_\eta+Z_{\bar\eta}-Z_{C\eta}-Z_{C\bar\eta})
\\
=\ & \chi^\cW_0 \bar\chi^\cW_{\frac{2}{3}} + \chi^\cW_{\frac{2}{3}} \bar\chi_0^\cW + \chi^\cW_{\frac{2}{5}} \bar\chi_{\frac{1}{15}}^\cW + \chi_{\frac{1}{15}}^\cW \bar\chi_{\frac{2}{5}}^\cW+ |\chi_{\frac{2}{3}}^\cW|^2 + |\chi_{\frac{1}{15}}^\cW|^2 - |\chi_{\frac{1}{8}}^{\cW'}|^2 - |\chi_{\frac{1}{40}}^{\cW'}|^2 \,.
\label{eqn:part_CM}
\fe
Here, we do not need to specify how the four-way junction is resolved because the $F$-symbol $F_{\tilde{C} M \tilde{C}}^M$ is trivial and thus there is no ambiguity in the resolution. From the graded partition function, we can read off the $\mathbb{Z}_2^{\tilde{C}}$ charge of operators in the $M$ defect Hilbert space. It is determined by whether the operators come from the $\eta,\bar\eta$ defect Hilbert space or $C\eta,C\bar\eta$ defect Hilbert space. This will be useful later when we check the chiral tube algebra action. An alternative way to compute this partition function is to solve the $A$-bimodule structures in tetracritical Ising CFT from gauging the $\mZ_2^C$ symmetry in three-state Potts CFT. This allows us to express the torus partition functions in the gauged theory in terms of those of the ungauged theory.

From $\tilde Z^{\tilde C}_M$ in \eqref{eqn:part_CM}, we can infer how $M$ acts on the spin 3 currents $V(z),\bar V(\bar z)$. Performing a modular $S$ transformation, denoted by $\mathcal{S}$, yields the $M$-graded $\tilde C$ defect partition function:
\ie
\tilde{Z}_{\tilde{C}}^M = \cS \cdot \tilde{Z}_M^{\tilde{C}} = 2 \chi_0 \bar\chi_3 + 2 \chi_3 \bar\chi_0 + 2 \chi_{\frac{2}{5}} \bar\chi_{\frac{7}{5}} + 2 \chi_{\frac{7}{5}} \bar\chi_{\frac{2}{5}} - |\chi_{\frac{2}{3}}|^2 - |\chi_{\frac{1}{15}}|^2 \,.
\fe
From this, we see that $M$ acts trivially on the spin 3 currents $V(z)$ and $\bar V(\bar z)$, which are represented by $\chi_3 \bar\chi_0$ and  $\chi_0 \bar\chi_3$ in the $\tilde C$ defect partition function, respectively. The coefficient $2$ in front of these terms reflects the quantum dimension of $M$. 

Since the $M$ line absorbs the $\tilde C$ line, the $\tilde C$ lines attached to the non-local currents $V(z)$ and $\bar V(\bar z)$ can end on the vertical $M$ line in the $M$-defect Hilbert space. The lasso operators are then constructed by pushing the currents onto the $\mathbb{Z}_2^{\tilde C}$ projectors $P_{\tilde C\pm}$:
\ie
V P_{\tilde C\pm}=\tikz[baseline=-0.75]{\draw[black] (0,0)--(0,1.5); \draw (0,0.75) arc (90:45:0.75); \draw[black] (0.53,0.53)--(0.73,0.73); \node[right] at (-0.05,1.35) {\small $M$}; \node[right] at (0.38,0.35) {\small $\tilde C$}; \node[right] at (0.73,0.73) {\small $V$}; \draw[black] (0,0.5) arc (90:-90:0.5); \node[right] at (0.35,-0.4) {\small $P_{\tilde C\pm}$}; \draw[black] (0,-0.5) arc (-90:-270:0.5)}  = \frac{1}{2} \le \ 
\tikz[baseline=-0.75]{\draw[black] (0,0)--(0,1.5); \draw (0,0.75) arc (90:45:0.75); \draw[black] (0.53,0.53)--(0.73,0.73); \node[right] at (-0.05,1.35) {\small $M$}; \node[right] at (0.38,0.35) {\small $\tilde C$};\node[right] at (0.73,0.73) {\small $V$}} \pm \ 
\tikz[baseline=-0.75]{\draw[black] (0,0)--(0,1.5); \draw (0,0.75) arc (90:405:0.75); \draw[black] (0.53,0.53)--(0.73,0.73); \node[right] at (0.73,0.73) {\small $V$}; \node[right] at (-0.05,1.35) {\small $M$}; \node[right] at (0.58,-0.4) {\small $\tilde C$}; } \ri \,.\label{eq:M_3way}
\fe
Since $M$ acts trivially on the spin 3 currents $V(z),\bar V(\bar z)$ have integer mode expansions on $P_{\tilde C+}$ and half-integer mode expansions on $P_{\tilde C-}$. Expanding the currents into modes, we obtain the following lasso operators from $V(z)$:
\ie
{[W_n]}_{M}^{M} = \tikz[baseline=-0.75]{\draw[black] (0,0)--(0,1.5); \draw (0,0) circle (0.75); \node[right] at (0,0.35) {\small $M$}; \node[right] at (0,1.1) {\small $M$};  
\node[right] at (0.6,-0.5) {\small $W_n P_{\tilde{C}+}$}} \,,\quad
[W_r]_{M}^{M} = \tikz[baseline=-0.75]{\draw[black] (0,0)--(0,1.5); \draw (0,0) circle (0.75); \node[right] at (0,0.35) {\small $M$}; \node[right] at (0,1.1) {\small $M$};  
\node[right] at (0.6,-0.5) {\small $W_r P_{\tilde{C}-}$}} \,.
\fe
There is a similar set of lasso operators built from $\bar V(\bar z)$. The lasso operators acting on $\mathbb{Z}_2^{\tilde C}$-even $M$ defect Hilbert space form an algebra isomorphic to $\cW_3\times \overline{\cW_3}$ algebra, while those acting on $\mathbb{Z}_2^{\tilde C}$-odd Hilbert space form an algebra isomorphic to $\cW_3'\times \overline{\cW_3'}$ algebra.
This agrees with what is displayed in the $\mathbb{Z}_2^{\tilde C}$-graded $M$ defect partition function \eqref{eqn:part_CM}:~in the $M$ defect Hilbert space, the $\tilde{C}$-even sector forms modules under the $\mathcal{W}_3\times \overline{\cW_3}$ algebra, while the $\tilde{C}$-odd sector forms modules under the $\mathcal{W}_3'\times\overline{\cW_3'}$ algebra.

Note that in the $\tilde{C}$-even sector, the $W_3$ modules with opposite $W_0$ eigenvalues are indistinguishable. It is because $[W_0]^M_M$, when expanding into explicit terms as in \eqref{eq:M_3way}, involves two types of three-ways junctions: $\text{Hom}(M\times\tilde C,M)$ and $\text{Hom}(\tilde C\times M,M)$.
Importantly, there exists a gauge transformation that redefines these three-way junctions simultaneously by a $-1$ phase without changing the $F$-symbols.\footnote{This is characterized by the lazy 2-cohomology $H_l^2(\cC, \mC^*)$ of a fusion category\cite{Perez-Lona:2024yih,panaite2010pseudosymmetric}. There is a naturally induced homomorphism from it into the Brauer-Picard group ${\rm BrPic}(\cC)$\cite{Etingof:2009yvg,Diatlyk:2023fwf}.} Thus, we can redefine $[W_0]^M_M$ by a $-1$ phase freely, and as a result, there are only 4 inequivalent $W_3$ modules in the $M$ defect Hilbert space. They are distinguished by their conformal weight $h\in\{0,\frac{1}{15},\frac{2}{3},\frac{2}{5}\}$. 

Finally, let us discuss the chiral tube algebra acting on the $N$ and $\tilde{C} N$ defect Hilbert space. The relevant defect partition function is 
\ie
\tilde{Z}_N =\ & \frac{1}{2}(Z_N+Z_N^C + Z_{NC}+Z_{NC}^C )
\\
=\ & \chi_0 \bar\chi_{\frac{1}{8}} + \chi_{\frac{1}{8}} \bar\chi_0 + \chi_3 \bar\chi_{\frac{13}{8}} +  \chi_{\frac{13}{8}} \bar\chi_3 + \chi_{\frac{2}{5}} \bar\chi_{\frac{1}{40}} + \chi_{\frac{1}{40}} \bar\chi_{\frac{2}{5}} + \chi_{\frac{7}{5}} \bar\chi_{\frac{21}{40}} + \chi_{\frac{21}{40}} \bar\chi_{\frac{7}{5}} \\
&+ \chi_{\frac{2}{3}} (\bar\chi_{\frac{1}{8}} + \bar\chi_{\frac{13}{8}}) + ( \chi_{\frac{1}{8}} + \chi_{\frac{13}{8}} ) \bar\chi_{\frac{2}{3}} + \chi_{\frac{1}{15}} ( \bar\chi_{\frac{1}{40}} + \bar\chi_{\frac{21}{40}} ) + ( \chi_{\frac{1}{40}} + \chi_{\frac{21}{40}} ) \bar\chi_{\frac{1}{15}} \,, \label{eq:ZN}
\fe
\ie
\tilde{Z}_{\tilde{C} N}=\ & \frac{1}{2}(Z_N-Z_N^C + Z_{NC}-Z_{NC}^C ) 
\\
=\ & \chi_0 \bar\chi_{\frac{13}{8}} + \chi_{\frac{13}{8}} \bar\chi_0 + \chi_3 \bar\chi_{\frac{1}{8}} +  \chi_{\frac{1}{8}} \bar\chi_3 + \chi_{\frac{7}{5}} \bar\chi_{\frac{1}{40}} + \chi_{\frac{1}{40}} \bar\chi_{\frac{7}{5}} + \chi_{\frac{2}{5}} \bar\chi_{\frac{21}{40}} + \chi_{\frac{21}{40}} \bar\chi_{\frac{2}{5}} \\
&+ \chi_{\frac{2}{3}} (\bar\chi_{\frac{1}{8}} + \bar\chi_{\frac{13}{8}}) + ( \chi_{\frac{1}{8}} + \chi_{\frac{13}{8}} ) \bar\chi_{\frac{2}{3}} + \chi_{\frac{1}{15}} ( \bar\chi_{\frac{1}{40}} + \bar\chi_{\frac{21}{40}} ) + ( \chi_{\frac{1}{40}} + \chi_{\frac{21}{40}} ) \bar\chi_{\frac{1}{15}} \,. \label{eq:ZCN}
\fe
The $\tilde{C}$-graded partition function can also be worked out in the same spirit as~\eqref{eqn:part_CM}:
\ie
\tilde{Z}_N^{\tilde{C}} =\ & \frac{1}{2}(Z_N+Z_N^C - Z_{NC}-Z_{NC}^C )
\\
=\ & - \chi_0 \bar\chi_{\frac{1}{8}} + \chi_{\frac{1}{8}} \bar\chi_0 - \chi_3 \bar\chi_{\frac{13}{8}} +  \chi_{\frac{13}{8}} \bar\chi_3 - \chi_{\frac{2}{5}} \bar\chi_{\frac{1}{40}} + \chi_{\frac{1}{40}} \bar\chi_{\frac{2}{5}} - \chi_{\frac{7}{5}} \bar\chi_{\frac{21}{40}} + \chi_{\frac{21}{40}} \bar\chi_{\frac{7}{5}} \\
&- \chi_{\frac{2}{3}} (\bar\chi_{\frac{1}{8}} + \bar\chi_{\frac{13}{8}}) + ( \chi_{\frac{1}{8}} + \chi_{\frac{13}{8}} ) \bar\chi_{\frac{2}{3}} - \chi_{\frac{1}{15}} ( \bar\chi_{\frac{1}{40}} + \bar\chi_{\frac{21}{40}} ) + ( \chi_{\frac{1}{40}} + \chi_{\frac{21}{40}} ) \bar\chi_{\frac{1}{15}} \,, 
\\
\tilde{Z}_{\tilde{C} N}^{\tilde{C}} =\ & \frac{1}{2}(Z_N-Z_N^C - Z_{NC}+Z_{NC}^C ) 
\\
=\ & -\chi_0 \bar\chi_{\frac{13}{8}} + \chi_{\frac{13}{8}} \bar\chi_0 - \chi_3 \bar\chi_{\frac{1}{8}} +  \chi_{\frac{1}{8}} \bar\chi_3 - \chi_{\frac{7}{5}} \bar\chi_{\frac{1}{40}} + \chi_{\frac{1}{40}} \bar\chi_{\frac{7}{5}} - \chi_{\frac{2}{5}} \bar\chi_{\frac{21}{40}} + \chi_{\frac{21}{40}} \bar\chi_{\frac{2}{5}} \\
&- \chi_{\frac{2}{3}} (\bar\chi_{\frac{1}{8}} + \bar\chi_{\frac{13}{8}}) + ( \chi_{\frac{1}{8}} + \chi_{\frac{13}{8}} ) \bar\chi_{\frac{2}{3}} - \chi_{\frac{1}{15}} ( \bar\chi_{\frac{1}{40}} + \bar\chi_{\frac{21}{40}} ) + ( \chi_{\frac{1}{40}} + \chi_{\frac{21}{40}} ) \bar\chi_{\frac{1}{15}}\,. \label{eq:ZCNC}
\fe
By performing a modular $S$ transformation, we can infer how $N$ and $N\tilde C$ acts on the $\tilde{C}$ defect Hilbert space:
\ie
&\tilde{Z}_{\tilde{C}}^N = \cS \cdot \tilde{Z}_N^{\tilde{C}} = \sqrt{3} (\chi_0 \bar\chi_3 - \chi_3 \bar\chi_0 - \chi_{\frac{2}{5}} \bar\chi_{\frac{7}{5}} + \chi_{\frac{7}{5}} \bar\chi_{\frac{2}{5}}) - \chi_{\frac{1}{8}} \bar\chi_{\frac{13}{8}} + \chi_{\frac{13}{8}} \bar\chi_{\frac{1}{8}} + \chi_{\frac{1}{40}} \bar\chi_{\frac{21}{40}} - \chi_{\frac{21}{40}} \bar\chi_{\frac{1}{40}} \,,
\\
&\tilde{Z}_{\tilde{C}}^{N\tilde{C}} = \cS \cdot \tilde{Z}_{N\tilde{C}}^{\tilde{C}} = \sqrt{3} (\chi_0 \bar\chi_3 -  \chi_3 \bar\chi_0 -  \chi_{\frac{2}{5}} \bar\chi_{\frac{7}{5}} +  \chi_{\frac{7}{5}} \bar\chi_{\frac{2}{5}}) + \chi_{\frac{1}{8}} \bar\chi_{\frac{13}{8}} - \chi_{\frac{13}{8}} \bar\chi_{\frac{1}{8}} - \chi_{\frac{1}{40}} \bar\chi_{\frac{21}{40}} + \chi_{\frac{21}{40}} \bar\chi_{\frac{1}{40}} \,.
\label{eqn:part_NC}
\fe
In particular, we find that both $N$ and $N\tilde C$ act on the spin 3 current $V(z)$ by a $-1$ phase and act trivially on $\bar V(\bar z)$. Because of this $-1$ phase, in the $N$ and $N\tilde C$ defect Hilbert space, the current $V(z)$ has opposite mode expansion compared to \eqref{eqn:tube_w3_1} and \eqref{eqn:tube_w3}. It has half-integer mode expansion on $P_{\tilde C+}$ and integer mode expansion on $P_{\tilde C-}$, which leads to the following lasso operators:
\ie
&[W_r]_{N}^{N\tilde{C}} = \tikz[baseline=-0.75]{\draw[black] (0,0)--(0,1.5); \draw (0,0) circle (0.75); \node[right] at (0,0.35) {\small $N$}; \node[right] at (0,1.1) {\small $N\tilde{C}$};  
\node[right] at (0.6,-0.5) {\small $W_r P_{\tilde{C}+}$}} \,,\quad
[W_r]_{N\tilde{C}}^{N} = \tikz[baseline=-0.75]{\draw[black] (0,0)--(0,1.5); \draw (0,0) circle (0.75); \node[right] at (0,0.35) {\small $N\tilde{C}$}; \node[right] at (0,1.1) {\small $N$};  
\node[right] at (0.6,-0.5) {\small $W_r P_{\tilde{C}+}$}} \,, \\
&[W_n]_{N}^{N\tilde{C}} = \tikz[baseline=-0.75]{\draw[black] (0,0)--(0,1.5); \draw (0,0) circle (0.75); \node[right] at (0,0.35) {\small $N$}; \node[right] at (0,1.1) {\small $N\tilde{C}$};  
\node[right] at (0.6,-0.5) {\small $W_n P_{\tilde{C}-}$}} \,,\quad
[W_n]_{N\tilde{C}}^{N} = \tikz[baseline=-0.75]{\draw[black] (0,0)--(0,1.5); \draw (0,0) circle (0.75); \node[right] at (0,0.35) {\small $N\tilde{C}$}; \node[right] at (0,1.1) {\small $N$};  
\node[right] at (0.6,-0.5) {\small $W_n P_{\tilde{C}-}$}} \,.
\fe
On the other hand, since $N$ and $N\tilde C $ act trivially on $\bar V(\bar z)$, it has the same mode expansion as in \eqref{eqn:tube_w3_1} and \eqref{eqn:tube_w3}. As a result, the lasso operators acting on the $\mathbb{Z}_2^{\tilde C}$-even $N,N\tilde C$ defect Hilbert spaces form an algebra isomorphic to $\cW_3'\times \overline{\cW_3}$ algebra, while those acting on $\mathbb{Z}_2^{\tilde C}$-odd Hilbert spaces form an algebra isomorphic to $\cW_3\times \overline{\cW_3'}$ algebra. This is consistent with the defect partition functions \eqref{eq:ZN}, \eqref{eq:ZCN} and \eqref{eq:ZCNC}.

\section{\texorpdfstring{$\mf{su}(2)_1$}{su(2)1} Kac-Moody Algebra and Orbifolds}
\label{sec:SU(2)}

Kac-Moody algebras are another important class of chiral algebras. They are generated by  the stress tensor $T(z)$ and chiral spin 1 currents $J^a(z)$. The existence of spin 1 currents implies a continuous Lie-group symmetry, whose associated TDLs can twist the chiral algebra. 
Such twisted Kac-Moody algebras have been studied in the literature (see e.g.~\cite{Goddard:1986bp}). In this section, we will focus on the $\mf{su}(2)_1$ Kac-Moody algebra. We will discuss its twisted algebras and use them to construct the associated $\mf{su}(2)_1$ chiral tube algebra and its orbifolds. We will then illustrate these constructions with concrete examples, including the $SU(2)_1$ WZW model and the $c=1$ free compact boson CFTs.

\subsection{\texorpdfstring{$\mf{su}(2)_1$}{su(2)1} Chiral Tube Algebra}
\label{subsec:su2_twisted_algebra}

We begin by reviewing the $\mf{su}(2)_1$ chiral algebra.
It consists of three spin 1 currents denoted by $J^a(z)$ with $a=1,2,3$. Their OPE is given by
\ie
J^a(z) J^b(w) \sim \frac{\delta^{ab}}{2(z-w)^2} + \frac{i \varepsilon^{abc} J^c(w)}{z-w} \,.
\fe
Expanding them in modes,
\ie
J^a(z) = \sum_{n\in\mZ} \frac{J^a_n}{z^{n+1}} \,,
\fe
we arrive at the Kac-Moody algebra:\footnote{In this convention, the currents are normalized such that the zero modes $J^a_0$ act as
$\frac{1}{2}\sigma^a$ on the fundamental $su(2)$ representation, with $\sigma^a$ denoting the Pauli matrices.}
\ie
{[J_n^a, J_m^b]} = \sum_c i\varepsilon_{abc} J_{n+m}^c + \frac{n}{2} \delta_{ab} \delta_{n+m,0} \,.
\label{eqn:su2_1_algebra}
\fe
We can choose a Cartan generator $J^3$ and define the corresponding raising and lowering operator $J^{\pm} = J^1 \pm iJ^2$, which will be convenient later. In this basis, the algebra becomes
\ie
{[J^3_n, J^3_m]} &= \frac{n}{2} \delta_{n+m,0} \,, \\
[J^+_n, J^+_m] &= [J^-_n, J^-_m] = 0 \,, \\
[J^+_n, J^-_m] &= n \delta_{n+m,0} + 2J^3_{n+m} \,, \\
[J^3_n, J^{\pm}_m] &= \pm J^{\pm}_{n+m} \,.
\label{eq:SU(2)_chiral_algebra}
\fe

The stress tensor is obtained via the Sugawara construction
\ie
T(z) = \frac{1}{3} \sum_a (J^aJ^a)(z) \,.
\label{eqn:sugawara}
\fe
The normalization constant is fixed from the $TJ$ OPE
\ie
T(z) J^a(w) \sim \frac{J^a(w)}{(z-w)^2} + \frac{\partial J^a(w)}{z-w} \,.
\fe
One can do a similar mode expansion for the stress tensor
\ie
T(z) = \sum_{n\in\mZ} \frac{L_n}{z^{n+2}} \,.\label{eqn:stress_mode}
\fe
For these modes, the Sugawara construction takes the form
\ie
L_n = \frac{1}{3} \sum_a \left\{ \sum_{m \leq -1} J^a_m J^a_{n-m} + \sum_{m \geq 0} J^a_{n-m} J^a_m \right\} \,.
\label{eqn:sugawara_mode}
\fe
It leads to the following commutation relations in addition to the Virasoro algebra
\ie
{[L_n, J_m^a]} = -m J^a_{n+m} \,.
\fe

\subsubsection{Twisted Algebra}

The existence of an $\mathfrak{su}(2)_1$ chiral  algebra implies an $SU(2)$ global symmetry. We will focus on the $U(1)$ Cartan subgroup generated by $J^3(z)$ since any other elements of the $SU(2)$ symmetry are related to this $U(1)$ Cartan subgroup by conjugation. The TDLs of this $U(1)$ Cartan subgroup take the form
\ie
\cL_{\thetaa} = \exp\left[{\thetaa \oint_C dz J^3(z)}\right].
\label{eqn:cartan}
\fe
The symmetry parameter $\thetaa$ has a periodicity of 2, with $\thetaa=1$ being the center element of $SU(2)$.
From the $JJ$ OPE, we find that the $U(1)$ Cartan symmetry acts on the currents as
\begin{equation} \cL_{\thetaa} \cdot J^\pm (z) = e^{\pm2\pi i \thetaa} J^\pm(z)\,, \qquad \cL_{\thetaa} \cdot J^3(z) = J^3(z)\,.
\label{eq:jjopetwisted}\end{equation} As a result, in the $\cL_{\thetaa}$ defect Hilbert space, the currents obey a twisted boundary condition:\footnote{There is a relative minus sign in the phases between (\ref{eq:jjopetwisted}) and (\ref{eq:s1twistedlx})  because $J^{\pm}(z)$ passes through the symmetry defect from a different direction.}
\ie
J^\pm(z e^{2\pi i}) = e^{\mp 2\pi i \thetaa} J^\pm(z) \,,\qquad J^3(z e^{2\pi i}) = J^3(z)\,,
\label{eq:s1twistedlx}
\fe
which modifies the mode expansion to 
\ie
J^\pm(z) = \sum_{n \in \mZ} \frac{J^\pm_{n\pm\thetaa}}{z^{n \pm \thetaa+1}}\,,\qquad J^3(z) = \sum_{n \in \mZ } \frac{J^3_{n}}{z^{n+1}}\,.\label{eq:J^a_mode}
\fe
Importantly, there is a shift of $\pm\thetaa$ in the Laurent expansion for $J^{\pm}(z)$. 

Using these modes, we can construct the following lasso operators that act on the $\cL_{\thetaa}$ defect Hilbert space: 
\ie
J^\pm_{ \thetaa,n} := \tikz[baseline=-0.75]{\draw[black] (0,0)--(0,1.5); \draw[black, -{Stealth[round, length=5pt, width=5pt, bend]}] (0,0)--(0,0.4); \draw[black, -{Stealth[round, length=5pt, width=5pt, bend]}] (0,0)--(0,1.2); \draw (0,0) circle (0.75); \node[right] at (0,0.35) {\small $\cL_{\thetaa}$}; \node[right] at (0,1.1) {\small $\cL_{\thetaa}$}; \draw[black,-{Stealth[round, length=5pt, width=5pt, bend]}] (-0.75,0) arc (180:360:0.75); \node[right] at (0.5,-0.5) {\small $J_{n\pm \thetaa}^\pm$}},
\qquad 
J^3_{x,n}  := \tikz[baseline=-0.75]{\draw[black] (0,0)--(0,1.5); \draw[black, -{Stealth[round, length=5pt, width=5pt, bend]}] (0,0)--(0,0.4); \draw[black, -{Stealth[round, length=5pt, width=5pt, bend]}] (0,0)--(0,1.2); \draw (0,0) circle (0.75); \node[right] at (0,0.35) {\small $\cL_{\thetaa}$}; \node[right] at (0,1.1) {\small $\cL_{\thetaa}$}; \draw[black,-{Stealth[round, length=5pt, width=5pt, bend]}] (-0.75,0) arc (180:360:0.75); \node[right] at (0.5,-0.5) {\small $J_{n}^3$}} \,.
\fe
They generate the $\mathfrak{su}(2)_1$ chiral tube algebra. They can be composed only if they share the same twist $x$. For a fixed $x$, they form a twisted $\mathfrak{su}(2)_1$ chiral algebra:
\ie
{[J^3_{\thetaa,n}, J^3_{\thetaa,m}]} &= \frac{n}{2} \delta_{n+m,0} \,, \\
[J^+_{\thetaa,n}, J^+_{\thetaa,m}] &= [J^-_{\thetaa,n}, J^-_{\thetaa,m}] = 0 \,, \\
[J^+_{\thetaa,n}, J^-_{\thetaa,m}] &= (n + \thetaa) \delta_{n+m,0} + 2J^3_{\thetaa,n+m} \,, \\
[J^3_{\thetaa,n}, J^{\pm}_{\thetaa,m}] &= \pm J^{\pm}_{\thetaa,n+m} \,.
\label{eqn:twisted_su2_chiral_algebra}
\fe
This algebra is isomorphic to the Kac-Moody algebra \eqref{eq:SU(2)_chiral_algebra} by the following redefinition
\ie
\Big(J^3_{x,n},J^\pm_{x,n}\Big)\rightarrow \left(J^3_n-\frac{x}{2}\delta_{n,0},J^\pm_{n}\right)\,.
\fe

The stress tensor $T(z)$ by definition is preserved by the TDLs. Thus, its mode expansion~\eqref{eqn:stress_mode} remains unmodified in the $\mathcal{L}_x$ defect Hilbert space. Using these modes, we can define the following stress tensor lasso operators that act on the $\mathcal{L}_x$ defect Hilbert space:
\ie
L_{\thetaa, n} := \tikz[baseline=-0.75]{\draw[black] (0,0)--(0,1.5); \draw[black, -{Stealth[round, length=5pt, width=5pt, bend]}] (0,0)--(0,0.4); \draw[black, -{Stealth[round, length=5pt, width=5pt, bend]}] (0,0)--(0,1.2); \draw (0,0) circle (0.75); \node[right] at (0,0.35) {\small $\cL_{\thetaa}$}; \node[right] at (0,1.1) {\small $\cL_{\thetaa}$}; \draw[black,-{Stealth[round, length=5pt, width=5pt, bend]}] (-0.75,0) arc (180:360:0.75); \node[right] at (0.5,-0.5) {\small $L_{n}$}} \,.
\fe
Same as the current lasso operators $J^a_{x,n}$, these stress tensor lasso operators compose with themselves and with the current lasso operators only if they share the same twist $x$. For a fixed $x$, the stress tensor lasso operators form an algebra isomorphic to the Virasoro algebra and their commutation relations with the current lasso operators are
\ie
{[L_{\thetaa,n}, J_{\thetaa,m}^3]} &= -m J^3_{\thetaa,n+m}\,,
\\
[L_{\thetaa,n}, J_{\thetaa,m}^\pm] &= -(m\pm x) J^\pm_{\thetaa,n+m}\,.\label{eq:TJ_algebra_twist}
\fe

By the Sugawara construction \eqref{eqn:sugawara_mode}, these stress tensor lasso operators $L_{x,n}$ should be expressible in terms of the current lasso operators $J^a_{x,n}$. However, the Sugawara construction generically has a normal ordering ambiguity. The most general form of this ambiguity is
\ie
L_{\thetaa,n} = \frac{1}{3}  \sum_{a}\le \sum_{m \leq -1} J^a_{\thetaa,m} J^a_{\thetaa,n-m} +  \sum_{m \geq 0} J^a_{\thetaa,n-m} J^a_{\thetaa,m}   \ri + A \delta_{n,0} + B_n J^3_{\thetaa,n}\,.
\label{eqn:sugawara_twist}
\fe
Here, the constant $A$ appears only for $n =0$ because only in this case, a central term is generated when we change the ordering using the commutation relations in \eqref{eqn:twisted_su2_chiral_algebra}.
In Appendix~\ref{app:normalorder}, we fix the coefficients in (\ref{eqn:sugawara_twist}) to be
\begin{equation}
    A = -\frac{1}{6}\thetaa^2\,,~~~~B_n = -\frac23 \thetaa\,,
\end{equation}
by imposing $[L_{\thetaa,n}, J^+_{\thetaa,0}] = -\thetaa J^+_{\thetaa,n}$ and $[L_{\thetaa,1}, L_{\thetaa,-1}] = 2L_{\thetaa,0}$. One can verify that with these coefficients, the twisted Sugawara construction \eqref{eqn:sugawara_twist} is consistent with the twisted algebras \eqref{eqn:twisted_su2_chiral_algebra} and \eqref{eq:TJ_algebra_twist}.

One can repeat the same construction for the $\mathfrak{su}(2)_1$ anti-chiral algebra 
\ie
\bar J^a(\bar z) \bar J^b(\bar w) \sim \frac{\delta^{ab}}{2(\bar z-\bar w)^2} + \frac{i \varepsilon^{abc} \bar J^c(\bar w)}{\bar z-\bar w} \,.
\fe
There are however a few subtleties regarding the sign conventions. First, the TDLs for the $U(1)$ Cartan subgroup are defined as
\ie
\cL_{\bar\thetaa} = \exp\left[-\bar\thetaa \oint_C d\bar{z} \bar{J}^3(\bar{z})\right] \,,
\fe
with an extra sign in the exponent. It acts on the currents in the same way as its chiral counterpart:
\begin{equation} \cL_{\bar\thetaa} \cdot \bar J^\pm (\bar z) = e^{\pm2\pi i \bar \thetaa} \bar J^\pm(\bar z)\,, \qquad \cL_{\bar\thetaa} \cdot \bar J^3(\bar z) = \bar J^3(\bar z)\,.
\end{equation}
Second, in the $\cL_{\bar\thetaa}$ defect Hilbert space, the anti-chiral algebra is twisted in the direction opposite to the chiral algebra, so that the mode numbers of $\bar{J}^{\pm}(\bar z)$ are shifted by $\mp \bar\thetaa$
\ie
\bar J^\pm(\bar z) = \sum_{n \in \mZ} \frac{\bar J^\pm_{n\mp\bar\thetaa}}{\bar z^{n \mp \bar \thetaa+1}}\,,\qquad \bar J^3(\bar z) = \sum_{n \in \mZ } \frac{\bar J^3_{n}}{\bar z^{n+1}}\,.
\fe
This is because the monodromy of the $\bar z$ coordinate has the opposite orientation compared to that of the $z$ coordinate. 

\subsubsection{Highest Weight Modules and Spectral Flow}

The $\mf{su}(2)_1$ Kac-Moody algebra has two irreducible highest weight modules. They can be constructed by applying $J_n^a$ with $n\leq 0$ on a highest weight state $|\phi\ra$, which satisfies
\ie
&J^+_0 |\phi\ra= 0 \,,\\
&J^3_n |\phi\ra = J^+_n |\phi\ra = J^-_n |\phi\ra = 0 \,,\ \text{for } n>0 \,.
\fe
These modules are labeled by their $SU(2)$ representation at grade $0$, namely the truncated $SU(2)$ representation generated by the action of $J^-_0$ on $|\phi\rangle$.
One of the two modules corresponds to the 1-dimensional trivial representation at grade $0$, while the other corresponds to the 2-dimensional fundamental representation.
Conventionally, they are also labeled by their affine Dynkin labels, $[1,0]$ and $[0,1]$, respectively. 
Their characters are given by
\ie
K^2_0(\tau) &=  \frac{1}{\eta(\tau)} \sum_{m\in\mZ} q^{m^2} = \frac{\theta_3(2\tau)}{\eta(\tau)} \,, \\
K^2_1(\tau) &=  \frac{1}{\eta(\tau)} \sum_{m\in\mZ} q^{(m+\frac{1}{2})^2} = \frac{\theta_2(2\tau)}{\eta(\tau)} \,.
\label{su2_character}
\fe 
For later use, we also introduce the characters refined by the $U(1)$ Cartan symmetry\,:
\ie
K^2_0(\tau, z) &= \frac{1}{\eta(\tau)} \sum_{m\in\mZ} q^{m^2} y^m \,, \\
K^2_1(\tau, z) &= \frac{1}{\eta(\tau)} \sum_{m\in\mZ} q^{(m+\frac{1}{2})^2} y^{m+\frac{1}{2}} \,,
\label{eqn:refined_su2}
\fe
where the states in the trace are weighted by $y^{J_0^3}=e^{2\pi i z J_0^3}$. 
Here, we adopt the notation that complies with the standard $\mf{u}(1)_N$ characters
\ie
K^N_{\lambda}(\tau) = \frac{1}{\eta(\tau)} \sum_{n\in\mZ} q^{\frac{(Nn + \lambda)^2}{2N}} \,.\label{eq:u(1)_N_character}
\fe
Recall that the $\mathfrak{u}(1)_N$ chiral algebra is generated by a $u(1)$ current $J(z)$ together with two vertex operators $V^\pm(z)$ of charge $\pm N$, whose OPE includes
\ie
J(z)J(w)\sim \frac{N}{(z-w)^2}\,,
\qquad
 J(z) V^\pm(w)\sim \pm\frac{N V^\pm(w)}{z-w}\,.
\fe
For $N=2$, the $\mathfrak{u}(1)_2$ chiral algebra is isomorphic to the $\mathfrak{su}(2)_1$ chiral algebra under the identification \ie
\Big(J^3(z), J^\pm(z)\Big)=\left(\frac{1}{2}J(z),V^\pm(z)\right)\,.
\fe
Thus, they share the same characters.

What are the highest weight modules of the twisted $\mf{su}(2)_1$ chiral algebra? 
Since the twisted algebra is isomorphic to the $\mf{su}(2)_1$ Kac-Moody algebra, their highest weight modules share the same module structure and differ only by a shift in conformal weights and $u(1)$ charges. Therefore, as in the untwisted case, we can label the modules in terms of their Dynkin labels:~$[1,0]$ and $[0,1]$. 

For these twisted modules, their highest weight state $|\phi\ra$  satisfies
\ie
J^a_{\thetaa,n} |\phi\ra  = 0 \,, \text{ for } n>0 \,.
\fe
In addition, $|\phi\ra$ is also a highest weight state of the twisted ${su}(2)$ algebra formed by $J^a_{\thetaa,0}$\,:
\ie
\relax [J^3_{\thetaa,0}, J^\pm_{\thetaa,0}] &= \pm J^\pm_{\thetaa,0}  \,,\quad
[J^+_{\thetaa,0}, J^-_{\thetaa,0}] &= \thetaa  + 2J^3_{\thetaa,0} \,.
\fe
We can therefore denote $|\phi\ra = |\lambda, Q\ra$, with $\lambda\in\mathbb{Z}_{\geq 0}$ the Dynkin label of the twisted $su(2)$ module and $Q=\frac{1}{2}(\lambda - \thetaa)$ the $J^3_{\thetaa,0}$ eigenvalue of $|\phi\ra$. The shift in $Q$ by $-\frac{1}{2}\thetaa$ follows from the fact that the twisted ${su}(2)$ algebra is isomorphic to the untwisted algebra by redefining
\ie
\Big(J^3_{x,0},J^\pm_{x,0}\Big)\rightarrow \left(J^3_0-\frac{x}{2},J^\pm_{0}\right)\,.
\fe
The full twisted $su(2)$ module is generated by acting $J^-_{\thetaa,0}$ on $|\phi\rangle$ repeatedly. This produces states with $J^3_{\thetaa,0}$ eigenvalues 
\ie
Q \in \left\{-\frac{1}{2}(\lambda + \thetaa),\, -\frac{1}{2}(\lambda + \thetaa) + 1,\, \dots,\, \frac{1}{2}(\lambda - \thetaa)\right\}\,.
\fe
Compared to the untwisted $su(2)$ modules, the spectrum of $J_{\thetaa,0}^3$ eigenvalue is uniformly shifted by $-\frac{1}{2}\thetaa$.
As a consequence, the TDL $\mathcal{L}_\alpha=e^{\alpha\oint dz J^3(z)}$ acts on the $\mathcal{L}_x$ defect Hilbert space by an additional phase $e^{-\pi i \alpha \thetaa}$. 
This reflects the self anomaly of the $U(1)$ Cartan symmetry, inherited from the nontrivial Kac-Moody level $k=1$ of $\mathfrak{su}(2)_1$.

The conformal weight of the highest weight states can be computed as follows:
\begin{align}
\la \lambda, Q |\, L_{\thetaa,0} \,| \lambda, Q \ra \nonumber
=\,& \la \lambda, Q | \left[\frac{1}{3}\le \frac{1}{2} J^+_{\thetaa,0} J^-_{\thetaa,0} + \frac{1}{2} J^-_{\thetaa,0} J^+_{\thetaa,0} + J^3_{\thetaa,0} J^3_{\thetaa,0} \ri - \frac{2}{3} \thetaa J^3_{\thetaa,0} - \frac{1}{6}\thetaa^2 \right] |\lambda, Q \ra \nonumber \\
=\,& \frac{Q(Q+1)}{3} - \frac{2}{3}\thetaa Q - \frac{\thetaa(\thetaa-1)}{6} \,,
\label{eq:finalhexpression}
\end{align}
where in the second line of (\ref{eq:finalhexpression}), we used the commutation relation between $J_{\thetaa,0}^+$ and $J_{\thetaa,0}^-$ and the fact that $J^+_{\thetaa,0}$ annihilates $|\lambda,Q\rangle$. 
We thus have
\ie
\left\la 0,-\frac{\thetaa}{2} \right| L_{\thetaa,0} \left| 0, - \frac{\thetaa}{2} \right\ra = \frac{\thetaa^2}{4}\,, \qquad \left\la 1,\frac{1-\thetaa}{2} \right| L_{\thetaa,0} \left| 1, \frac{1-\thetaa}{2} \right\ra = \frac{(1 - \thetaa)^2}{4} \,.
\fe

Using this information, we can construct the corresponding twisted $\mf{su}(2)_1$ characters from the refined characters~\eqref{eqn:refined_su2}.
Observe that $J^{\pm}$ change the $U(1)$ charge by $\pm 1$, while shifting the conformal weight by $\pm \thetaa$ relative to the original $\mathfrak{su}(2)_1$ module. Meanwhile, $J^3$ does not change the $U(1)$ charge and does not shift the conformal weight. Thus, the shift in the conformal weight is tied to the $U(1)$ charge. Using this observation, we can derive the refined twisted $\mf{su}(2)_1$ character of the $[1,0]$ and $[0,1]$ modules from the refined $\mf{su}(2)_1$ characters~\eqref{eqn:refined_su2}:
\ie
K^2_{\thetaa}(\tau,z) &= \frac{q^{\frac{\thetaa^2}{4}} y^{-\frac{\thetaa}{2}}}{\eta(\tau)} \sum_{m\in\mZ} q^{m^2 - m\thetaa} y^m 
\,, 
\\
K^2_{\thetaa + 1}(\tau, z) &= \frac{q^{\frac{\thetaa^2}{4}} y^{-\frac{\thetaa}{2}}}{\eta(\tau)} \sum_{m\in\mZ} q^{(m+\frac{1}{2})^2 - (m+ \frac{1}{2})\thetaa} y^{m+ \frac{1}{2}} 
\,.
\label{eq:characterswithu1bla}
\fe
The sum in (\ref{eq:characterswithu1bla}) 
can be understood in terms of spectral flow, which maps a state at zero twist $x=0$ to a state at non-zero twist $x$ \cite{Schwimmer:1986mf}. Under this spectral flow, the conformal weight and $U(1)$ charge transform as 
\ie
h\rightarrow h-x Q+\frac{1}{4}x^2\,,\quad Q\rightarrow Q-\frac{1}{2}x\,.
\fe
This precisely matches with the spectral flow reviewed in Appendix \ref{app:sf}, which is derived using modular transformation. Specializing \eqref{eq:characterswithu1bla} to $y=1$, we obtain the twisted characters
\ie
K^2_{\thetaa}(\tau) &= \frac{1}{\eta(\tau)} \sum_{m\in\mZ} q^{\frac{(2m + \thetaa)^2}{4}}\,, 
\\
K^2_{\thetaa + 1}(\tau) &= \frac{1}{\eta(\tau)} \sum_{m\in\mZ} q^{\frac{(2m + \thetaa + 1)^2}{4}} \,.
\label{eqn:twisted_character}
\fe
The twisted characters \eqref{eqn:twisted_character} have periodicity $2$ in the twist parameter $\thetaa$ instead of $1$. This is because the $U(1)$ symmetry defect~\eqref{eqn:cartan} generated by $J^3$ is defined on $\thetaa \in [0,2)$ in our normalization, corresponding to choosing half-integer charges (see App.~\ref{app:sf} for more discussions). Note that although shifting $\thetaa$ by $1$ induces a nontrivial spectral flow that permutes the two modules, it preserves the twisted chiral algebra. As we will discuss later, this is consistent with the fact that in the $SU(2)_1$ WZW model, $x=1$ corresponds to the Verlinde line that preserves the $\mathfrak{su}(2)_1$ chiral algebra. 

\subsubsection{\texorpdfstring{$SU(2)_1$}{SU(2)1} WZW model}
\label{subsec:su2_1_wzw}

We now discuss a concrete realization of the $\mathfrak{su}(2)_1$ chiral tube algebra in the $SU(2)_1$ WZW model. The model has an $\mf{su}(2)_1\otimes \overline{\mf{su}(2)_1}$ chiral algebra. Its torus partition function can be decomposed into $\mf{su}(2)_1$ characters $K^2_0(\tau)$ and $K^2_1(\tau)$ as
\ie
Z(\tau,\bar\tau) = \frac{1}{|\eta(\tau)|^2} \sum_{e,m \in \mZ} q^{\frac{1}{4}(e + m)^2} \bar{q}^{\frac{1}{4} (e - m)^2} = K^2_0(\tau)K^2_0(-\bar\tau) + K^2_1(\tau)K^2_1(-\bar\tau) \,.\label{eq:WZW_Z}
\fe
The chiral and anti-chiral currents generate a global symmetry,
\ie
G=\frac{SU(2)_L \times SU(2)_R}{\mZ_2}\,,
\fe
where the unfaithful diagonal $\mZ_2$ center of the two $SU(2)$ factors is quotiented out. This theory is equivalent to the self-dual $c=1$ free compact boson, which we take to have radius $R = \sqrt{2}$. The $G$ global symmetry contains a $U(1)_m \times U(1)_w$ subgroup that corresponds to the shift symmetries of the compact boson and its dual boson. Without loss of generality, we take the associated symmetry defects to be generated by $J^3(z)$ and $\bar J^3(\bar z)$
\ie
\cL_{\varthetaa, \phii} = \exp\left[{(\varthetaa + \phii)\oint_C dz J^3(z) - (\varthetaa-\phii)\oint_C d\bar{z} \bar{J}^3(\bar z)}\right] \,,
\fe
where $\varthetaa\in [0,1)$ parametrizes $U(1)_m$ and $\phii\in[0,1)$ parametrizes $U(1)_w$. In this convention, $J^{\pm}(z)$ carries both $U(1)_m$ and $U(1)_w$ charge $\pm 1$, while $J^3(z)$ is neutral under both $U(1)$ factors. On the other hand, $\bar{J}^{\pm}(\bar z)$ carries $U(1)_m$ charge $\pm 1$ and $U(1)_w$ charge $\mp 1$, while $\bar J^3(z)$ is neutral under both $U(1)$ factors. 

Let us now consider defect Hilbert spaces twisted by a $U(1)_m\times U(1)_w$ TDL $\cL_{\varthetaa,\phii}$. The corresponding torus partition function is given by
\ie
Z_{\varthetaa,\phii}(\tau,\bar\tau) = \frac{1}{|\eta(\tau)|^2} \sum_{e,m \in \mZ} q^{\frac{1}{4}(e + m - \varthetaa-\phii)^2} \bar{q}^{\frac{1}{4} (e - m + \varthetaa-\phii)^2} \,.
\label{eqn:defect_selfdual}
\fe
Unlike \eqref{eq:WZW_Z}, this defect partition function cannot be decomposed into $\mathfrak{su}(2)_1$ characters. This is because $\cL_{\varthetaa,\phii}$ does not preserve the full $\mf{su}(2)_1 \otimes \overline{\mf{su}(2)_1}$ chiral algebra. Instead, it acts nontrivially on $J^\pm(z)$ and $\bar J^\pm(\bar z)$ and thereby twists their boundary conditions. The defect Hilbert space thus preserves a twisted $\mf{su}(2)_1 \otimes \overline{\mf{su}(2)_1}$ chiral algebra with independent twists $\thetaa = \varthetaa + \phii$ and $\bar\thetaa = \varthetaa - \phii$. 
This twisted chiral algebra is generated by $J^{a}_{x,n}$, $\bar{J}^{a}_{-\bar x,n}$, $L_{x,n}$ and $\bar{L}_{-\bar x,n}$. 
The defect partition function~\eqref{eqn:defect_selfdual} can therefore be decomposed into the twisted characters~\eqref{eqn:twisted_character} as
\ie
Z_{\varthetaa, \phii}(\tau,\bar\tau)=K^2_{ \phii+\varthetaa}(\tau) K^2_{\phii-\varthetaa }(-\bar\tau) + K^2_{1 +  \phii+\varthetaa }(\tau) K^2_{1 + \phii-\varthetaa }(-\bar\tau)\,.
\fe

The most general TDLs of the $G$ global symmetry are specified by elements 
\ie
(g,\bar g)\in \frac{SU(2)_L\times SU(2)_R}{\mathbb{Z}_2}\,.
\fe
Using independent chiral and anti-chiral $SU(2)$ conjugations, both $g$ and $\bar g$ can be brought into the respective $U(1)$ Cartan subgroups generated by $J^3$ and $\bar J^3$\,:
\ie
(g,\bar{g}) = \left(h\, e^{2\pi i x J^3} \,h^{-1},\,\bar{h}\, e^{2\pi i \bar x \bar{J}^3}\, \bar{h}^{-1}\right)\,, \quad h,\bar{h} \in SU(2)\,.
\fe
As a result, the $(g,\bar g)$ TDL shares the same defect partition function $Z_{u,v}$ as the $U(1)_m \times U(1)_w$ TDL $\cL_{\varthetaa, \phii}$ with $\varthetaa = \frac{1}{2}(x + \bar x)$ and $\phii = \frac{1}{2}(x - \bar x)$. Moreover, the defect Hilbert space of the $(g,\bar g)$ TDL likewise preserves a twisted $\mf{su}(2)_1 \otimes \overline{\mf{su}(2)_1}$ algebra with twists $x$ and $\bar x$. The only difference is that the preferred Cartan generators are now given $h\, J^3\, h^{-1}$ and $\bar h \,\bar J^3 \,\bar h^{-1}$ instead of $J^3$ and $\bar J^3$.

\subsection{Orbifold of \texorpdfstring{$\mf{su}(2)_1$}{su(2)1} Chiral Tube Algebra}
\label{subsec:su2_1_orbifold}

In this section, we show that the $\mf{su}(2)_1$ chiral algebra survives under discrete gauging in the form of a chiral tube algebra. Without loss of generality, we consider a non-anomalous $\mZ_N$ global symmetry under which the currents $J^{\pm}(z)$ carry charge $\pm 1$, while $J^3(z)$ is neutral. After gauging the $\mZ_N$ global symmetry, $J^{\pm}(z)$ are no longer genuine local operators. Instead, $J^+(z)$ is attached by the  $\mathbb{Z}_N$ Wilson line $\hat\eta$, which is the generator of the dual $\hat\mZ_N$ symmetry, while $J^-(z)$ is attached by $\hat\eta^{-1}$. The current $J^3(z)$ on the other hand remains a local operator.

As in Section~\ref{sec:tetracritical_Ising}, we need to first determine the monodromy of the currents in order to fix their mode expansions. Consider an operator $\cO_{Q,\cL}(0)$ which carries $\hat\mZ_N$ charge $Q$ and is attached by the TDL $\cL$. The monodromy of the non-local currents $J^{\pm}(z)$ around $\cO_{Q,\cL}(0)$ depends on two pieces of data. First, it depends on the $\hat\mZ_N$ charge $Q$ of $\cO_{Q,\cL}(0)$, since winding $J^{\pm}(z)$ around $\cO_{Q,\cL}(0)$ sweeps the
TDL $\hat\eta^{\pm 1}$ attached to the current across $\cO_{Q,\cL}(0)$. 
Second, it depends on how the TDL $\mathcal{L}$ attached to $\cO_{Q,\cL}(0)$ acts on the currents, since the currents cross this TDL once when winding around $\cO_{Q,\cL}(0)$. 
We will focus on a specific class of TDLs $\cL$ that commute with $\hat \eta$ and act on the currents as
\ie
\cL \cdot J^3(z) = J^3(z) \,,\qquad \cL \cdot J^{\pm}(z) = e^{\pm 2\pi i\alpha} J^{\pm}(z) \,.\label{eq:passing_action}
\fe
Examples of such TDLs include the $U(1)$ TDLs generated by the local current $J^3(z)$.
We will leave the analysis of more general TDLs to future work.

To construct the orbifold chiral tube algebra, we first define the following projector
\ie
P_{Q,\cL} = \frac{1}{N} \sum_{n=0}^{N-1} e^{-\frac{2\pi i nQ}{N}} \tikz[baseline=-0.75]{\draw[black] (0,0)--(0,1.5); \draw[black, -{Stealth[round, length=5pt, width=5pt, bend]}] (0,0)--(0,0.4); \draw[black, -{Stealth[round, length=5pt, width=5pt, bend]}] (0,0)--(0,1.2); \draw (0,0) circle (0.75); \node[right] at (0,0.35) {\small $\cL$}; \node[right] at (0,1.1) {\small $\cL$}; \draw[black,-{Stealth[round, length=5pt, width=5pt, bend]}] (-0.75,0) arc (180:360:0.75); \node[right] at (0.5,-0.5) {\small $\hat\eta^n$}} \,,\label{eq:projection_Q,L}
\fe
which projects onto the charge $Q$ sector with $Q \in [0,N)$. Here, $Q$ need not be integral and may in general be fractional, allowing us to incorporate the effect of mixed 't Hooft anomaly between $\cL$ and $\eta$, which will play a role later. Note that in the presence of mixed 't Hooft anomaly, we need to choose a resolution of the four-way junction, and different choices are related nontrivially by $F$-symbols.

With the projectors, we can construct the lasso operators of the chiral tube algebra by fusing the non-local currents onto the projectors as
\ie
J^+ P_{Q,\cL}=
\tikz[baseline=-0.75]{\draw[black] (0,0)--(0,1.5); 
\draw[black,-{Stealth[round, length=5pt, width=5pt, bend]}] (0.707,0.707) arc (45:70:1.0);
\draw[black] (0,1.0) arc (90:45:1.0); 
\draw[black, -{Stealth[round, length=5pt, width=5pt, bend]}] (0,0)--(0,0.4); 
\draw[black, -{Stealth[round, length=5pt, width=5pt, bend]}] (0,0)--(0,1.3);
\draw[black] (0.707,0.707)--(0.9,0.9); 
\node[right] at (-0.05,1.35) {\small $\cL\hat\eta$}; 
\node[right] at (0.58,0.55) {\small $\hat\eta$}; 
\node[right] at (0.8,1.0) {\small $J^+$}; 
\draw[black] (0,0.73) arc (90:-90:0.73); 
\node[right] at (0.55,-0.5) {\small $P_{Q,\cL}$}; 
\draw[black] (0,-0.73) arc (-90:-270:0.73);
\node[right] at (-0.05,0.35) {\small $\cL$}}
\,,\qquad 
J^- P_{Q,\cL}=\tikz[baseline=-0.75]{\draw[black] (0,0)--(0,1.5); 
\draw (0,1.0) arc (90:45:1.0); 
\draw[black,-{Stealth[round, length=5pt, width=5pt, bend]}] 
  (0.707,0.707) arc (45:70:1.0);
\draw[black] (0.707,0.707)--(0.9,0.9); 
\node[right] at (-0.05,1.35) {\small $\cL\hat{\eta}^{-1}$}; 
\node[right] at (0.58,0.55) {\small $\hat{\eta}^{-1}$}; 
\draw[black, -{Stealth[round, length=5pt, width=5pt, bend]}] (0,0)--(0,0.4); 
\draw[black, -{Stealth[round, length=5pt, width=5pt, bend]}] (0,0)--(0,1.3);
\node[right] at (0.8,1.0) {\small $J^-$}; 
\draw[black] (0,0.73) arc (90:-90:0.73); 
\node[right] at (0.55,-0.5) {\small $P_{Q,\cL}$}; 
\draw[black] (0,-0.73) arc (-90:-270:0.73);
\node[right] at (-0.05,0.35) {\small $\cL$}}\,.
\fe
Because the currents commute with the projectors, we can fuse the currents onto the projectors either from the outside, as illustrated above, or from the inside. Both constructions define the same operator. This is, however, not the case when we study the bosonization of the superconformal algebra in Section \ref{subsec:tricritical}, where the ordering matters.

When we wind $J^{\pm}(z)$ around $\cO_{Q,\cL}(0)$, we drag the attached $\hat\eta^{\pm 1}$ TDL around the projector $P_{Q,\mathcal L}$. Fusing this TDL onto $P_{Q,\mathcal L}$ generates a phase 
\ie
\hat\eta^{\pm 1} P_{Q,\cL} = e^{\mp \frac{2\pi i Q}{N}} P_{Q,\cL} \,.\label{eq:phase_1}
\fe
There is an extra minus sign in the exponent because the dragged TDL has the opposite orientation to the $\hat\eta$ TDLs used in the definition of the projector.
Note that when $Q$ is fractional, because of the mixed 't Hooft anomaly, the TDLs in the $\mathcal{L}$ defect Hilbert space fuse as
\ie
\hat\eta^a \times \hat \eta^b = \hat \eta^{a+b} e^{\frac{2\pi i}{N} Q(a+b - [a+b])}\,,
\fe
where $[a]=a\pmod N$.
The phase generated in \eqref{eq:phase_1}, along with the phase generated by the action of $\cL$ on the non-local currents \eqref{eq:passing_action} when the latter crosses the former, determines the monodromy of the currents around $\cO_{Q,\cL}(0)$: 
\ie
J^3(z e^{2\pi i})P_{Q,\mathcal L}&=J^3(z)P_{Q,\mathcal L}\,.
\\
J^\pm(z e^{2\pi i})P_{Q,\mathcal L}&=e^{\mp 2\pi i(\frac{Q}{N}+\alpha)}J^\pm(z)P_{Q,\mathcal L}\,.
\fe
One can show that the resulting monodromy is independent of the gauge choice used to represent the 't Hooft anomaly, and depends only on the TDL $\cL$ and the charged sector.\footnote{The charge $Q$ is by itself gauge dependent, so to unambiguously specify a charged sector in a given theory, one should specify the spectrum in that sector.}

With the monodromy determined, we can fix the mode expansion of the currents on the $\cL$ defect Hilbert space as 
\ie
J^{\pm}(z) = \sum_{n\in\mZ}\sum_{Q \in \mZ_N + \beta} \frac{J^{\pm}_{n \pm x} P_{Q,\cL}}{z^{n \pm x + 1}}\,,\qquad \thetaa = \frac{Q}{N} + \alpha\,,
\fe
where $\beta \in [0,1)$ denotes the fractional part of $Q$.
Using these modes, we define the following lasso operators\footnote{In principle, there should also be a label of $Q$ for the chiral tube algebra, characterizing which charged sector we project onto, but this is fully determined by $\cL$ and $\thetaa$ by definition. }
\ie
{[J^3_{\thetaa,n}]}^{\cL}_\cL = \tikz[baseline=-0.75]{\draw[black] (0,0)--(0,1.5); \draw[black, -{Stealth[round, length=5pt, width=5pt, bend]}] (0,0)--(0,0.4); \draw[black, -{Stealth[round, length=5pt, width=5pt, bend]}] (0,0)--(0,1.2); \draw (0,0) circle (0.75); \node[right] at (0,0.35) {\small $\cL$}; \node[right] at (0,1.1) {\small $\cL$};  
\node[right] at (0.5,-0.5) {\small $J^3_{n} P_{Q,\cL}$}} \,,\quad
[J^{\pm}_{\thetaa,n}]_\cL^{\cL\hat\eta^{\pm1}} = \tikz[baseline=-0.75]{\draw[black] (0,0)--(0,1.5); \draw[black, -{Stealth[round, length=5pt, width=5pt, bend]}] (0,0)--(0,0.4); \draw[black, -{Stealth[round, length=5pt, width=5pt, bend]}] (0,0)--(0,1.2); \draw (0,0) circle (0.75); \node[right] at (0,0.35) {\small $\cL$}; \node[right] at (0,1.1) {\small $\cL\hat\eta^{\pm 1}$}; 
\node[right] at (0.5,-0.5) {\small $J^{\pm}_{n \pm \thetaa} P_{Q,\cL}$}} \,.
\fe
The operator $[J^{3}_{\thetaa,n}]^{\cL}_\cL$ acts within a fixed defect Hilbert space, whereas $[J^{\pm}_{\thetaa,n}]_\cL^{\cL\hat\eta^{\pm1}}$ maps between different ones. This distinction arises because $J^3(z)$ is a local operator, whereas $J^{\pm}(z)$ is a non-local current attached by $\hat \eta^\pm$.

The chiral tube algebra formed by these lasso operators is 
\ie
{[J^3_{\thetaa,n}]}^{\cL}_\cL\, [J^3_{\thetaa,m}]^{\cL}_\cL-[J^3_{\thetaa,m}]^{\cL}_\cL\, [J^3_{\thetaa,n}]^{\cL}_\cL&= \frac{n}{2} \delta_{n+m,0} \,, \\
[J^+_{\thetaa,n}]^{\cL}_{\cL\hat{\eta}^{-1}} [J^-_{\thetaa,m}]_\cL^{\cL\hat\eta^{-1}} - [J^-_{\thetaa,m}]^\cL_{\cL\hat\eta} \,[J^+_{\thetaa,n}]_\cL^{\cL\hat\eta} &= \le n + \thetaa \ri \delta_{n+m,0} + 2[J^3_{\thetaa,n+m}]_\cL^{\cL} \,, 
\\
[J^3_{\thetaa,n}]^{\cL\hat\eta^{\pm 1}}_{\cL\hat\eta^{\pm 1}} [J^{\pm}_{\thetaa,m}]_\cL^{\cL\hat\eta^{\pm1}}  - [J^{\pm}_{\thetaa,m}]_\cL^{\cL\hat\eta^{\pm1}} [J^3_{\thetaa,n}]^{\cL}_\cL &= \pm [J^{\pm}_{\thetaa,n+m}]_\cL^{\cL\hat\eta^{\pm1}} \,.
\label{eq:su(2)_orbifold_chiral_tube_algebra}
\fe
Note that only lasso operators with the same twist $x$ compose and the dual $\hat\mZ_N$ charge $Q$ is conserved by the chiral tube algebra. This follows from the fact that the TDL $\hat\eta$ acts trivially on the currents. This property no longer holds when we study the bosonization of the superconformal algebra in Section \ref{subsec:tricritical}.  

The orbifold chiral tube algebra \eqref{eq:su(2)_orbifold_chiral_tube_algebra} is isomorphic to the $\mathfrak{su}(2)_1$ chiral tube algebra~\eqref{eqn:twisted_su2_chiral_algebra}, so they share the same module structures and characters. For twist $x$, the characters are given by
\ie
K_{\thetaa+2j}^2(\tau)= \frac{1}{\eta(\tau)} \sum_{m\in\mZ} q^{\le m + j + \frac{\thetaa}{2} \ri^2}\,,
\fe
where $j = 0$ corresponds the $[1,0]$ module and $j=\frac{1}{2}$ corresponds the $[0,1]$ module. 
There is however an important distinction. Since the orbifold chiral tube algebra maps between defect Hilbert spaces associated with $\cL, \cL\hat\eta, \dots, \cL\hat\eta^{N-1}$, each of its modules contain states from all of these defect Hilbert spaces. 
To account for this feature, we introduce refined characters that sum over states in a single defect Hilbert space. Suppose the highest weight state is in the $\mathcal{L}$ defect Hilbert space. Then, the refined character in the $\mathcal{L}\hat\eta^k$ defect Hilbert space is
\ie
\chi_{\thetaa,j}^{\cL\hat\eta^k}= \frac{1}{\eta(\tau)} \sum_{m\in N\mZ + k} q^{\le m + j + \frac{\thetaa}{2} \ri^2}\,.
\label{eqn:su2_character_gauging}
\fe
It coincides with the twisted $\mathfrak{u}(1)_{2N^2}$ characters $K^{2N^2}_{Nx+2N(j+k)}(\tau)$ that appeared in \eqref{eq:u(1)_N_character}.

\subsubsection{Compact Boson}
\label{subsec:su2/zn}

We now discuss the $c=1$ free compact boson CFT as a concrete realization of the orbifold chiral tube algebra.
Specifically, we consider gauging the $\mZ_N$ subgroup of the $U(1)_w$ symmetry in the $SU(2)_1$ WZW model. 
Recall that the $SU(2)_1$ WZW model is equivalent to the self-dual compact boson theory at radius $R=\sqrt{2}$. Gauging the $\mathbb{Z}_N$ winding symmetry then produces the compact boson theory at radius $R = \sqrt{2}N$, and the corresponding dual symmetry is the $\hat\mZ_N$ subgroup of $U(1)_m$ symmetry in the gauged theory. 
Following the same convention as in the previous section, we denote the generator of the gauged $\mZ_N$ symmetry by $\eta$ and the generator of the dual $\hat\mZ_N$ symmetry by $\hat\eta$. With this convention, $\hat\eta$ is identified with the $U(1)_m \times U(1)_w$ TDL $\mathcal{L}_{u,v}$ with $(u,v)=(-\frac{1}{N},0)$.

Since the $SU(2)_1$ WZW model has the $\mathfrak{su}(2)_1\times \overline{\mathfrak{su}(2)_1}$ chiral tube algebra, we expect that the gauged theory, i.e.~the compact boson theory at radius $R=\sqrt 2 N$, realizes an orbifold of the $\mathfrak{su}(2)_1\times \overline{\mathfrak{su}(2)_1}$ chiral tube algebra. We will demonstrate this explicitly by studying the action of the chiral tube algebra on the defect Hilbert spaces of the $U(1)_m \times U(1)_w$ TDLs $\cL_{\varthetaa,\phii}$. The compact boson theory also contains more general TDLs \cite{Fuchs:2007tx,Thorngren:2021yso}, whose defect Hilbert spaces are expected to host a more intricate chiral tube algebra. We leave the analysis of this chiral tube algebra to future work.

At a generic radius $R$, the torus partition function of the compact boson theory is 
\ie
Z(\tau,\bar\tau) = \frac{1}{|\eta(\tau)|^2} \sum_{e,m \in \mZ} q^{\frac{1}{2}(\frac{e}{R} + \frac{mR}{2})^2} \bar{q}^{\frac{1}{2}(\frac{e}{R} - \frac{mR}{2})^2} \,.
\fe
The defect partition function with a TDL $\cL_{\varthetaa,\phii}$ inserted along the time direction and a projector $P_{Q,\mathcal{L}_{u,v}}$ inserted along the spatial direction is
\ie
Z_{\varthetaa,\phii}^{Q}(\tau,\bar\tau) = \frac{1}{|\eta(\tau)|^2} \sum_{e\in N\mZ - Q - \phii,\,m\in\mZ - \varthetaa} q^{\frac{1}{2}\le \frac{e}{R} + \frac{mR}{2} \ri^2} \bar{q}^{\frac{1}{2}\le \frac{e}{R} - \frac{mR}{2} \ri^2} \,.
\fe
Here, the projector $P_{Q,\mathcal{L}_{u,v}}$ is constructed from $\hat\eta$ as in \eqref{eq:projection_Q,L}.
Because of the mixed 't Hooft anomaly between $U(1)_m$ and $U(1)_w$, there is an ambiguity in assigning $U(1)_m$ charges to states in the $U(1)_w$ defect Hilbert space and vice versa. 
Here, we choose a gauge in which the $U(1)_m$ charge $-Q$ is quantized as integers. In this gauge, the vertex operator in the $\mathcal{L}_{u,v}$ defect Hilbert space,
\ie
e^{\frac{ie\varphi}{R}} e^{\frac{imR\tilde\varphi}{2}}\,,\qquad e\in \mZ - \phii,\ \, m \in \mZ - \varthetaa\,,
\fe
has $U(1)_m$ charge $Q=-(e + \phii)$, where $\varphi$ is the compact boson and $\tilde\varphi$ is the dual boson. 

In the gauged theory, i.e.~the compact boson theory at radius $R=\sqrt{2} N$, the currents $J^-(z)$ and $\bar J^+(\bar z)$ become non-local operators attached by $\hat\eta^{-1}=\mathcal{L}_{\frac{1}{N},0}$, whereas $J^+(z)$ and $\bar J^-(\bar z)$ are attached by $\hat\eta=\mathcal{L}_{-\frac{1}{N},0}$. They are realized as the vertex operators 
\ie
J^{\pm}(z)=e^{\pm\frac{i \varphi}{\sqrt{2}}} e^{\pm\frac{i\tilde\varphi}{\sqrt{2}}}\,,\qquad\bar J^{\pm}(\bar z)=e^{\pm\frac{i \varphi}{\sqrt{2}}} e^{\mp\frac{i\tilde\varphi}{\sqrt{2}}}\,.
\fe
In the gauge we choose, $J^{\pm}(z)$ carries $U(1)_m \times U(1)_w$ charges $(\pm N, \pm \frac{1}{N})$, while $\bar{J}^{\pm}(\bar z)$ carries charges $(\pm N, \mp \frac{1}{N})$. 
From these non-local currents, we can construct the corresponding chiral tube algebra. 
To this end, we need to first determine the monodromy of the currents in the $\cL_{u,v}$ defect Hilbert space in the presence of the projector $P_{Q,\cL_{u,v}}$. Since $J^+(z)$ is attached by $\hat \eta$, winding it around the projector drags the attached TDL along with it. Fusing this TDL onto the projector then produces a phase $e^{-\frac{2\pi iQ}{N}}$. Similarly, because $\bar J^+(\bar z)$ is attached by $\hat \eta^{-1}$, it picks up the opposite phase $e^{\frac{2\pi iQ}{N}}$. 
The action of the TDL $\cL_{u,v}$ on the currents also contributes to the monodromy. This generates a phase $e^{-2\pi i(Nu+\frac{v}{N})}$ for $J^+(z)$ and $e^{-2\pi i(Nu-\frac{v}{N})}$ for $\bar J^+(\bar z)$.
Combining the two contributions, the total monodromy is $e^{-2\pi i x}$ for $J^+(z)$ and  $e^{2\pi i\bar x}$ for $\bar J^+(\bar z)$, with $x$ and $\bar x$ the twists on the chiral and anti-chiral algebra, respectively:
\ie
\thetaa = N\varthetaa + \frac{Q+\phii}{N}\,,\qquad \bar\thetaa =  - N\varthetaa + \frac{Q+\phii}{N}\,.\label{eq:twist_u,v}
\fe
The chiral tube algebra is then generated by the following lasso operators
\ie
&[J^3_{x,n}]_{\cL_{\varthetaa,\phii}}^{\cL_{\varthetaa,\phii}} = \tikz[baseline=-0.75]{\draw[black] (0,0)--(0,1.5); \draw[black, -{Stealth[round, length=5pt, width=5pt, bend]}] (0,0)--(0,0.4); \draw[black, -{Stealth[round, length=5pt, width=5pt, bend]}] (0,0)--(0,1.2); \draw (0,0) circle (0.75); \node[right] at (0,0.35) {\small $\cL_{\varthetaa,\phii}$}; \node[right] at (0,1.1) {\small $\cL_{\varthetaa,\phii}$}; 
\node[right] at (0.5,-0.5) {\small $J^3_n P_{Q,\cL_{\varthetaa,\phii}}$}} \,,
\quad
[J^{\pm}_{x,n}]^{\cL_{\varthetaa,\phii}\hat\eta^{\pm1}}_{\cL_{\varthetaa,\phii}} = \tikz[baseline=-0.75]{\draw[black] (0,0)--(0,1.5); \draw[black, -{Stealth[round, length=5pt, width=5pt, bend]}] (0,0)--(0,0.4); \draw[black, -{Stealth[round, length=5pt, width=5pt, bend]}] (0,0)--(0,1.2); \draw (0,0) circle (0.75); \node[right] at (0,0.35) {\small $\cL_{\varthetaa,\phii}$}; \node[right] at (0,1.1) {\small $\cL_{\varthetaa,\phii}\hat\eta^{\pm 1}$}; 
\node[right] at (0.5,-0.5) {\small $J^\pm_{n \pm \thetaa} P_{Q,\cL_{\varthetaa,\phii}}$}} \,, \\
&[\bar{J}^3_{\bar x,n}]^{\cL_{\varthetaa,\phii}}_{\cL_{\varthetaa,\phii}} = \tikz[baseline=-0.75]{\draw[black] (0,0)--(0,1.5); \draw[black, -{Stealth[round, length=5pt, width=5pt, bend]}] (0,0)--(0,0.4); \draw[black, -{Stealth[round, length=5pt, width=5pt, bend]}] (0,0)--(0,1.2); \draw (0,0) circle (0.75); \node[right] at (0,0.35) {\small $\cL_{\varthetaa,\phii}$}; \node[right] at (0,1.1) {\small $\cL_{\varthetaa,\phii}$}; 
\node[right] at (0.5,-0.5) {\small $\bar{J}^3_n P_{Q,\cL_{\varthetaa,\phii}}$}} \,,
\quad
[\bar{J}^{\pm}_{\bar x,n}]_{\cL_{\varthetaa,\phii}}^{\cL_{\varthetaa,\phii}\hat\eta^{\mp 1}} = \tikz[baseline=-0.75]{\draw[black] (0,0)--(0,1.5); \draw[black, -{Stealth[round, length=5pt, width=5pt, bend]}] (0,0)--(0,0.4); \draw[black, -{Stealth[round, length=5pt, width=5pt, bend]}] (0,0)--(0,1.2); \draw (0,0) circle (0.75); \node[right] at (0,0.35) {\small $\cL_{\varthetaa,\phii}$}; \node[right] at (0,1.1) {\small $\cL_{\varthetaa,\phii}\hat\eta^{\mp 1}$}; 
\node[right] at (0.5,-0.5) {\small $\bar{J}^\pm_{n \pm \bar\thetaa} P_{Q,\cL_{\varthetaa,\phii}}$}} \,.\label{eq:lasso_compact_boson}
\fe
For a fixed $\mathcal{L}_{u,v}$, these lasso operators form an orbifold $\mf{su}(2)_1$ chiral and anti-chiral tube algebra \eqref{eq:su(2)_orbifold_chiral_tube_algebra}. We can thus organize the spectrum into modules of these chiral tube algebra.

First, consider the local Hilbert space. A lasso operator \eqref{eq:lasso_compact_boson} of the chiral tube algebra generally maps a state from the local Hilbert space to the defect Hilbert spaces. For the state to remain in the local Hilbert space, we need to balance the chiral and anti-chiral lasso operators. In particular, if the chiral lasso operators map the state from the local Hilbert space to the $\hat\eta^k$ defect Hilbert space, then the anti-chiral lasso operators need to map it back from the $\hat\eta^k$ defect Hilbert space to the local Hilbert space. We can therefore decompose the torus partition function in terms of the refined characters \eqref{eqn:su2_character_gauging} of the orbifold chiral tube algebra as
\ie
&Z_{0,0}(\tau,\bar\tau) = 
\sum_{Q\in\mZ_N} Z_{0,0}^{Q}(\tau,\bar\tau)\,,
\\
&Z_{0,0}^{Q}(\tau,\bar\tau)= \frac{1}{|\eta(\tau)|^2} \sum_{e,m \in \mZ} q^{\frac{1}{4}\le e - \frac{Q}{N} + Nm \ri^2} \bar{q}^{\frac{1}{4}\le e - \frac{Q}{N} - Nm \ri^2} 
\\
&\qquad\quad \ \ \, = \sum_{j=0,\frac{1}{2}} \sum_{k\in\mZ_N} \chi_{\frac{Q}{N},j}^{\hat\eta^k}(\tau) \,\chi_{\frac{Q}{N},j}^{\hat\eta^{-k}}\!(-\bar\tau)\,.
\fe
Here, we first decompose the torus partition function $Z_{0,0}(\tau,\bar\tau)$ into the $\hat{\mathbb{Z}}_N$ refined partition function $Z^Q_{0,0}(\tau,\bar\tau)$. This is because the $\hat{\mathbb{Z}}_N$ charge is preserved by the lasso operators and determines the twist of the corresponding modules.
The defect partition function of the $U(1)_m \times U(1)_w$ TDLs can similarly be decomposed into the refined characters \eqref{eqn:su2_character_gauging} of the orbifold chiral tube algebra as
\ie
&Z_{\varthetaa, \phii}(\tau,\bar\tau)=\sum_{Q\in\mathbb{Z}_N} Z_{\varthetaa,\phii}^{Q}(\tau,\bar\tau)\,,
\\
&Z_{\varthetaa,\phii}^{Q}(\tau,\bar\tau) = \frac{1}{|\eta(\tau)|^2} \sum_{e,m\in\mZ } q^{\frac{1}{4}\le e-\frac{Q + \phii}{N} + Nm-Nu \ri^2} \bar{q}^{\frac{1}{4}\le e-\frac{Q + \phii}{N} - Nm+Nu \ri^2} 
\\
&\qquad \quad \ \ \ = \sum_{j=0,\frac{1}{2}} \sum_{k\in\mZ_N} \chi_{ \frac{Q+\phii}{N}+N\varthetaa,j}^{\hat\eta^k}(\tau) \,\chi_{ \frac{Q+\phii}{N}- N\varthetaa,j}^{\hat\eta^{-k}}(-\bar\tau)\,,
\fe
where the twist parameters exactly match with the ones determined in \eqref{eq:twist_u,v}.

\section{\texorpdfstring{$\mathcal{N}=1$}{N=1}  Superconformal Algebra and Bosonization}
\label{sec:susy}

Consider a 2d fermionic CFT that contains a fermionic chiral current $G(z)$ of conformal weight $(h,\bar h)=(\frac{3}{2},0)$, which is a Virasoro primary. When its self-OPE closes on the stress tensor $T(z)$, the chiral current $G(z)$ can be interpreted as a supercurrent and together with $T(z)$
generates the $\cN=1$ superconformal algebra \cite{Friedan:1984rv}:
\ie
{[L_m,L_n]} &= (m-n) L_{m+n} + \frac{c}{12}(m^3-m) \delta_{m+n,0} \,, \\
\{G_m, G_n\} &= 2L_{m+n} + \frac{c}{3} \le m^2 - \frac{1}{4} \ri \delta_{m+n,0} \,, \\
[L_m, G_n] &= \le \frac{1}{2}m-n \ri G_{m+n} \,,
\label{eqn:n=1_algebra}
\fe
where $G_n$ is the mode of the supercurrent $G(z)$ defined as
\ie
G(z) = \sum_{n} \frac{G_n}{z^{n+\frac{3}{2}}} \,,\quad G_n=\frac{1}{2\pi i}\oint dz\, z^{n+\frac{1}{2}} G(z) \,.
\label{eq:G_mode_expansion}
\fe
Here, we have not yet specified the allowed values of the mode index $n$ because as we will explain below, different choices correspond to different $\cN=1$ superconformal algebras. 

If there are more than one supercurrent, the superconformal algebra is extended to a larger superconformal algebra, such as $\cN=2$ or $\cN=4$ superconformal algebra. Our analysis in this section generalizes to those cases as well.

By definition, a fermionic SCFT\footnote{More generally, whenever we have a chiral operator $\Psi(z)$ with half-integer spin in a fermionic CFT, we can define a $(-1)^{F_L}$ symmetry that acts nontrivially on that operator. By assigning trivial charges on all the highest weight states of the chiral tube algebra generated by $\Psi(z)$, one can show that this is a symmetry of all the correlation functions. } has a  $\mZ_2^L$ chiral fermion-parity symmetry $(-1)^{F_L}$ that counts the number of chiral fermionic fields. The supercurrent $G(z)$ is charged under this symmetry. As a result, there are two $\cN=1$ superconformal algebras, corresponding to the two possible monodromies of $G(z)$, which are related by a $(-1)^{F_L}$ twist in the radial quantization. 
When $G(z)$ has a trivial monodromy, the mode indices in the mode expansion \eqref{eq:G_mode_expansion} take values in $n\in \mZ + \frac{1}{2}$, which defines the Neveu-Schwarz (NS) version of the $\cN=1$ superconformal algebra. When $G(z)$ has a $-1$ monodromy, the mode indices take values in $n\in \mZ$, which defines the Ramond (R) version of the $\cN=1$ superconformal algebra.\footnote{Note that the periodicity of $G(z)$ around a local operator, or around the $S^1$ for radial quantization, is the opposite. This comes from the fact that under the conformal map $z=e^{w}$ from the cylinder $(w=\tau+i\phi)$ to the plane $(z)$, a holomorphic field $\psi$ with weight $h$ transforms as 
\ie
\psi_{\text{cyl}}(w)=(\partial_w z)^h \psi(z) = z^h \psi(z)\,,
\fe 
The factor of $z^h$ has $h=\frac12 \text{ mod } 1$, so the NS sector, though defined by trivial monodromy around the origin, describes a field that is anti-periodic on $S^1$.}

Analogously, there is a $\mZ_2^R$ anti-chiral fermion-parity symmetry $(-1)^{F_R}$ that counts the number of anti-chiral fermionic fields. When the theory contains an anti-chiral supercurrent $\bar{G}(\bar z)$ of conformal weight $(h,\bar h)=(0,\frac{3}{2})$, it admits an anti-chiral copy of the $\cN=1$  superconformal algebra~\eqref{eqn:n=1_algebra}, which also has the NS and R versions depending on the monodromy of the supercurrent $\bar G(\bar z)$.
Importantly, there is a subtlety in constructing the spectrum in the R-R sector, where both the chiral and anti-chiral superconformal algebras are of the R version. Normally, one would expect that the chiral and anti-chiral algebras factorize and we only need to sew their representations together in a modular covariant way. This is however not the case here because the zero modes, $G_0$ and $\bar{G}_0$, do not commute; rather, they anti-commute. 
The algebra generated by these zero modes admits a 2-dimensional irreducible module if the states are not BPS, i.e.~$G_0$ and $\bar G_0$ do not annihilate them (from the algebra, non-BPS implies  $h, \bar h>\frac c{24}$).  Consequently, the R-R sector spectrum is necessarily at least doubly degenerate if the state has $h, \bar h>\frac c{24}$. 

Given a fermionic theory, we can bosonize it to a bosonic theory by summing over spin structures. Historically, this procedure is also known as the GSO projection\cite{Gliozzi:1976qd} (see e.g.\cite{Karch:1902.05550,Ji:2019ugf} for recent reviews). In this section, we are interested in understanding what happens to the superconformal algebra after bosonization.

\subsection{Review on Bosonization and Fermionization} 

We start by reviewing bosonization and fermionization. On a genus $g$ surface $\Sigma_g$, the partition function of the bosonized theory $Z_B$ is related to that of the original fermionic theory $Z_F$ by
\ie
Z_B[a] = \frac{1}{2^g} \sum_{\rho} Z_F[\rho ] (-1)^{{\rm Arf}(\rho+a)} \,,
\label{eq:bosonization}
\fe
where $a$ denotes the background gauge field for the dual $\mZ_2$ symmetry of the bosonic theory. The Arf invariant $\text{Arf}(\rho)$ is a non-singular quadratic form of the spin structure $\rho$\cite{atiyah1971riemann}. Physically, the spin structure determines the periodicity of fermionic fields around the non-contractible cycles. For our purpose, on a torus with Euclidean signature, the Arf invariant takes the values
\ie
{\rm Arf}(\rho) = \begin{cases}
    1\,, & \rho = PP\,, \\
    0\,, & \rho = AA\,,AP\,,PA\,,
\end{cases}
\fe
where $P$ denotes periodic and $A$ denotes anti-periodic boundary conditions. Here, we choose not to use the NS and R notation, which denotes the periodicity of a fermionic field when we quantize it on $S^1$ and is more naturally defined in the Lorentzian signature. 

Since the spin structure $\rho$ is a $H^1(\Sigma_g, \mZ_2)$ torsor, we can split it into a reference spin structure $\rho_0$ and a $\mathbb{Z}^F_2$ gauge field $b\in H^1(\Sigma_g, \mZ_2)$ as $\rho=\rho_0+b$. For a non-chiral fermionic CFT, $Z_F[\rho]=Z_F[\rho_0+b]$ can be interpreted as the partition function with reference spin structure $\rho_0$ and the $\mZ^F_2$ background gauge field $b$ for the $\mZ_2^F$ fermion-parity symmetry $(-1)^F = (-1)^{F_L + F_R}$.\footnote{This interpretation of the partition function however does not hold for chiral CFTs, where it can lead to a discrepancy between bosonization and gauging of the $(-1)^F$ symmetry~\cite{BoyleSmith:2403.03953}.} Thus, summing over spin structure is equivalent to summing over the $\mZ^F_2$ gauge field $b$, and 
bosonization is equivalent to gauging the non-anomalous $\mZ_2^F$ symmetry. This involves the process of projecting to the $(-1)^F$ neutral states, and including states from the $(-1)^F$-twisted sector. Under bosonization, the local and defect operator spectrum is reshuffled as
\ie
\text{$(-1)^F$-even NS-NS sector}\ &\longrightarrow\ \text{$\mathbb{Z}_2$-even untwisted sector}\,,
\\
\text{$(-1)^F$-odd NS-NS sector}\ &\longrightarrow \ \text{$\mathbb{Z}_2$-odd $\eta$-twisted sector}\,,
\\
\text{$(-1)^F$-even R-R sector}\ &\longrightarrow \ \text{$\mathbb{Z}_2$-odd untwisted sector}\,,
\\
\text{$(-1)^F$-odd R-R sector}\ &\longrightarrow\ \text{$\mathbb{Z}_2$-even $\eta$-twisted sector}\,.\label{eq:bosonization_operator}
\fe
We can recover the original fermionic theory by fermionizing the bosonized theory. Given a 2d bosonic theory with a non-anomalous $\mZ_2$ symmetry, we can fermionize it to a fermionic theory as
\ie
Z_F(\rho) = \frac{1}{2^g} \sum_{a \in H^1(\Sigma_g,\mZ_2)} Z_B(a) (-1)^{{\rm Arf}(\rho + a)} \,.
\fe
This fermionization procedure is the reversal of the bosonization procedure in \eqref{eq:bosonization}.

\begin{figure}[h]
    \centering
    \includegraphics[width=0.5\linewidth]{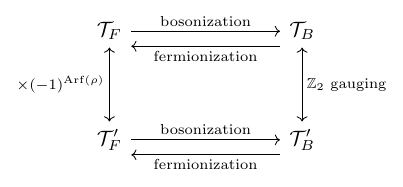}
    \caption{The commutative diagram among fermionic theories $\cT_F$, $\cT_F'$ and bosonic theories $\cT_B$, $\cT_B'$ under bosonization, fermionization, the stacking of fermionic SPT and $\mathbb{Z}_2$ gauging. }
    \label{fig:commutative}
\end{figure}

In fact, there exists another pair of bosonization and fermionization procedures:
\ie
Z_B(a) &= \frac{1}{2^g} \sum_{b \in H^1(\Sigma_g, \mZ_2)} Z_F(\rho ) (-1)^{{\rm Arf}(\rho + a + b)+{\rm Arf}(\rho  + b)} \,,
\\
Z_F(\rho)&=\frac{1}{2^g} \sum_{a \in H^1(\Sigma_g,\mZ_2)} Z_B(a) (-1)^{{\rm Arf}(\rho + a)+{\rm Arf}(\rho)}\,,\label{eq:BF2}
\fe
which are also the reversal of each other. This bosonization procedure is related to the one in \eqref{eq:bosonization} by stacking a fermionic SPT $(-1)^{{\rm Arf}(\rho)}$ before summing over spin structures. This stacking swaps the $(-1)^F$-even R-R sector with the $(-1)^F$-odd R-R sector. Upon bosonization, it amounts to swapping the $\mathbb{Z}_2$-odd untwisted sector with the $\mathbb{Z}_2$-even $\eta$-twisted sector, which is essentially gauging the dual $\mZ_2$ symmetry in the bosonic theory. This then leads to the commutative diagram in Figure~\ref{fig:commutative}.

In summary, bosonization is a topological manipulation that reshuffles the operator spectrum. As a result, we expect that the superconformal algebra survives in the form of chiral tube algebra in the bosonized theory. In what follows, we will make precise how it is realized in the bosonization of the simplest superconformal field theory, the $A$-series $\cN=1$ minimal model with central charge $c = \frac{7}{10}$, which produces the tricritical Ising CFT \cite{Friedan:1984rv}. 

\subsection{Fermionic Theory: \texorpdfstring{$\mathcal{N}=1$}{N=1} Minimal Model}

The Virasoro algebra at $c = \frac{7}{10}$ has 6 irreducible modules, that combine into 4 extended irreducible modules of the $\mathcal{N}=1$ superconformal algebra as listed in Table~\ref{tab:SUSY710} (see e.g. \cite{DiFrancesco:1997nk}).

\begin{table}[h] 
\centering
\renewcommand{\arraystretch}{1.4}
\begin{tabular}{|c|cc|c c|c|c  |} 
 \hline
 Sector & \multicolumn{2}{|c|}{NS} & \multicolumn{2}{|c|}{NS} & R & R
 \\
 \hline
 Symbol & $1$ &   $\varepsilon''$ & $\varepsilon$ & $\varepsilon'$ & $\sigma$   & $\sigma'$ \\
 \hline
 Conformal weight & $0$ & $\frac32$& $\frac1{10}$ & $\frac35$  &  $\frac3{80}$ &  $\frac7{16}$  
  \vspace{-18pt}\\
         &&&&&&\\
 \hline
\end{tabular}
\caption{Irreducible modules of the Virasoro and the $\cN=1$ superconformal algebra at  $c=\frac{7}{10}$. Virasoro modules grouped within the same box combine into a single $\cN=1$ module, whose primaries are connected by modes of supercurrent $G(z)$. }
\label{tab:SUSY710}
\end{table}

The $\cN=1$ minimal model at $c=\frac{7}{10}$ is a diagonal CFT of the $\cN=1$ superconformal algebra. Depending on the spin structure, its torus partition functions  are given by
\ie
Z_{\rm NS-NS} &= |\chi_{0} + \chi_{\frac{3}{2}}|^2 + |\chi_{\frac{1}{10}} + \chi_{\frac{3}{5}}|^2 \,, \\
Z_{\rm R-R} &= 2|\chi_{\frac{3}{80}}|^2 + 2|\chi_{\frac{7}{16}}|^2 \,,
\fe
which manifestly can be decomposed into characters of the $\cN=1$ superconformal algebra.
From the torus partition functions, we can read off the operator content, which is summarized in Table~\ref{tab:n=1_operator}. The first line in the table is obtained by decomposing the NS-NS partition function into integer-spin and half-integer-spin parts. Integer-spin operators are bosonic and charge neutral under $(-1)^F$, while the half-integer ones are fermionic and carry a nontrivial $(-1)^F$ charge. In the R-R sector, as mentioned before, the spectrum is necessarily doubly degenerate because of the anti-commutativity of the zero modes. 
The assignment of the $(-1)^F$ charge within each pair of degenerate states is a matter of convention, which can be switched by stacking a fermionic SPT phase $(-1)^{{\rm Arf}(\rho)}$. 

\begin{table}[h]
    \centering
    \renewcommand{\arraystretch}{1.4}
    \begin{tabular}{|c|c|c|}
    \hline
         & $(-1)^F$-even & $(-1)^F$-odd \\
         \hline
       NS-NS  &  $1_{0,0}, \varepsilon_{\frac{1}{10}, \frac{1}{10}}, \varepsilon'_{\frac{3}{5}, \frac{3}{5}}, \varepsilon''_{\frac{3}{2}, \frac{3}{2}}$ & $G_{\frac{3}{2},0}, \bar G_{0,\frac{3}{2}}, \psi'_{\frac{1}{10}, \frac{3}{5}}, \bar\psi'_{\frac{3}{5}, \frac{1}{10}}$ \vspace{-18pt}\\
         &&\\
       \hline
       R-R & $\sigma_{\frac{3}{80}, \frac{3}{80}}, \sigma'_{\frac{7}{16}, \frac{7}{16}}$ & $\mu_{\frac{3}{80}, \frac{3}{80}}, \mu'_{\frac{7}{16}, \frac{7}{16}}$
       \vspace{-18pt}\\
         &&\\
       \hline
    \end{tabular}
    \caption{The operator spectrum of the $\cN=1$ minimal model with $c = \frac{7}{10}$. }
    \label{tab:n=1_operator}
\end{table}

As a diagonal $\cN=1$ minimal model, there are simple TDLs that are Verlinde lines that preserve the chiral $\cN=1$ superconformal algebra. They are in one-to-one correspondence with the irreducible modules of the superconformal algebra and form a super fusion category \cite{Gu:2010na,Aasen:2017ubm,Chang:2022hud}. Specifically, we have four simple TDLs that correspond to the tensor product of a unitary Fibonacci category and a fermionic $\mZ_2^L$ symmetry generated by $(-1)^{F_L}$\cite{Chang:2022hud,Kikuchi:2022jbl}. Here we adopt the convention that $(-1)^{F_L}$ is a q-type TDL that satisfies
\ie
(-1)^{F_L}\times (-1)^{F_L} = (1_b + 1_f) I \,.
\fe
Physically, it means that on a q-type TDL there is a fermionic topological operator. The same holds for the $(-1)^{F_R} = (-1)^{F + F_L}$ symmetry. As a result, the quantum dimensions of these TDLs are $\sqrt{2}$.
Finally, along with the $\mZ_2^F$ fermion parity $(-1)^F$, which is a non-anomalous and has a nontrivial mixed anomaly with $(-1)^{F_R}$, we have in total 8 TDLs.

The action of $(-1)^{F_L}$ and $(-1)^{F_R}$ on local operators is readily determined from the chiral and anti-chiral fermion number, with an additional factor of $\sqrt{2}$ in our convention. The $(-1)^{F_L}$ defect partition functions can then be obtained by performing a modular $S$ transformation on the graded partition functions $Z^{(-1)^{F_L}}$ and $Z^{(-1)^{F_R}}$:
\ie
Z_{\rm R-NS} = 2\chi_{\frac{7}{16}} (\bar\chi_0 + \bar\chi_{\frac{3}{2}}) + 2\chi_{\frac{3}{80}} (\bar\chi_{\frac{1}{10}} + \bar\chi_{\frac{3}{5}} )  \,, \quad Z_{\rm NS-R} = Z_{\rm R-NS}^* \,.
\label{eqn:tri_chiral_fermion_parity_part}
\fe
The double degeneracy in the spectrum originates from the existence of a fermionic zero mode on the q-type TDL.\footnote{Regarding the $(-1)^{F_L}$ symmetry, there is another natural definition of the corresponding TDL that rescales it by a factor of $\frac{1}{\sqrt{2}}$. With this normalization, the invertibility of the fusion rules become manifest, and the actions on local operators become simply $\pm 1$. However, the price to pay is that the associated defect partition function is no longer properly quantized, and thus does not admit a Hilbert space interpretation. }

\subsection{Bosonic Theory: Tricritical Ising CFT}\label{subsec:tricritical}
Bosonization of the $\cN=1$ minimal model at $c=\frac{7}{10}$ gives the tricritical Ising CFT, which is the $A$-series Virasoro minimal model with $(p,q)=(5,4)$. As a diagonal minimal model, its the local primary operators are in one-to-one correspondence with the irreducible modules of the Virasoro algebras, and we will label them in the same way as in Table~\ref{tab:SUSY710}.
Furthermore, the simple TDLs only contain the Verlinde lines that are in one-to-one correspondence with the irreducible modules:
\ie
\begin{array}{cccccc}
I & \eta & N & W & \eta W & N W 
\\
1 & \varepsilon'' & \sigma' & \varepsilon' & \varepsilon & \sigma
\end{array}\,.
\fe
The fusion category formed by these TDLs is given by the tensor product of the Ising category and the unitary Fibonacci category.
We summarize their fusion rules as follows:
\ie
\eta^2 = I\,, \quad N^2 = I+\eta \,,\quad W^2 = I + W\,,\label{eq:fusion_tricritical}
\fe
where $\eta$ generates the dual $\mZ_2$ symmetry after bosonization, $N$ is the duality line that forms the Ising category with the $\mathbb{Z}_2$ lines, and $W$ is the Fibonacci line. 

The duality line $N$ can be interpreted as the image of both $(-1)^{F_L}$ and $(-1)^{F_R}$ under bosonization. This agrees with the fact that they all share the same quantum dimension $\sqrt{2}$, and furthermore, the $N$ defect Hilbert space~\eqref{eqn:tri_N_part} is the direct sum of the $(-1)^{F_L}$ and $(-1)^{F_R}$ defect Hilbert spaces~\eqref{eqn:tri_chiral_fermion_parity_part}, up to the fermionic double degeneracy.
Physically, the bosonization map of the TDLs can be understood as follows.
Since the $(-1)^{F_L}$ and $(-1)^{F_R}$ symmetries each have a mixed anomaly with $(-1)^F$, stacking the fermionic SPT $(-1)^{\text{Arf}(\rho)}$ on the $\cN=1$ minimal model can be undone by applying either symmetry transformation. This is reflected in the fact that the $(-1)^F$-even R-R sector and the $(-1)^F$-odd R-R sector are identical in the $\cN=1$ minimal model. Consequently, the two bosonization procedures lead to the same bosonic theory and the resulting bosonic theory, i.e.~the tricritical Ising CFT, is invariant under gauging the dual $\mZ_2$ symmetry generated by $\eta$. This self-duality then gives rise to the duality line $N$. The Fibonacci category is left unaffected under bosonization, since it does not interact with any fermion parity symmetries.

For Verlinde lines, the defect partition functions are given by
\ie
Z_{\cL_i} = \sum_{j,k} N_{ik}^j \chi_j \bar\chi_k \,,
\fe
where $N_{ik}^j$ is the fusion coefficient appearing in \eqref{eq:fusion_tricritical}.
For later convenience, we summarize the operator spectrum related to the $\mathbb{Z}_2$  line $\eta$ in Table~\ref{tab:tricritical_Z2}. It is consistent with how bosonization reshuffles the operator spectrum in the $\cN=1$ minimal model (Table~\ref{tab:n=1_operator}) following \eqref{eq:bosonization_operator}.
\vspace{-12pt}
\begin{table}[h]
    \centering
    \renewcommand{\arraystretch}{1.4} 
    \begin{tabular}{|c|c|c|}
    \hline
         & $\mZ_2$-even & $\mZ_2$-odd \\
         \hline
        {\rm untwisted} & $1_{0,0}, \varepsilon_{\frac{1}{10}, \frac{1}{10}}, \varepsilon'_{\frac{3}{5}, \frac{3}{5}}, \varepsilon''_{\frac{3}{2}, \frac{3}{2}}$ & $\sigma_{\frac{3}{80}, \frac{3}{80}}, \sigma'_{\frac{7}{16}, \frac{7}{16}}$ 
        \vspace{-18pt}\\
         &&\\
        \hline
        $\eta$-{\rm twisted} & $\mu_{\frac{3}{80}, \frac{3}{80}}, \mu'_{\frac{7}{16}, \frac{7}{16}}$ & $G_{\frac{3}{2},0}, \bar G_{0,\frac{3}{2}}, \psi'_{\frac{1}{10}, \frac{3}{5}}, \bar\psi'_{\frac{3}{5}, \frac{1}{10}}$
        \vspace{-18pt}\\
         &&\\
        \hline
    \end{tabular}
    \caption{The operator spectrum of the tricritical Ising CFT in different sectors related to the $\mZ_2$ symmetry generated by $\eta$. }
    \label{tab:tricritical_Z2}
\end{table}
\vspace{-10pt}

\noindent We also explicitly write down the $N$ defect partition function
\vspace{-6pt}
\ie
Z_N = \chi_{\frac{7}{16}} (\bar\chi_{0} +  \bar\chi_{\frac{3}{2}}) + \chi_{\frac{3}{80}} (\bar\chi_{\frac{3}{5}} + \bar\chi_{\frac{1}{10}}) + (\chi_0  + \chi_{\frac{3}{2}}) \bar\chi_{\frac{7}{16}} + (\chi_{\frac{3}{5}} + \chi_{\frac{1}{10}}) \bar\chi_{\frac{3}{80}} \,.
\label{eqn:tri_N_part}
\fe

\subsection{Bosonization of Superconformal Algebra}
We are now ready to discuss what happens to the superconformal algebra after bosonization. We will show that it becomes a chiral tube algebra in the bosonized theory.

The supercurrents $G(z),\bar G(\bar z)$ live in the NS-NS sector and are $(-1)^F$ odd. Thus, after bosonization, they become defect operators attached by the dual $\mZ_2$ symmetry line $\eta$ and carry a nontrivial $\mZ_2$ charge, as listed in Table~\ref{tab:tricritical_Z2}. The construction of the chiral tube algebra from these non-local supercurrents is analogous to the one in Section \ref{sec:orbifold_W3}
, except for an important subtle difference:~now the  currents $G(z),\bar G(\bar z)$ are charged under the TDL $\eta$ that is attached to them. This  will affect the lasso operators of the chiral tube algebra.

In order to build these lasso operators, we define the following projectors
\ie
P_{\eta,\pm} = \frac{1}{2} (1 \pm \eta) \,,
\fe
that absorb the $\eta$ line. We can then push the non-local supercurrent $G(z),\bar G(\bar z)$ onto the projectors and construct an operator from the local Hilbert space $\cH$ to the $\eta$ defect Hilbert space $\cH_\eta$ as follows
\ie
G P_{\eta,\pm}=\tikz[baseline=-0.75]{\draw[black] (0,0.75)--(0,1.5); \draw (0,0.75) arc (90:45:0.75); \draw[black] (0.53,0.53)--(0.73,0.73); \node[right] at (-0.05,1.35) {\small $\eta$}; \node[right] at (0.73,0.73) {\small $G$}; \draw[black] (0,0.5) arc (90:-90:0.5); \node[right] at (0.35,-0.4) {\small $P_{\eta,\pm}$}; \draw[black] (0,-0.5) arc (-90:-270:0.5)} = \frac{1}{2} \le \ \tikz[baseline=-0.75]{\draw[black] (0,0.75)--(0,1.5); \draw (0,0.75) arc (90:45:0.75); \draw[black] (0.53,0.53)--(0.73,0.73); \node[right] at (-0.05,1.35) {\small $\eta$};\node[right] at (0.73,0.73) {\small $G$}} \pm \ 
\tikz[baseline=-0.75]{\draw[black] (0,0.75)--(0,1.5); \draw (0,0.75) arc (90:45:0.75); \draw[black] (0.53,0.53)--(0.73,0.73); \node[right] at (-0.05,1.35) {\small $\eta$};\node[right] at (0.73,0.73) {\small $G$}; \draw[black] (0,0.5) arc (90:-90:0.5); \draw[black] (0,-0.5) arc (-90:-270:0.5); \node[right] at (0.35,-0.4) {\small ${\eta}$}; } \ri = \frac{1}{2} \le \ 
\tikz[baseline=-0.75]{\draw[black] (0,0.75)--(0,1.5); \draw (0,0.75) arc (90:45:0.75); \draw[black] (0.53,0.53)--(0.73,0.73); \node[right] at (-0.05,1.35) {\small $\eta$};\node[right] at (0.73,0.73) {\small $G$}} \pm \ 
\tikz[baseline=-0.75]{\draw[black] (0,0.75)--(0,1.5); \draw (0,0.75) arc (90:405:0.75); \draw[black] (0.53,0.53)--(0.73,0.73); \node[right] at (0.73,0.73) {\small $G$}; \node[right] at (-0.05,1.35) {\small $\eta$}; } \ri \,,\label{eq:GP1}
\fe
and similarly an operator from the $\eta$ defect Hilbert space $\cH_\eta$ to the local Hilbert space $\cH$
\vspace{-4pt}
\ie
G P_{\eta,\pm}=\tikz[baseline=-0.75]{\draw[black] (0,0.75)--(0,0); \draw (0,0.75) arc (90:45:0.75); \draw[black] (0.53,0.53)--(0.73,0.73); \node[right] at (-0.4,1.05) {\small $\eta$}; \node[right] at (0.73,0.73) {\small $G$}; \draw[black] (0,0.5) arc (90:-90:0.5); \node[right] at (0.35,-0.4) {\small $P_{\eta,\pm}$}; \draw[black] (0,-0.5) arc (-90:-270:0.5)} = \frac{1}{2} \le \ \tikz[baseline=-0.75]{\draw[black] (0,0.75)--(0,0); \draw (0,0.75) arc (90:45:0.75); \draw[black] (0.53,0.53)--(0.73,0.73); \node[right] at (-0.4,1.05) {\small $\eta$};\node[right] at (0.73,0.73) {\small $G$}} \pm \ 
\tikz[baseline=-0.75]{\draw[black] (0,0.75)--(0,0); \draw (0,0.75) arc (90:45:0.75); \draw[black] (0.53,0.53)--(0.73,0.73); \node[right] at (-0.4,1.05) {\small $\eta$};\node[right] at (0.73,0.73) {\small $G$}; \draw[black] (0,0.5) arc (90:-90:0.5); \draw[black] (0,-0.5) arc (-90:-270:0.5); \node[right] at (0.35,-0.4) {\small ${\eta}$}; } \ri = \frac{1}{2} \le \ 
\tikz[baseline=-0.75]{\draw[black] (0,0.75)--(0,0); \draw (0,0.75) arc (90:45:0.75); \draw[black] (0.53,0.53)--(0.73,0.73); \node[right] at (-0.4,1.05) {\small $\eta$};\node[right] at (0.73,0.73) {\small $G$}} \pm \ 
\tikz[baseline=-0.75]{\draw[black] (0,0.75)--(0,0); \draw (0,0.75) arc (90:405:0.75); \draw[black] (0.53,0.53)--(0.73,0.73); \node[right] at (0.73,0.73) {\small $G$}; \node[right] at (-0.4,1.05) {\small $\eta$}; } \ri \,. \label{eq:GP2}
\fe
Expanding these operators into modes give us the lasso operators of the chiral tube algebra. 

In order to determine the appropriate mode expansion, we need to fix the monodromy of the supercurrents $G(z), \bar G(\bar z)$, which in general receives three contributions:~one from the projector, one from the action of the vertical TDLs and one from the spin of the current. As an example, consider the monodromy of the supercurrent $G(z)$ around the projector $P_{\eta,-}$ in \eqref{eq:GP1}. When we wind $G(z)$ by $2\pi$, the projector $P_{\eta,-}$ contributes a $-1$ phase as in the case of the $\cW_3$ chiral tube algebra,~\eqref{eqn:tube_w3_1} and~\eqref{eqn:tube_w3}, in the tetracritical Ising CFT. In addition, there is a $-1$ phase from the half-integer spin of $G(z)$:
\ie
\tikz[baseline=-0.75]{\draw[black] (0,0.75)--(0,1.5); \draw (0,0) circle (0.75); \draw[black] (0.53,0.53)--(0.73,0.73); \node[right] at (0,1.1) {\small $\eta$}; \node[right] at (0.73,0.73) {\small $G$} ;\filldraw [black] (0.73,0.73) circle (2pt)
} \rightarrow   \quad
\tikz[baseline=-0.75]{\draw[black] (0,0.75)--(0,1.5); \draw (0,0) circle (0.75); \draw[black] (-0.53,0.53) .. controls (-0.53,1) and (-0.85,1) .. (-0.93,0.73); \draw[black] (-0.93,0.73) arc (180:315:0.15); \node[right] at (0,1.1) {\small $\eta$}; \node[above] at (-0.73,0.83) {\small $G$}; \filldraw [black] (-0.69,0.68) circle (2pt); \draw[blue,->] (0.73,0.73) arc (45:-210:1.03)
} = -
\tikz[baseline=-0.75]{\draw[black] (0,0.75)--(0,1.5); \draw (0,0) circle (0.75); \draw[black] (-0.69,0.68) arc (135:260:0.15); \node[right] at (0,1.1) {\small $\eta$}; \draw[]; \node[above] at (-0.73,0.83) {\small $G$}; \filldraw [black] (-0.69,0.68) circle (2pt)
}\,.
\fe
Lastly, we have to pass $G(z)$ through the vertical $\eta$ line once and this generates another $-1$ phase. Hence, in total, $G(z)$ has a $-1$ monodromy around the projector $P_{\eta,-}$ and thus has integer mode expansion with $n\in\mathbb{Z}$ in \eqref{eq:G_mode_expansion}. In comparison, when we wind $G(z)$ around the projector $P_{\eta,-}$ in \eqref{eq:GP2}, its monodromy receives only contributions from the projector and the spin of the current. The vertical $\eta$ line does not intersect the path of $G(z)$ and hence does not contribute a $-1$ phase. All together, $G(z)$ has a trivial monodromy in this case and therefore has a half-integer mode expansion with $n\in\mathbb{Z}+\frac{1}{2}$ in \eqref{eq:G_mode_expansion}.

In summary, the lasso operators of the chiral tube algebra are
\vspace{-6pt}
\ie
{[G_n]}^{\eta}_I = \tikz[baseline=-0.75]{\draw[black] (0,0.75)--(0,1.5); \draw (0,0) circle (0.75); \node[right] at (0,1.1) {\small $\eta$};
\node[right] at (0.5,-0.5) {\small $G_{n} P_{\eta,-}$}} ,\quad
[G_r]^{\eta}_I = \tikz[baseline=-0.75]{\draw[black] (0,0.75)--(0,1.5); \draw (0,0) circle (0.75); \node[right] at (0,1.1) {\small $\eta$};
\node[right] at (0.5,-0.5) {\small $G_r P_{\eta,+}$}} ,
\label{eqn:lasso_susy_1}
\fe
\ie
{[G_n]}^I_{\eta} = \tikz[baseline=-0.75]{\draw[black] (0,0)--(0,0.75); \draw (0,0) circle (0.75); \node[right] at (0,0.375) {\small $\eta$}; 
\node[right] at (0.5,-0.5) {\small $G_{n} P_{\eta,+}$}} ,\quad
[G_r]^I_{\eta} = \tikz[baseline=-0.75]{\draw[black] (0,0)--(0,0.75); \draw (0,0) circle (0.75); \node[right] at (0,0.375) {\small $\eta$}; 
\node[right] at (0.5,-0.5) {\small $G_r P_{\eta,-}$}} ,
\label{eqn:lasso_susy_2}
\fe
where we adopt the convention in the string theory literature that $n\in \mZ $ and $r \in \mZ + \frac{1}{2}$. Since the supercurrent $G(z)$ carries a nontrivial $\eta$-charge, we have the following relation:
\ie
GP_{\eta,\pm} = P_{\eta,\mp} G \,,\quad\bar{G}P_{\eta,\pm} = P_{\eta,\mp} \bar{G} \,.
\fe
It implies that the chiral tube algebra maps $\mZ_2$-even operators to $\mZ_2$-odd operators and vice versa. As a result, the lasso operators split into two independent sets, $[G_n]_I^\eta,[G_n]^I_\eta$ and $[G_r]_I^\eta,[G_r]^I_\eta$, which do not interact with each other. These two sets of lasso operators each generate an algebra isomorphic to the R and NS versions of the $\cN=1$ superconformal algebra, respectively. There is a similar anti-chiral $\cN=1$ superconformal chiral tube algebra built from the supercurrent $\bar G(\bar z)$.
The action of the lasso operators in~\eqref{eqn:lasso_susy_1} and~\eqref{eqn:lasso_susy_2}, as well as their anti-chiral counterparts, are consistent with the operator content of the tricritical Ising CFT listed in Table \ref{tab:tricritical_Z2}.

\subsection{Chiral Tube Algebra in Other Defect Hilbert Spaces}
Next, we discuss the chiral tube algebra acting on the other defect Hilbert spaces. Since the TDL $W$ does not interact with $\eta$ at all, the chiral tube algebra acting on the $W$ and $\eta W$ defect Hilbert spaces is essentially the same as that acting on the local and $\eta$ defect Hilbert spaces, except that the lasso operators now include an additional $W$ line inserted in the vertical direction. We will not elaborate further on this case.

By contrast, the chiral tube algebra in the $N$ defect Hilbert space is considerably more intricate, and we will analyze it in detail below. We start by determining the $N$ actions on supercurrents $G(z),\bar G(\bar z)$ and the $\eta$ actions on $N$-defect operators. 

Firstly, because the $F$-symbol $F_{N\eta N}^{\eta} = -1$ is nontrivial, the action of $\eta$ line on the $N$ defect Hilbert space admits two different resolutions. We choose the following one
\vspace{-5pt}
\ie
\eta^- = \tikz[baseline=0.9cm]{\draw (-1,0)--(1,0); \draw (-1,2)--(1,2); \draw (-1,0)--(-1,2); \draw (1,0)--(1,2); \draw[red] (0,0)--(0,2); \draw (-1,1)--(0,1.2); \draw (0,0.8)--(1,1); \node[below] at (-0.5,1.15) {\small $\eta$}; \node[above] at (0.5,0.85) {\small $\eta$}}\ \,,
\fe
where the superscript on $\eta$ indicates the chosen resolution. The other resolution will be denoted by $\eta^+$.
Using the $F$-move, one can show that $( \eta^-)^2=-1$ :
\vspace{-4pt}
\ie
(\eta^-)^2 = \tikz[baseline=0.9cm]{\draw (-1,0)--(1,0); \draw (-1,2)--(1,2); \draw (-1,0)--(-1,2); \draw (1,0)--(1,2); \draw[red] (0,0)--(0,2); \draw (-1,1.3)--(0,1.5); \draw (0,1.1)--(1,1.3); \node[below] at (-0.5,1.45) {\small $\eta$}; \node[above] at (0.5,1.15) {\small $\eta$};
\draw (-1,0.6)--(0,0.8); \draw (0,0.4)--(1,0.6); \node[below] at (-0.5,0.75) {\small $\eta$}; \node[above] at (0.5,0.45) {\small $\eta$}
}=-\ \tikz[baseline=0.9cm]{\draw (-1,0)--(1,0); \draw (-1,2)--(1,2); \draw (-1,0)--(-1,2); \draw (1,0)--(1,2); \draw[red] (0,0)--(0,2); \draw (-1,1.3)--(0,1.5); \draw (0,0.8)--(1,1.3); \node[below] at (-0.5,1.45) {\small $\eta$}; \node[above] at (0.5,1.15) {\small $\eta$};
\draw (-1,0.6)--(0,1.1); \draw (0,0.4)--(1,0.6); \node[below] at (-0.5,0.75) {\small $\eta$}; \node[above] at (0.5,0.45) {\small $\eta$}
}
= -\
\tikz[baseline=0.9cm]{\draw (-1,0)--(1,0); \draw (-1,2)--(1,2); \draw (-1,0)--(-1,2); \draw (1,0)--(1,2); \draw[red] (0,0)--(0,2)
}\,.
\fe
Therefore, the $\eta$ line acts on the $N$ defect Hilbert space with eigenvalues $\pm i$. Consequently, the appropriate projector in the $N$ defect Hilbert space should take the form
\vspace{-8pt}
\ie
P_{N,\pm i} = \frac{1}{2} \sum_{n=0}^{1} (\mp i)^n\,  \tikz[baseline=-0.75]{\draw[red] (0,0)--(0,1.5); \draw (0,0) circle (0.75); \node[right] at (0.6,-0.5) {\small $(\eta^-)^n$}} \,,
\fe
where we use the red lines to denote the $N$ line and suppress the explicit label $N$. Because of the nontrivial $F$-symbols, the fusion of the projector with the supercurrents $G(z),\bar G(\bar z)$ is more subtle. Fusing the supercurrents from the top gives 
\ie
G P_{N,\pm i} = \frac{1}{2} \le \ \tikz[baseline=-0.75]{\draw[red] (0,0)--(0,1.5); \draw (0,0.75) arc (90:45:0.75); \draw[black] (0.53,0.53)--(0.73,0.73); \node[right] at (0.73,0.73) {\small $G$}} \mp i \ 
\tikz[baseline=-0.75]{\draw[red] (0,0)--(0,1.5); \draw (0,0.75) arc (90:45:0.75); \draw[black] (0.53,0.53)--(0.73,0.73); \node[right] at (0.73,0.73) {\small $G$}; \draw[black] (0,0.5) arc (90:-90:0.5); \draw[black] (0,-0.5) arc (-90:-270:0.55)} \ri = \frac{1}{2} \le \ 
\tikz[baseline=-0.75]{\draw[red] (0,0)--(0,1.5); \draw (0,0.75) arc (90:45:0.75); \draw[black] (0.53,0.53)--(0.73,0.73); \node[right] at (0.73,0.73) {\small $G$}} \pm i \ 
\tikz[baseline=-0.75]{\draw[red] (0,0)--(0,1.5); \draw (0,0.75) arc (90:405:0.75); \draw[black] (0.53,0.53)--(0.73,0.73); \node[right] at (0.73,0.73) {\small $G$}} \ri \,,
\fe
while fusing them from the bottom gives
\ie
P_{N,\pm i} G &= \frac{1}{2} \le \ \tikz[baseline=-0.75]{\draw[red] (0,0)--(0,1.5); \draw (0,0.75) arc (90:45:0.75); \draw[black] (0.53,0.53)--(0.73,0.73); \node[right] at (0.73,0.73) {\small $G$}} \mp i \ 
\tikz[baseline=-0.75]{\draw[red] (0,0)--(0,1.5); \draw (0,0.75) arc (90:45:0.75); \draw[black] (0.53,0.53)--(0.73,0.73); \node[below] at (0.73,0.73) {\small $G$}; \draw[black] (0,1.2) arc (90:-90:1.2); \draw[black] (0,-1.2) arc (-90:-270:1.3)} \ri = \frac{1}{2} \le \ 
\tikz[baseline=-0.75]{\draw[red] (0,0)--(0,1.5); \draw (0,0.75) arc (90:45:0.75); \draw[black] (0.53,0.53)--(0.73,0.73); \node[right] at (0.73,0.73) {\small $G$}} \mp i \ 
\tikz[baseline=-0.75]{\draw[red] (0,0)--(0,1.5); \draw (0,1.2) arc (90:425:1.2); \draw[black] (0.53,0.53)--(0.73,0.73); \node[below] at (0.73,0.73) {\small $G$}; \draw[black] (0.53,0.53) .. controls (0.32,0.7) and (0.32,1) .. (0.507,1.088)} \ri \\
&= \frac{1}{2} \le \ 
\tikz[baseline=-0.75]{\draw[red] (0,0)--(0,1.5); \draw (0,0.75) arc (90:45:0.75); \draw[black] (0.53,0.53)--(0.73,0.73); \node[right] at (0.73,0.73) {\small $G$}} \pm i \ 
\tikz[baseline=-0.75]{\draw[red] (0,0)--(0,1.5); \draw (0,0.75) arc (90:405:0.75); \draw[black] (0.53,0.53)--(0.73,0.73); \node[right] at (0.73,0.73) {\small $G$}} \ri \,.
\fe
In the last equality, we use the fact that $G(z)$ has half-integer spin. Comparing the two fusions, we find that
\ie
GP_{N, \pm i} = P_{N, \pm i} G \,, \quad \bar{G}P_{N, \pm i} = P_{N, \pm i} \bar{G} \,,
\fe
implying that the chiral tube algebra does not change the $\eta^-$ charge of the states in $\cH_{N}$. 

On the other hand, to fully determine the monodromy of the supercurrents, we also need to determine the action of $N$ on them. As before, there are two choices of resolution of actions due to the nontrivial $F$-symbol $F_{N\eta N}^{\eta} = -1$. Note that the actions of $\eta^-$ on $N$ defect operators are completely determined by their spins\cite{Chang:2018iay}:
\ie
\eta^- = \begin{cases}
    +i \ & s = \frac{1}{16} + \frac{\mZ}{2} \,,\\
    -i \ & s = - \frac{1}{16} + \frac{\mZ}{2} \,.
\end{cases}
\fe
Using this relation, we can obtain the $\eta^-$-graded  $N$ defect partition function
\ie
Z_{N}^{\eta^-} = -i (\chi_{\frac{7}{16}} \bar\chi_{0} + \chi_{\frac{7}{16}} \bar\chi_{\frac{3}{2}} + \chi_{\frac{3}{80}} \bar\chi_{\frac{3}{5}} + \chi_{\frac{3}{80}} \bar\chi_{\frac{1}{10}}) + i(\chi_0 \bar\chi_{\frac{7}{16}} + \chi_{\frac{3}{2}} \bar\chi_{\frac{7}{16}} + \chi_{\frac{3}{5}} \bar\chi_{\frac{3}{80}}  + \chi_{\frac{1}{10}} \bar\chi_{\frac{3}{80}} ) \,.
\fe
Applying a modular $S$ transformation, we obtain the $N^+$-graded  $\eta$ defect partition function
\ie
Z_{\eta}^{N^+} := \tikz[baseline=0.9cm]{\draw (-1,0)--(1,0); \draw (-1,2)--(1,2); \draw (-1,0)--(-1,2); \draw (1,0)--(1,2); \draw[red] (-1,1)--(1,1); \draw (0,0)--(-0.2,1); \draw (0.2,1)--(0,2); \node[left] at (0.3,0.5) {\small $\eta$}; \node[right] at (-0.3,1.5) {\small $\eta$}; \node[below] at (0.8,1) {\small $N$}} = i\sqrt{2} ( - \chi_0 \bar\chi_{\frac{3}{2}} +\chi_{\frac{1}{10}} \bar\chi_{\frac{3}{5}} - \chi_{\frac{3}{5}} \bar\chi_{\frac{1}{10}} + \chi_{\frac{3}{2}} \bar\chi_0 ) \,.
\fe
From this, we read off the $N^+$ action on the supercurrents: 
\ie
{N}^+ \cdot G = i\sqrt{2} G\,,\quad {N}^+ \cdot \bar{G} = -i\sqrt{2} \bar{G}\,.
\fe

Now, using the actions of $N$ lines on supercurrents, we can derive their monodromy around the projectors. As an illustration, consider the monodromy of $G(z)$ around $P_{N,-i}$:
\ie
\tikz[baseline=-0.75]{\draw[red] (0,0)--(0,1.5); \draw (0,0) circle (0.5); \node[below] at (0,0) {\tiny $P_{N,-i}$};  \draw (0,0.75) arc (90:45:0.75); \draw[black] (0.53,0.53)--(0.73,0.73); \node[right] at (0.73,0.73) {\small $G$}} \rightarrow
\tikz[baseline=-0.75]{\draw[red] (0,0)--(0,1.5); \draw (0,0.75) arc (90:-225:0.75); \draw[black] (-0.53,0.53) .. controls (-0.53,1) and (-0.85,1) .. (-0.93,0.73); \draw[black] (-0.93,0.73) arc (180:315:0.15); \node[above] at (-0.73,0.83) {\small $G$}; \filldraw [black] (-0.69,0.68) circle (2pt); \draw (0,0) circle (0.5); \node[below] at (0,0) {\tiny $P_{N,-i}$}; } = -\ 
\tikz[baseline=-0.75]{\draw[red] (0,0)--(0,1.5); \draw (0,0.75) arc (90:-225:0.75); \draw[black] (-0.53,0.53)--(-0.33,0.73); \node[above] at (-0.33,0.73) {\small $G$}; \filldraw [black] (-0.33,0.73) circle (2pt); \draw (0,0) circle (0.5); \node[below] at (0,0) {\tiny $P_{N,-i}$}; } = -i \  
\tikz[baseline=-0.75]{\draw[red] (0,0)--(0,1.5); \draw (0,0.75) arc (90:-90:0.75); \draw (0,-0.75) arc (-90:-270:0.85); \node[above] at (1,1) {\small $G$}; \draw (0,0) circle (0.5); \node[below] at (0,0) {\tiny $P_{N,-i}$}; \draw (0,1.2) arc (90:45:1.2); \draw (0.849,0.849)--(1,1)} = -\ \tikz[baseline=-0.75]{\draw[red] (0,0)--(0,1.5); \draw (0,0) circle (0.5); \node[below] at (0,0) {\tiny $P_{N,-i}$};  \draw (0,0.75) arc (90:45:0.75); \draw[black] (0.53,0.53)--(0.73,0.73); \node[right] at (0.73,0.73) {\small $G$}}  \,.
\fe
In the first equality, we obtain a $-1$ phase from the half-integer spin of $G(z)$. In the second equality, we apply the inverse ${N}^-$ action on $G(z)$, which produces a $i$ phase. This differs from the inverse ${N}^+$ action by a sign. Lastly, we absorb the $\eta^-$ line into the projector, which leads to a $-i$ phase. In the end, the derivation shows that $G(z)$ obeys a anti-periodic boundary condition around $P_{N,-i}$ and thus has an integer mode expansion. The monodromy of $G(z)$ around the other projectors can be derived similarly.

With the monodromy information, we can construct the following lasso operators:
\ie
{[G_n]}^N_N = \tikz[baseline=-0.75]{\draw[red] (0,0)--(0,1.5); \draw (0,0) circle (0.75); 
\node[right] at (0.5,-0.5) {\small $G_{n} P_{N,-i}$}} ,\;\quad
[G_r]^N_N = \tikz[baseline=-0.75]{\draw[red] (0,0)--(0,1.5); \draw (0,0) circle (0.75); 
\node[right] at (0.5,-0.5) {\small $G_r P_{N,i}$}} \,, \\
[\bar{G}_r]^N_N = \tikz[baseline=-0.75]{\draw[red] (0,0)--(0,1.5); \draw (0,0) circle (0.75); 
\node[right] at (0.5,-0.5) {\small $\bar{G}_{r} P_{N,-i}$}} ,\;\quad
[\bar{G}_n]^N_N = \tikz[baseline=-0.75]{\draw[red] (0,0)--(0,1.5); \draw (0,0) circle (0.75); 
\node[right] at (0.5,-0.5) {\small $\bar{G}_n P_{N,i}$}} \,.
\fe
Since $G(z)$ commutes with the projectors, the lasso operators $[G_n]^{N}_N$ and $[G_r]^N_N$ form two independent algebras isomorphic to the $\cN=1$ superconformal algebra in the R and NS sector, respectively. The same holds for the anti-chiral tube algebra. This structure is consistent with the defect partition function \eqref{eqn:tri_N_part} of the $N$ defect Hilbert space.

\section{Discussions and Outlook}
\label{sec:discussion}

We highlight a few future directions and open questions.

\paragraph{Representation theory of chiral tube algebras.}
For a strongly regular chiral algebra $\mathcal{V}$, the category of its representations $\text{Rep}(\mathcal{V})$ is known to form a modular tensor category \cite{Moore:1988qv,Huang:2005gs}. Physically, this modular tensor category describes the anyons in the 3d topological quantum field theory whose boundary realizes the chiral algebra $\mathcal{V}$. It would be interesting to systematically develop a representation theory of chiral tube algebras and to identify the corresponding mathematical structure.

\paragraph{SymTFT for chiral tube algebras.}
The representations of tube algebras are encoded compactly in the 3d SymTFT \cite{Barrett:1993ab,TURAEV1992865,kirillov2010turaevviroinvariantsextendedtqft,Gaiotto:2014kfa,kong2015boundarybulkrelationtopologicalorders,Kong_2017,Pulmann:2019vrw,Thorngren:2019iar,Ji:2019jhk,Lichtman:2020nuw,Kong_2020,Gaiotto:2020iye,Aasen:2020jwb,Apruzzi:2021nmk,Chatterjee:2022kxb,Freed:2022qnc,Kaidi:2022cpf} associated with the TDLs. They are in one-to-one correspondence with the anyons in the SymTFT \cite{evans1995ocneanu,Izumi:2000qa,Mueger:2001crc,Lin:2022dhv,Bartsch:2023wvv,Bhardwaj:2023ayw}. It would be interesting to develop an analogous SymTFT perspective for chiral tube algebras. The continuous spacetime SymTFTs developed in \cite{Apruzzi:2025hvs} may provide a useful starting point.

\paragraph{Chiral strip algebras.}
Recent works have generalized tube algebras to strip algebras \cite{Copetti:2408.01490,Cordova:2408.11045,Choi:2409.02159,Choi:2409.02806,Bhardwaj:2409.02166}, which describe the interplay between topological defects and boundaries. It would be interesting to extend chiral tube algebras in a similar direction by developing ``chiral strip algebras". Such algebras could perhaps be used to bootstrap the conformal data associated with the conformal boundaries in rational CFTs.

\paragraph{Chiral tube algebras in 4d SCFTs.} There is an intriguing correspondence between 4d $\mathcal{N}=2$ superconformal field theories (SCFTs) and chiral algebras \cite{Beem:1312.5344}. It would be interesting to connect chiral tube algebras with this SCFT/VOA correspondence. In particular, under the correspondence, a superconformal surface defect gives rise to a module of the corresponding chiral algebra, whose character computes the surface defect Schur index \cite{Cordova:2017mhb}. Similarly, a twisted superconformal surface defects attached by 3d topological defects gives rise to a twisted module of the corresponding chiral algebra. From the perspective of the chiral algebra plane, an ordinary surface defect appears as a local point operator, whereas a twisted surface defect appears as a non-local point operator attached by a TDL. This suggests that one might be able to define a chiral tube algebra acting simultaneously on superconformal surface defects and their twisted counterparts. A natural setup for exploring this idea is the compactification of a 4d SCFT on $S^2\times\Sigma$, where the resulting two-dimensional non-unitary CFT on $\Sigma$ carries the chiral algebra associated with the original 4d SCFT \cite{Rastelli:2025nyn}. It would also be interesting to study twisted defects attached by non-invertible symmetries \cite{Choi:2021kmx,Kaidi:2021xfk,Choi:2022zal,Kaidi:2022uux,Bashmakov:2022uek} in 4d SCFTs, and to explore their interactions with the corresponding chiral algebras. An example of such non-invertible twisted defects has been studied in Maxwell theory \cite{Shao:2025qvf}.
Another hint that chiral tube algebras might arise naturally in the SCFT/VOA correspondence is that the characters of twisted modules appear in the lens space index of 4d $\mathcal{N}=2$ SCFTs \cite{Fluder:1710.06029}.

\section*{Acknowledgments}
We thank Federico Ambrosino, Yu-Jui Chen, Yichul Choi, Clay C\'{o}rdova, Davide Gaiotto, Yuya Kusuki, 
Krzysztof Pilch, Brandon Rayhaun, Hubert Saleur, Nathan Seiberg, Shu-Heng Shao,  Steve Shenker, David Simmons-Duffin, Jaewon Song, 
Ryan Thorngren, Yifan Wang, Yi-Nan Wang, and Nicholas Warner for helpful discussions. We thank Federico Ambrosino, Diego Delmastro, Justin Kaidi, Liang Kong, Yuya Kusuki, Kantaro Ohmori, Brandon Rayhaun, Sakura Schafer-Nameki, Shu-Heng Shao, Adar Sharon, Yuji Tachikawa, and Yunqin Zheng for the comments on a draft. N.B. and C.L. are supported by the U.S. Department of Energy, Office of Science, Office of High Energy Physics under grant Contract Number DE-SC0026324. H.T.L. is supported by the U.S. Department of Energy, Office of Science, Office of High Energy Physics of U.S. Department of Energy under grant Contract Number DE-SC0012567 (High Energy Theory research), the Packard Foundation award for Quantum Black Holes from Quantum Computation and Holography, and the Simons Investigator Award No. 926198. This manuscript benefited from language
improvements assisted by ChatGPT 5.5.

\appendix

\section{Local Currents in Virasoro Minimal Models}
\label{app:generalminimalmodel}

In this appendix, we consider local chiral currents in Virasoro minimal models in general. The minimal model $\mathcal{M}(p,q)$ has central charge
\begin{equation}
    c = 1-\frac{6(p-q)^2}{pq}\,,
\end{equation}
where $p, q$ are coprime. Since there is a symmetry between $p$ and $q$, we will always without loss of generality assume $p$ is odd in this appendix.

Minimal models have an ADE classification. For every $(p,q)$, there is always an $A$-series minimal model. On the other hand, a $D$-series minimal model exists whenever $q$ is even and $q\geq 6$, and an $E$-series minimal model exists whenever $q=12$, $18$, or $30$, corresponding to $E_6$, $E_7$, and $E_8$, respectively.

The $A$-series minimal model contains only scalar primaries, so it has no enhanced chiral algebra.
However, it always has a $\mZ_2$ Verlinde line with a non-local chiral current of spin 
\ie
h_{q-1,1} = \frac{(p-2)(q-2)}{4}
\fe
living at its end. The $D$-series minimal models are obtained by gauging this $\mZ_2$ symmetry.

If $q \equiv 2 ~(\text{mod}~4)$, this non-local  chiral current becomes a local chiral current in the $D$-series minimal model. Its chiral algebra is therefore $\mathcal{W}(2,\frac{(p-2)(q-2)}4)$. Moreover, because the modular invariant takes a block-diagonal form, this $\cW$ algebra is the maximally extended algebra. For example, the $D$-series minimal model $\mathcal{M}(6,7)$ (which has $c=\frac 67$), has chiral algebra $\mathcal{W}(2,5)$ (see e.g. \cite{Blumenhagen:1990jv,Kausch:1990bn, Bouwknegt:1992wg} for discussion on $\mathcal{W}(2,5)$ in general). By $\mathbb{Z}_2$ gauging, we expect there is a corresponding chiral tube algebra in the $A$-series minimal models that takes a form very similar to that in Section \ref{sec:w3}.

If $q \equiv 0~(\text{mod}~4)$, the non-local  chiral current in the $A$-series minimal model remains a non-local chiral current in the $D$-series minimal model, so the $D$-series minimal model does not have an enhanced chiral algebra. However, in its fermionization, this non-local chiral current becomes local and leads to an enhanced $\mathcal{W}(2,\frac{(p-2)(q-2)}4)$ algebra. Similarly, the fermionization of the $A$-series minimal models also have the same enhanced chiral algebra. This implies that, by bosonization, both the $A$ series and the $D$ series minimal model have a chiral tube algebra generated by these fermionic non-local chiral currents, similar to what happens in Section \ref{sec:susy}.

The $E$-series minimal models also have enhanced chiral algebras. In particular, the bosonic theories have chiral algebras $\mathcal{W}(2,p-3)$, $\mathcal{W}(2,4p-8)$, $\mathcal{W}(2,p-5, 3p-9,7p-14)$ for $E_6$, $E_7$, and $E_8$, (i.e.~$q=12, 18, 30$), respectively. Note also that the $E_7$ minimal models have the same chiral algebra as the $D$-series minimal models at $q=18$, namely $\mathcal{W}(2,4p-8)$. This is reminiscent of the fact that the $D$-series and $A$-series minimal models have the same chiral algebra when $q \equiv 0~(\text{mod}~4)$.

\section{Derivation of Spectral Flow}
\label{app:sf}

In this appendix, we give a quick review and derivation of spectral flow \cite{Schwimmer:1986mf} for convenience. First, let us for simplicity assume we have a compact $U(1)$ charge purely left-moving. If we ignore the charge quantization and normalize the level of the $U(1)$ current $k$ to be $1$, then the partition function $Z(\tau,\bar\tau,z)$ defined as
\begin{equation}
    Z(\tau,\bar\tau,z) = \text{Tr}(e^{2\pi i \tau (L_0-\frac c{24})} e^{-2\pi i \bar\tau (\bar{L_0}-\frac{c}{24})} e^{2\pi i z J_0})
\end{equation} obeys
\begin{equation}
    Z\left(\frac{a \tau+b}{c\tau+d}, \frac{a\bar\tau+b}{c\bar\tau+d},\frac{z}{c\tau+d}\right) = e^{\pi i \frac{ cz^2}{c\tau+d}} Z(\tau,\bar \tau,z).
    \label{eq:modularcharge}
\end{equation}
See Appendix B of \cite{Benjamin:2016fhe} for a derivation. (\ref{eq:modularcharge}) makes no assumption on the level or compactness of the $u(1)$. It is true even if the $u(1)$ is noncompact.

Now, let us assume the $U(1)$ is compact at level $k$.  A compact $U(1)$ means we can choose to rescale the operator $J_0$ so that the charges are integral with gcd $1$. Suppose this rescaling gives the operator $J_0$ a level $k$, then (\ref{eq:modularcharge}) becomes
\begin{equation}
    Z\left(\frac{a \tau+b}{c\tau+d}, \frac{a\bar\tau+b}{c\bar\tau+d},\frac{z}{c\tau+d}\right) = e^{\pi i k \frac{ cz^2}{c\tau+d}} Z(\tau,\bar \tau,z)
    \label{eq:modularcharge2}
\end{equation}
Charge integrality means that 
\begin{equation}
    Z(\tau,\bar\tau, z+1) = Z(\tau,\bar\tau, z)
\end{equation}
This implies for any integer $\ell$:
\begin{align}
     e^{\pi i k \frac{z^2}{\tau}} Z(\tau,\bar\tau,z)=Z\left(-\frac1\tau, -\frac1{\bar\tau},\frac{z}{\tau}\right) =  Z\left(-\frac1\tau, -\frac1{\bar\tau},\frac{z}{\tau}+\ell\right) 
     = e^{\pi i k \frac{(z+\ell\tau)^2}{\tau}} Z(\tau,\bar\tau,z+\ell\tau) \,.
     \label{eq:derivbla}
\end{align}
In the third equality of (\ref{eq:derivbla}), we simply took the general transformation property (\ref{eq:modularcharge2}) and plugged in $z+\ell\tau$ for $z$, and $a=0,b=-1,c=1,d=0$. Now set the first and last expression of (\ref{eq:derivbla}) equal. Expanding we get
\begin{equation}
    e^{\pi i k (2\ell z+ \ell^2\tau)} Z(\tau,\bar\tau, z+\ell\tau) = Z(\tau,\bar\tau,z)\,.
    \label{eq:sfprop}
\end{equation}
This means the following spectral flow transformation
\begin{align}
    L_0 \rightarrow L_0 + \ell J_0 + \frac{\ell^2 k}{2}\,,\qquad J_0 \rightarrow J_0 + \ell k\,, \nonumber
\end{align}
leaves the spectrum invariant.

We can also consider a general global $U(1)$ symmetry generated by a (non-holomorphic) current $J_\mu(z,\bar z)$. From current conservation, $J_z$ (and $J_{\bar z}$) is a (anti-)holomorphic current that we denote by $J(z)$ (and $\bar{J}(\bar z)$). We denote the levels of $J(z)$ and $\bar{J}(\bar z)$ by $k$ and $\bar k$. The partition function
\begin{equation}
    Z(\tau,\bar\tau,z,\bar z) = \text{Tr}(e^{2\pi i \tau (L_0-\frac c{24})} e^{-2\pi i \bar\tau (\bar{L_0}-\frac{c}{24})} e^{2\pi i z J_0} e^{-2\pi i \bar z \bar{J_0}})\end{equation}
    now obeys the modular transformation property
\begin{equation}
    Z\left(\frac{a \tau+b}{c\tau+d}, \frac{a\bar\tau+b}{c\bar\tau+d},\frac{z}{c\tau+d}, \frac{\bar z}{c\bar\tau+d}\right) = e^{\pi i k \frac{ cz^2}{c\tau+d}} e^{-\pi i \bar k \frac{c {\bar z}^2}{c\bar\tau+d}}Z(\tau,\bar \tau,z,\bar z)
    \label{eq:modularchargeNonHolo}
\end{equation}
and the charge integrality condition
\begin{equation}
    Z(\tau,\bar\tau, z+1,\bar z -1) = Z(\tau,\bar\tau, z, \bar z)
\end{equation}
(since $J_0 + \bar{J_0} \in \mathbb Z$). This implies then that for any integer $\ell$,
\begin{align}
    e^{\pi i \left( \frac{k z^2}{\tau} - \frac{\bar k \bar{z}^2}{\bar\tau} \right)}Z(\tau,\bar\tau,z,\bar z) &= Z\left(-\frac1\tau,-\frac1{\bar\tau},\frac{z+\ell\tau}{\tau}, \frac{\bar z-\ell \bar\tau}{\bar \tau}\right) \nonumber \\
    &= e^{\pi i k \frac{(z+\ell\tau)^2}{\tau}} e^{-\pi i \bar k \frac{(\bar z - \ell \bar\tau)^2}{\bar\tau}} Z(\tau,\bar\tau,z+\ell\tau,\bar z-\ell\bar\tau).
\end{align}
Finally this implies
\begin{equation}
    e^{\pi i k (2\ell z+\ell^2\tau)} e^{-\pi i \bar k(-2\ell \bar z+\ell^2\bar\tau)}  Z(\tau,\bar\tau,z+\ell\tau,\bar z - \ell\bar\tau) = Z(\tau,\bar\tau,z,\bar z).
\end{equation}
This means the following spectral flow transformation
\vspace{-2pt}
\begin{align}
    L_0 \rightarrow L_0 + \ell J_0 + \frac{\ell^2 k}{2}\,,\qquad J_0 \rightarrow J_0 + \ell k\,, \nonumber \\
    \bar{L}_{0} \rightarrow \bar{L}_{0} -\ell \bar{J}_0 + \frac{\ell^2 \bar k}{2}\,,\qquad \bar J_0 \rightarrow \bar J_0 - \ell \bar k\,
\end{align}
\vspace{-2pt}
leaves the spectrum invariant. See \cite{Benjamin:2020swg} for more discussion on this.

Note that we are free to choose a different normalization of the charges $J_0, \bar{J}_0$. If we rescale the charges by some constant $\alpha$, then the levels $k, \bar k$ get rescaled by $\alpha^2$, and the spectral flow maps local operators to local operators whenever $\ell$ is a multiple of $\alpha^{-1}$. Our convention for the $su(2)_1$ characters in (\ref{eq:characterswithu1bla}) have half-integer charges. Hence the spectral flow parameter $\thetaa$ has periodicity $2$, not $1$. The convention used in the main text is the related to the one used in this appendix by
\vspace{-2pt}
\ie
J_0=2Q\,,\quad \ell=\frac{x}{2}\,.
\fe
\vspace{-2pt}
The $\mathfrak{u}(1)$ Cartan of the $\mathfrak{su}(2)_1$ chiral algebra has level $k=2$. This gives the spectral flow
\vspace{-2pt}
\ie
h\rightarrow h - xQ + \frac{1}{4}x^2\,,\qquad Q\rightarrow Q - \frac{1}{2}x\,,
\fe
\vspace{-2pt}
which agrees with the one derived in the main text.

\section{Normal ordering ambiguity}
\label{app:normalorder}

In Section \ref{sec:SU(2)}, we consider the twisted $\mathfrak{su}(2)_1$ chiral algebra. The Sugawara stress tensor, by a general twist $\thetaa$, had a normal order ambiguity which we wrote in (\ref{eqn:sugawara_twist}). We now fix the coefficients $A$, $B_n$ in (\ref{eqn:sugawara_twist}) as follows.

We first fix $B_n$ by requiring that $[L_{\thetaa,n}, J^+_{\thetaa,0}] = -\thetaa J^+_{\thetaa,n}$. From  (\ref{eqn:twisted_su2_chiral_algebra}), we get:
\vspace{-8pt}
\ie
{[L_{\thetaa,n}, J^+_{\thetaa,0}]} =\,& \frac{1}{3}  \left[\sum_{m \leq -1} \le J^+_{\thetaa,m} J^3_{\thetaa,n-m} + J^3_{\thetaa,m} J^+_{\thetaa,n-m} \ri  + \sum_{m \geq 0} \le J^+_{\thetaa,n-m} J^3_{\thetaa,m} + J^3_{\thetaa,n-m} J^+_{\thetaa,m} \ri \right] +\\
&\frac{1}{6} \left[ \sum_{m \leq -1} J^+_{\thetaa,m} (-2J^3_{\thetaa,n-m} - \thetaa \delta_{n-m, 0}) + \sum_{m \geq 0} (-2J^3_{\thetaa,n-m} - \thetaa \delta_{n-m,0}) J^+_{\thetaa,m} \right] +\\
&\frac{1}{6} \left[ \sum_{m \leq -1} (-2J^3_{\thetaa,m} - \thetaa\delta_{m,0}) J^+_{\thetaa,n-m} + \sum_{m \geq 0} J^+_{\thetaa,n-m} (-2J^3_{\thetaa,m} - \thetaa \delta_{m,0}) \right]  + B_n J^+_{n;\thetaa}  \\ 
=\, &
-\frac{1}{3} \thetaa J^+_{n;\thetaa} + B_n J^+_{n;\thetaa} \,.
\fe
which leads to $B_n = -\frac{2}{3} \thetaa$. We then fix $A$ by requiring that $[L_{\thetaa,1}, L_{\thetaa,-1}] = 2L_{\thetaa,0}$:
\vspace{-8pt}
{\small
\ie
&[L_{\thetaa,1}, L_{\thetaa,-1}] 
\\
=&  \left[L_{\thetaa,1}, 
\frac{1}{3}  \left(\sum_{m \leq -1} J^3_{\thetaa,m} J^3_{\thetaa,-1-m} +  \sum_{m \geq 0} J^3_{\thetaa,-1-m} J^3_{\thetaa,m}\right) \right] +\\
& \left[L_{\thetaa,1},\frac{1}{6}\le \sum_{m \leq -1} J^+_{\thetaa,m} J^-_{\thetaa,-1-m} + \sum_{m \geq 0} J^-_{\thetaa,-1 -m} J^+_{\thetaa,m} + \sum_{m \leq -1 } J^-_{\thetaa,m} J^+_{\thetaa,-1-m} + \sum_{m \geq 0} J^+_{\thetaa,-1 -m} J^-_{\thetaa,m} \ri  
- \frac{2}{3} \thetaa J^3_{\thetaa,-1}\right] \\
=\,& \frac{1}{3} \left[\sum_{m \leq -1} \le -m J^3_{\thetaa,m + 1} J^3_{
\thetaa,-1-m} + (1+m) J^3_{\thetaa,m} J^3_{\thetaa,-m} \ri  + \sum_{m \geq 0} \le (1+m) J^3_{\thetaa,-m} J^3_{\thetaa,m} -m  J^3_{\thetaa,-1-m} J^3_{\thetaa,m + 1} \ri \right] \\
& + \frac{1}{6}  \sum_{m \leq -1} (-(m + \thetaa) J^+_{\thetaa,m+1} J^-_{\thetaa,-1-m} + (1+m + \thetaa) J^+_{\thetaa,m} J^-_{\thetaa,-m} ) \, \\
&+\frac{1}{6}\sum_{m \geq 0} ((1+m+\thetaa) J^-_{\thetaa,-m} J^+_{\thetaa,m} -(m + \thetaa) J^-_{\thetaa,-1-m} J^+_{\thetaa,m+1} )\,  \\
&+\frac{1}{6}  \sum_{m \leq -1} ((-m+\thetaa) J^-_{\thetaa,m+1} J^+_{\thetaa,-1-m} + (1+m-\thetaa) J^-_{\thetaa,m} J^+_{\thetaa,-m} ) \,\\
&+\frac{1}{6} \sum_{m \geq 0} ((1+m - \thetaa) J^+_{\thetaa,-m} J^-_{\thetaa,m} - (m - \thetaa) J^+_{\thetaa,-1-m} J^-_{\thetaa,m+1} ) -\frac{2}{3} \thetaa J^3_{\thetaa,0}  \\
 =\,& \frac{2}{3} \left( \sum_{m \leq -1} J^3_{\thetaa,m} J^3_{\thetaa,-m} +  \sum_{m \geq 0} J^3_{\thetaa,-m} J^3_{\thetaa,m}\right) +
\\
&\frac{1}{3}\left(\sum_{m \leq -1} J^+_{\thetaa,m} J^-_{\thetaa,-m} + \sum_{m \geq 0} J^-_{\thetaa,-m} J^+_{\thetaa,m} + \sum_{m \leq -1} J^-_{\thetaa,m} J^+_{\thetaa,-m} + \sum_{m \geq 0} J^+_{\thetaa,-m} J^-_{\thetaa,m}  \right)   - \frac{\thetaa}{3} [J^+_{\thetaa,0}, J^-_{\thetaa,0}]-\frac{2}{3} \thetaa J^3_{\thetaa,0} \\
=\,&2L_{\thetaa,0}-2A - \frac{\thetaa^2}{3} \,,
\fe
}
which leads to $A = - \frac{1}{6}\thetaa^2$.

\bibliographystyle{JHEP}
\bibliography{refs}
 \end{document}